%% file: main.tex
\documentclass[twocolumn]{aastex631}

\usepackage{bm}
\usepackage{tikz}
\usepackage{CJK}
\usepackage{mathtools}
\usepackage{mhchem}
\usepackage{xcolor}

\shorttitle{Optimizing ANN Emulators for Galaxy SED Fitting}
\shortauthors{Mathews et al.}

\graphicspath{{./}{figures/}}

\begin{document}
\begin{CJK*}{UTF8}{gbsn}

\title{As Simple as Possible but No Simpler:\\Optimizing the Performance of Neural Net Emulators for Galaxy SED Fitting}


\include{authors.tex}

\correspondingauthor{Elijah P. Mathews}
\email{apj@elijahmathews.com}


\begin{abstract}

Artificial neural network emulators have been demonstrated to be a very computationally efficient method to rapidly generate galaxy spectral energy distributions (SEDs), for parameter inference or otherwise. Using a highly flexible and fast mathematical structure, they can learn the nontrivial relationship between input galaxy parameters and output observables. However, they do so imperfectly, and small errors in flux prediction can yield large differences in recovered parameters. In this work, we investigate the relationship between an emulator's execution time, uncertainties, correlated errors, and ability to recover accurate posteriors. We show that emulators can recover consistent results to traditional fits, with precision of $25\!-\!40\%$ in posterior medians for stellar mass, stellar metallicity, star formation rate, and stellar age. We find that emulation uncertainties scale with an emulator's width $N$ as $\propto N^{-1}$ while execution time scales as $\propto N^2$, resulting in an inherent tradeoff between execution time and emulation uncertainties. We also find that emulators with uncertainties smaller than observational uncertaities are able to recover accurate posteriors for most parameters without a significant increase in catastrophic outliers. Furthermore, we demonstrate that small architectures can produce flux residuals that have significant correlations, which can create dangerous systematic errors in colors. Finally, we show that the distributions chosen for generating training sets can have a large effect on emulators' ability to accurately fit rare objects. Selecting the optimal architecture and training set for an emulator will minimize the computational requirements for fitting near-future large-scale galaxy surveys.

\end{abstract}


\section{Introduction} \label{sec:intro}

The study of galaxy formation aims to answer questions such as how galaxies evolve and are structured, and how and when they form their stars. To answer these questions, we collect measurements of the light emitted by them in the form of photometry and/or spectra, which collectively is a galaxy's spectral energy distribution (SED). However, due to the complex physics that govern a galaxy's emission of light, it can often be very difficult to properly interpret these observations \citep{walcher11,conroy13}.

In order to interpret observed galaxy SEDs, we first require stellar population synthesis (SPS) methods that can model and predict a galaxy's SED for a given set of physical parameters. In the simplest cases, these can be a dedicated stellar population spectral library \citep[e.g.][]{bc03,maraston05}, or alternatively a joint combination of a stellar spectral library \citep[e.g. MILES,][]{sanchezblazquez06} and a stellar isochrone library \citep[e.g. MIST,][]{dotter16,choi16} from which an SED for a simple or composite stellar population (SSP or CSP, respectively) can be predicted assuming a given initial mass function \citep[IMF, e.g.][]{salpeter55,kroupa01,chabrier03}, dust attenuation law \citep[e.g.][]{calzetti00}, and star formation history (SFH).\footnote{This is not an exhaustive list of galaxy properties that influence a galaxy's SED. Refer to the \citet{conroy13} review article for a more in-depth discussion on the topic of SPS.} From these, physical parameters can be inferred using techniques such as $\chi^2$-minimization or maximum likelihood estimation. Recent codes, such as \texttt{FSPS} \citep{conroy09,conroy10}, also allow modifications of stellar physics (e.g. changes to horizontal branch morphology) or the addition of other expected contributions to galaxy SEDs such as active galactic nuclei (AGN), such as in the case of \texttt{MAGPHYS} \citep{dacunha08}, \texttt{BayeSED} \citep{han12,han14,han19}, and \texttt{X-CIGALE} \citep{yang20}.

SPS codes like \texttt{FSPS} typically require about $\sim\! 10-100\ \textrm{ms}$ to compute a prediction for a galaxy SED depending on the amount of features used in the prediction.\footnote{CPU benchmarks throughout based on an AMD Ryzen 7 2700X CPU.} Furthermore, new fitting codes have been created that combine SPS routines with more advanced fitting techniques, often combining SPS with sampling routines to perform Bayesian inference on galaxy SEDs, leading to an ensemble of possible solutions (i.e. posteriors) for a galaxy's physical properties as opposed to only providing a single best-fit model. Some Bayesian SED-fitting codes employing MCMC sampling include \texttt{CIGALE} and its X-ray module \texttt{X-CIGALE} \citep{burgarella05,noll09,boquien19,yang20}, \texttt{BayeSED} \citep{han12,han14,han19}, \texttt{Prospector} \citep{leja17,johnson21b}, \texttt{Bagpipes} \citep{carnall18,carnall19}, and \texttt{MCSED} \citep{bowman20a,bowman20b}. Generating an estimate of the posterior typically requires a very large number of executions of the SPS code, of order $10^5 - 10^6$ for photometric SEDs \citep{leja19}.\footnote{This work focuses on addressing timing issues that face on-the-fly SED-fitting codes like \texttt{Prospector}. Grid-based SED-fitting codes perform better in this regard by utilizing pre-computed grids, but face multiplicative issues when adding more parameters (i.e. more grid axes) to their physical model.}

The $\sim\!10-100\ \textrm{ms}$ execution time for SPS actually presents a significant problem with the enormous amount of data we are soon to be faced with: it is too slow. In fact, going forward, these existing SED-fitting procedures are \emph{orders of magnitude} too slow to feasibly fit large catalogs of SEDs with Bayesian SED-fitting codes. This problem is most obvious when fitting photometric surveys, which has been shown to consume about $\sim\! 10-100$ CPU hours per galaxy fit for a 14-parameter physical model \citep[e.g.][]{leja19}. Within the next few years, the Vera C.\ Rubin Observatory (VRO) will begin science operations and initiate its Legacy Survey of Space and Time \citep[LSST,][]{ivezic19}, which is expected to deliver photometry for $\sim\! 10^{10}$ galaxies. This alone would overwhelm current computational resources, as fitting the full sample would require on the order of $\sim\! 10^{11}$ CPU hours, equivalent to a $\sim\!\$10\ \textrm{billion USD}$\footnote{Assuming a typical cost per CPU hour of $\$0.10\ \textrm{USD}$ and that each object is only fit \emph{once}, which is unlikely to be the case as additional fits will likely be desired to determine how new VRO observations of each object further constrain objects' physical parameters.} investment, an amount that is comparable to the cost of the James Webb Space Telescope \citep[JWST,][]{witze18}. The environmental impact would also be significant, producing an estimated $\sim\!10^8\ \textrm{kg}\ \ce{CO2}$ with current electrical energy sources, comparable to the emissions of a modern widebody aircraft operating continuously for $\sim\!200\ \textrm{days}$.\footnote{Environmental impact based on an assumed $\sim\!1\ \textrm{g}\ \ce{CO2}$ equivalent emissions per CPU hour. Aircraft comparison based on fuel capacity and range data provided by \citet{boeing21} for the Boeing 787-9, an assumed cruise speed of $900\ \textrm{km}\ \textrm{hr}^{-1}$, and an assumed $10\ \textrm{kg}\ \ce{CO2}$ equivalent emissions per US gallon of jet fuel burned.} Furthermore, spectroscopic surveys will be even more computationally expensive to fit than photometric surveys of similar size due to their increased spectral resolution over multi-band photometry and their need for significantly more likelihood function calls when fitting \citep[e.g. $\sim\!100$ hours per fit,][]{tacchella22}, with several large spectroscopic surveys being either planned or underway. For example, the zCOSMOS survey \citep{lilly07,lilly09,knobel12} provides spectra for $\sim\!10^4$ galaxies at $z < 1$. A sample of this size would be difficult enough to fit on its own, but ever larger spectroscopic surveys are on the way, including the Prime Focus Spectroscopic Galaxy Evolution Survey \citep[PFS,][]{greene22}, Hobby-Eberly Telescope Dark Energy Experiment \citep[HETDEX,][]{hill04}, and the Dark Energy Spectroscopic Instrument \citep[DESI,][]{desi16}, which will deliver spectra for approximately $10^5$, $10^6$, and $10^7$ galaxies, respectively.

Thus, faster SED-fitting methods are clearly needed in order to fit large samples of galaxy SEDs. One such potential solution is to replace existing SPS routines altogether with an artificial neural network (ANN) that is trained to mimic an SPS code's behavior. ANNs are highly flexible mathematical structures that can mimic the behavior of arbitrarily complex functions (i.e. behaving as a highly nonlinear regression), both in their arbitrary width \citep{cybenko89,hornik91} and arbitrary depth \citep{zhou17,hanin17,kidger19} cases. Crucially, ANNs are typically very fast to execute, $\sim10^{3}-10^{4}$ times faster than SPS routines, with execution times for typical ANN architectures of $1-100\ \mu\textrm{s}$. ANNs have been used to model highly nonlinear relationships in astrophysics for several decades \citep[e.g. to predict a galaxy's morphological type based on observations,][]{storrielombardi92}. They were first used to emulate SPS by \citet{han12} in the \texttt{BayeSED} SED-fitting code, which used an ANN with a single 20-node hidden layer to emulate a $5-6$ physical parameter model. \citet{alsing20} further developed on ANN SPS emulation in the \texttt{Speculator} code, which uses wider and deeper ANNs to produce photometric and spectroscopic \texttt{FSPS} emulators for the 14-parameter Prospector-$\alpha$ physical model as used by \citet{leja19}.

However, it should be noted that while SPS routines are typically \emph{at worst} interpolations of their underlying physical models (and are thus guaranteed to provide sensible output based on their underlying physics), SPS emulators are \emph{at best} approximations of these SPS routines, providing no guarantee to the user of supplying sufficiently accurate or precise output for a given set of inputs. In this sense, they produce flux uncertainties that are of order $< 0.01 - 0.1$ mag depending on their ANN architecture, training set, and optimization hyperparameters. These inaccuracies in flux predictions, even if they are small and similar to assumed systematic uncertainties in the SPS methods (e.g. $\sim\!5\%$), could then result in highly incorrect posterior distributions when used to infer galaxy properties due to the highly complicated and degenerate nature of likelihood space, along with incorrect inferred colors if they vary as a function of wavelength. Furthermore, due to the black-box nature of ANNs, it may be extremely difficult to know when an emulator's rare but catastrophic errors lead to these incorrect solutions without comparing to a fit using the original SPS code it replaces (which voids the purpose of using the emulator). This effect has not yet been thoroughly investigated in the literature; \citet{kwon22} found that spectroscopic emulators have the ability to recover accurate posteriors, but fit only a small sample of $\sim\!100$ mock galaxy SEDs and are not sensitive to rare, catastrophic (e.g. $\geq 3\sigma$) outliers in posterior estimation.

In this work, we use a similar technique as \citet{alsing20} to develop a set of photometric emulators that replace an SPS code for an 18-parameter physical model, which we then use to fit a large sample of $\sim\!10^4$ photometric galaxy SEDs from the 3D-HST photometric catalog \citep{brammer12,skelton14}. In doing so, we investigate the performance of each emulator, including the speed at which an emulator can produce a predicted SED and the accuracy and precision in which it is able to do so. We then examine their ability to recover accurate posterior distributions when combined with an MCMC-based SED-fitting code, and compare the rate of extreme posterior discrepancies. Using this information, we develop a method for selecting a suitable architecture for a given SED-fitting task that minimizes emulation execution time while not sacrificing posterior accuracy. Section \ref{sec:methods} describes in detail the codes which our emulators are trained to mimic, the data we train them on, the ANN architectures used, and the methods we use to fit SEDs with them, while Section \ref{sec:results} displays the performance of our emulators. Section \ref{sec:discuss} includes the discussion, before stating our conclusions in Section \ref{sec:conclusion}. Throughout the work, we assume a flat $\Lambda$CDM cosmology using the \citet{hinshaw13} parameter values.\footnote{$h = 0.6932$, $\Omega_{\textrm{m},0} = 0.2865$, $\Omega_{\textrm{b},0} = 0.04628$, $T_{\textrm{CMB},0} = 2.725\ \textrm{K}$, $N_\textrm{eff}=3.04$, $m_\nu = 0\ \textrm{eV}/c^2$.} The emulators trained in this work along with the code used to fit SEDs with them is available on GitHub.\footnote{\href{https://www.github.com/elijahmathews/MathewsEtAl2023}{https://www.github.com/elijahmathews/MathewsEtAl2023}}

\section{Methods and Data}\label{sec:methods}

In this section, we detail the methods we used in this work to develop SPS emulators, fit galaxy SEDs using them, and quantify their performance. In Section \ref{sec:sps}, we describe the generation of our model spectra (from which photometry that we will train on can be computed), along with the chosen physical model. Then, in Section \ref{sec:emul}, we describe the methods used for our ANN emulation, including information about the networks' architectures and training procedures. Finally, in Section \ref{sec:photdata}, we describe the photometric data that we will be fitting as a test for the emulators, in addition to the specifics of the fitting routine used.

\subsection{Stellar Population Synthesis}\label{sec:sps}

To generate galaxy SEDs, we use \texttt{FSPS}\footnote{The version of \texttt{FSPS} used in this work was the state of the Git repository at commit \href{https://github.com/cconroy20/fsps/commit/5327f501ee71c8e40aa00d322d84d65294936946}{\texttt{5327f501ee71c8e40aa00d322d84d65294936946}}.} \citep{conroy09,conroy10}, accessed using the \texttt{python-fsps}\footnote{The version of \texttt{python-fsps} used in this work was the state of the Git repository at commit \href{https://github.com/dfm/python-fsps/commit/d689181f6dd609d64f3c22b63be4520be017a657}{\texttt{d689181f6dd609d64f3c22b63be4520be017a657}}.} Python package \citep{johnson21c}. Within \texttt{FSPS}, we select the MIST isochrones \citep{dotter16,choi16} and the MILES spectral library \citep{sanchezblazquez06}.

We begin our physical model with the Prospector-$\alpha$ model assumed by \citet{leja19}, consisting of 14 parameters quantifying a galaxy's mass, stellar and gas-phase metallicity, SFH, dust properties, and AGN contribution. We then make a handful of modifications to the fiducial Prospector-$\alpha$ model. First, we extend the model's lower limit on the cumulative stellar mass formed to $10^{6}\ M_\odot$ to allow for lower mass solutions when fitting objects at very low redshift. Furthermore, as we will be fitting photometry that has coverage in the rest-frame mid-IR, we allow 3 parameters governing the SED from thermal dust emission \citep[i.e.][]{draine07} to vary (whereas Prospector-$\alpha$ holds these 3 parameters to fixed values), which includes the PAH mass fraction $Q_\textrm{PAH}$, minimum radiation field strength for dust emission $U_\textrm{min}$, and the fraction $\gamma$ of starlight exposed to radiation field strengths $U_\textrm{min} < U \leq U_\textrm{max}$ (where $U_\textrm{max}$ is fixed). Finally, while redshift will remain a fixed parameter for the vast majority of the fits we will perform, each galaxy requires a different fixed redshift. Thus, redshift must also be included in the emulator, although it is a particularly difficult parameter to emulate in photometric emulators and may warrant special treatment in the future (see Section \ref{sec:zred} for further discussion). These collectively assemble an 18 parameter physical model, where the 18 parameters will be used as inputs to our ANN emulators. The 18 parameters, in addition to the distributions used for generating training sets (and test sets) and the distributions used as priors when fitting to photometry, are listed in Table \ref{tab:physmod}.

\begin{deluxetable*}{p{4cm}p{2cm}p{5cm}p{5cm}}\label{tab:physmod}
\centerwidetable
\tabletypesize{\scriptsize}
\tablecaption{Physical Parameters}
\tablehead{\colhead{Parameter} & \colhead{Symbol} & \colhead{Train/Test Distribution} & \colhead{Prior Distribution}}
\startdata
Cumulative stellar mass formed & $\log_{10}M_{*,\textrm{formed}}$ & $\textrm{Uniform}\,(6, 12.5)$ & $\textrm{Uniform}\,(6, 12.5)$ \\
Stellar metallicity & $\log_{10}(Z_*/Z_\odot)$ & $\textrm{Uniform}\,(-1.98, 0.19)$ & $\textrm{Normal}\,(\mu,\sigma)$ with $\mu$ and $\sigma$ from \citet{gallazzi05} mass-metallicity relation, truncated to $\left[-1.98, 0.19\right]$ \\
Adjacent SFR ratios (6 parameters) & $\log_{10}\left(\frac{\textrm{SFR}_i}{\textrm{SFR}_{i+1}}\right)$ & $t(2)$, scaled to $\mu=0$ and $\sigma=3$, truncated to $\left[-5, 5\right]$ & $t(2)$, scaled to $\mu=0$ and $\sigma=0.3$, truncated to $\left[-5, 5\right]$ \\
Diffuse dust optical depth & $\tau_2$ & $\textrm{Normal}\,(0.3, 3)$, truncated to $\left[0,4\right]$ & $\textrm{Normal}\,(0.3, 1)$, truncated to $\left[0,4\right]$ \\
Power law modifier for \citet{calzetti00} dust attenuation law & $n$ & $\textrm{Uniform}\,(-1.2,0.4)$ & $\textrm{Uniform}\,(-1.2,0.4)$ \\
Ratio of birth cloud dust to diffuse dust & $\tau_1/\tau_2$ & $\textrm{Normal}\,(1,1)$, truncated to $\left[0,2\right]$ & $\textrm{Normal}\,(1,0.3)$, truncated to $\left[0,2\right]$ \\
Ratio of AGN luminosity to bolometric luminosity & $\log_{10}{f_\textrm{AGN}}$ & $\textrm{Uniform}\,(-5, \log_{10}{3})$ & $\textrm{Uniform}\,(-5, \log_{10}{3})$ \\
AGN torus dust optical depth & $\log_{10}{\tau_\textrm{AGN}}$ & $\textrm{Uniform}\,(\log_{10}{5}, \log_{10}{150})$ & $\textrm{Uniform}\,(\log_{10}{5}, \log_{10}{150})$ \\
Gas-phase metallicity & $\log_{10}(Z_\textrm{gas}/Z_\odot)$ & $\textrm{Uniform}\,(-2.0, 0.5)$ & $\textrm{Uniform}\,(-2.0, 0.5)$ \\
Redshift & $z$ & $\textrm{Uniform}\,(0.5, 3)$ & Fixed to each object's $z_\textrm{best}$ from \citet{momcheva16} \\
PAH mass fraction & $Q_\textrm{PAH}$ & $\textrm{Normal}\,(2,4)$, truncated to $\left[0,7\right]$ & $\textrm{Normal}\,(2,2)$, truncated to $\left[0,7\right]$ \\
Minimum radiation field for dust emission & $U_\textrm{min}$ & $\textrm{Normal}\,(1,20)$, truncated to $\left[0.1, 25\right]$ & $\textrm{Normal}\,(1,10)$, truncated to $\left[0.1, 25\right]$ \\
Fraction of dust mass exposed to $U_\textrm{min}$ & $\log_{10}{\gamma_\textrm{e}}$ & $\textrm{Normal}\,(-2,2)$, truncated to $\left[-4,0\right]$ & $\textrm{Normal}\,(-2,1)$, truncated to $\left[-4,0\right]$
\enddata
\end{deluxetable*}

We then select a set of 137 photometric filters, whose spectral flux densities will collectively be the outputs of the emulator. We base our initial filter set on the filters necessary for fitting SEDs from the 3D-HST photometric catalog \citep{brammer12,skelton14}, which we then extend by adding additional filters, including filters from 2MASS, GALEX, Herschel, JWST, SDSS, VISTA, and WISE. Since the filters used in the 3D-HST catalog are a subset of our full filter set, only a subset of our emulator's outputs will actually be used when fitting SEDs in this work. All of these aforementioned filters are included in \texttt{sedpy}\footnote{The version of \texttt{sedpy} used in this work was v0.3.0.} \citep{johnson21a}, a Python package for dealing with spectral fluxes and photometric filters.

Natively, \texttt{sedpy} outputs the spectral flux density observed through each filter in units of maggies; however, due to the fact that the fluxes covered by this physical model span $\sim\!40$ orders of magnitude, we elect to put these on a logarithmic scale to provide the emulator with a simpler output, a method that has also been employed in previous works. The natural choice for doing so would be to use the AB magnitude system as originally defined by \citet{oke83}; however, for this particular problem, we encounter an issue with the outputs of \texttt{FSPS}; namely, \texttt{FSPS} tends to bottom out in the UV at very small values of spectral flux density due to strong IGM absorption, with $m_\textrm{AB} > +100$ in some cases. With few exceptions, spectral flux densities that are fainter than about $m_\textrm{AB} \approx +35$ exceed the capabilities of most of the detectors in use today.

\begin{figure}
    \plotone{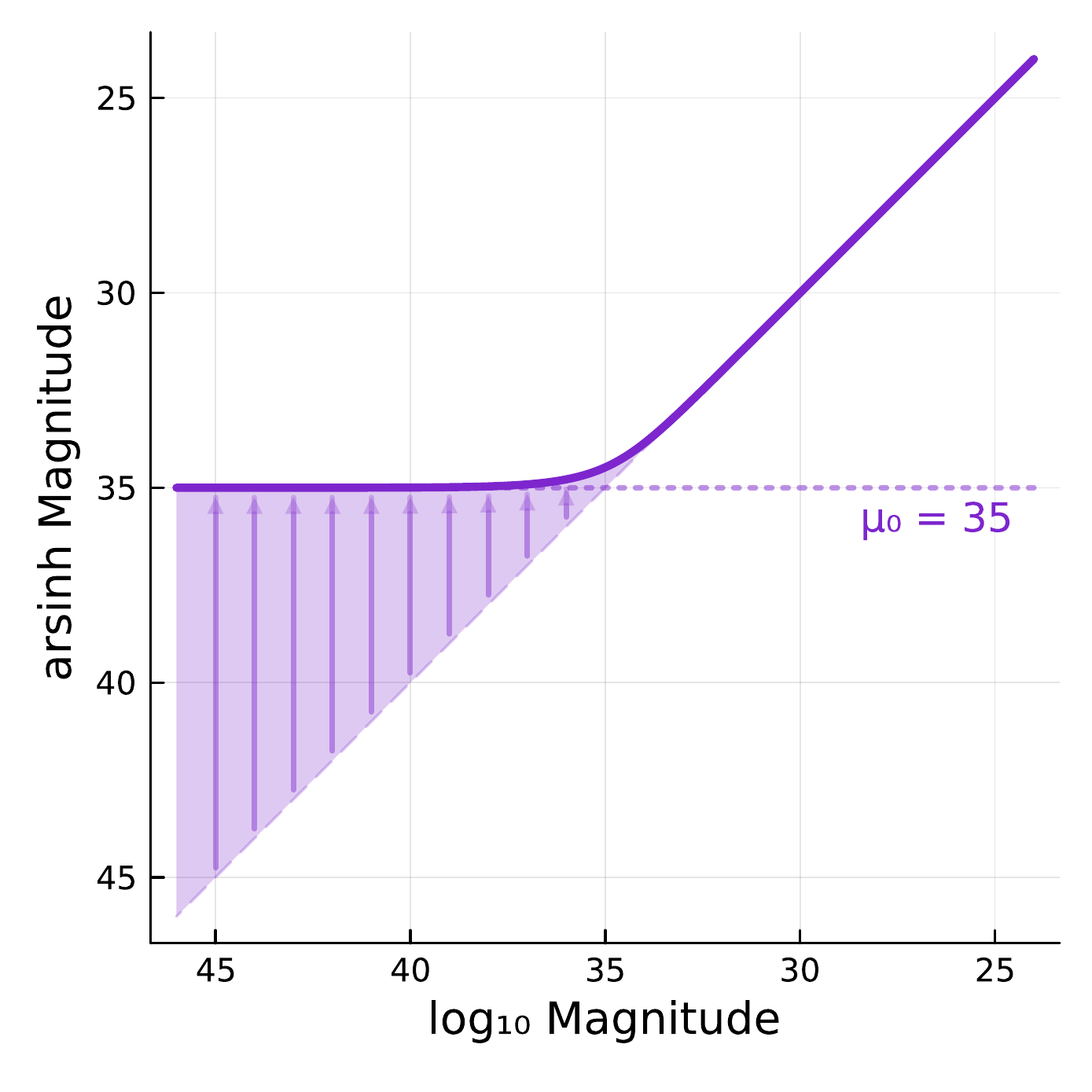}
    \caption{The difference in behavior between ordinary logarithmic magnitudes (e.g. AB magnitudes) and the $\textrm{arsinh}$ magnitude system used in the emulators in this work. For logarithmic magnitudes significantly brighter than $\mu_0$, the $\textrm{arsinh}$ system behaves identically to the logarithmic system. However, for logarithmic magnitudes $\geq \mu_0$, the $\textrm{arsinh}$ system asymptotically approaches a constant value $\mu_0$ (equal to $35$ in this work).}
    \label{fig:arsinh}
\end{figure}

The simplest way to solve this would be to raise all extremely faint spectral flux densities in our training set to a minimum spectral flux density of $m_\textrm{AB} = +35$, such that
\begin{equation}
    m_\textrm{AB,mod}(m_\textrm{AB}) = \begin{cases}
    m_\textrm{AB} & m_\textrm{AB} \leq +35 \\
    +35 & m_\textrm{AB} > +35 \\
    \end{cases}.
\end{equation}
However, with this method, a discontinuity is introduced into the derivative of the emulator's output. Preserving an emulator's ability to be continuously differentiable is desirable to leave open the possibility of combining an emulator with a gradient-based sampler in the future (see Section \ref{sec:gradsamp} for further discussion). Thus, we choose to avoid this differentiability issue by instead using an inverse hyperbolic sine (i.e. $\textrm{arsinh}$) magnitude system that is nearly identical to the one proposed by \citet{lupton99}, where for an observed spectral flux density $E_{\textrm{e},\nu}$, one can define an ``AB'' $\textrm{arsinh}$ magnitude $\mu_\textrm{AB}$ as
\begin{equation}
    \mu_\textrm{AB}(E_{\textrm{e},\nu}; a, \mu_0) = -a\,\textrm{arsinh}\left[\frac{E_{\textrm{e},\nu}}{2}\exp{\left(\frac{\mu_0}{a}\right)}\right] + \mu_0,
\end{equation}
where $a$ is set to $2.5\log_{10}e$ to preserve the same scaling at high values of spectral flux density as the ordinary AB magnitude system and $\mu_0$ is a parameter that sets the location of the low-end ``cutoff'' in this magnitude system, which we set to $\mu_0 = +35$ in this work. The behavior of this magnitude system is shown in Figure \ref{fig:arsinh}. In the bright limit, this magnitude system asymptotically approximates the ordinary $m_\textrm{AB}$ magnitude system as one would expect, while in the faint regime, the system approaches a constant value of $\mu_\textrm{AB} = +35$ regardless of the actual input spectral flux density. However, importantly, it does so smoothly and with a continuous derivative, thus preserving continuous differentiability.

\subsection{Artificial Neural Network Emulator}\label{sec:emul}

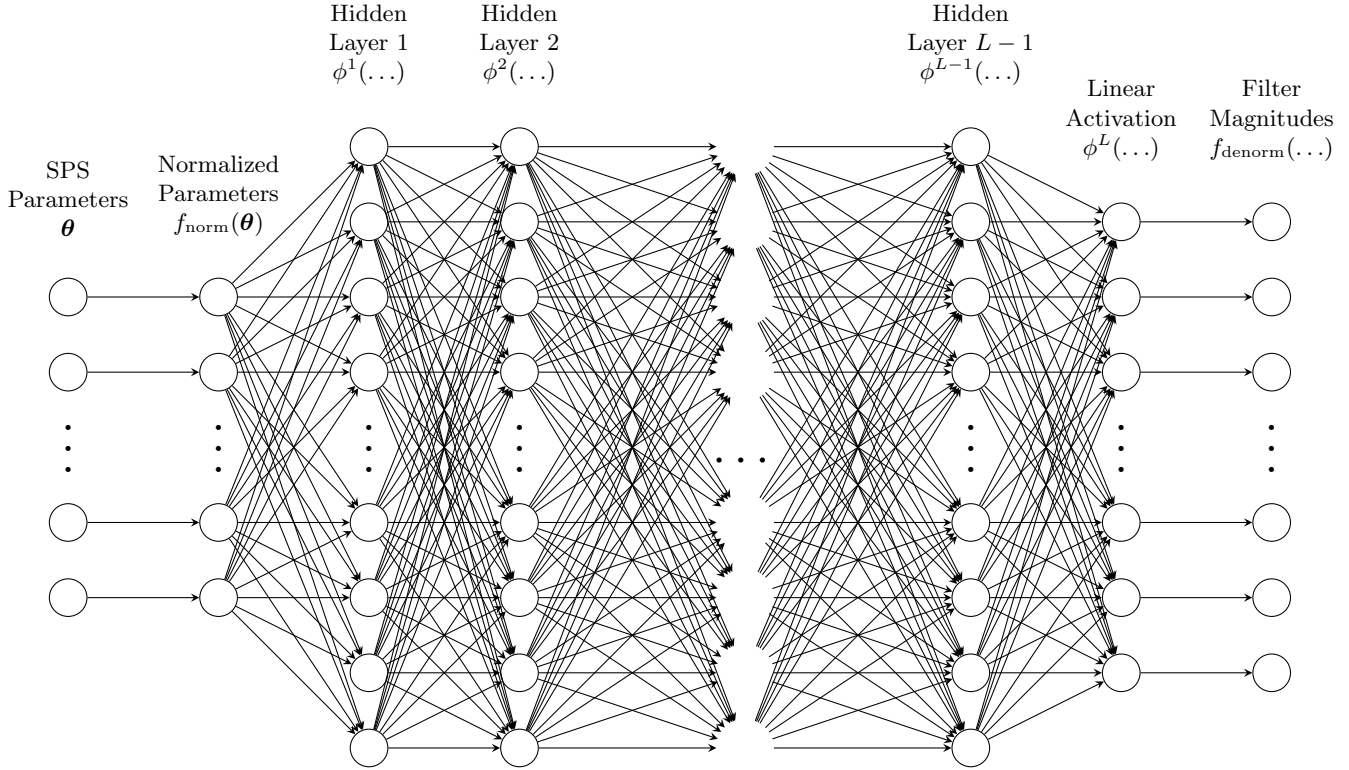
\begin{figure*}[htb]
    \centering
    \tikzset{%
        every neuron/.style={
            circle,
            draw,
            minimum size=0.5cm
        },
        every missneuron/.style={
            circle,
            draw=none,
            minimum size=0.75cm
        },
        neuron missing/.style={
            draw=none, 
            scale=2,
            text height=0.333cm,
            execute at begin node=$\vdots$
        },
        missneuron missing/.style={
            draw=none, 
            scale=2,
            text height=0.333cm,
            execute at begin node=$\cdots$
        },
    }
    \begin{tikzpicture}[x=2.0cm, y=1.0cm, >=stealth]
        \foreach \m/\l [count=\y] in {1,2,missing,3,4}
            \node [every neuron/.try, neuron \m/.try] (input-\m) at (0,2.5-\y) {};
        \foreach \m/\l [count=\y] in {1,2,missing,3,4}
            \node [every neuron/.try, neuron \m/.try] (inputnorm-\m) at (1,2.5-\y) {};
        \foreach \m [count=\y] in {1,2,3,4,missing,5,6,7,8}
            \node [every neuron/.try, neuron \m/.try ] (hiddenone-\m) at (2,4.5-\y) {};
        \foreach \m [count=\y] in {1,2,3,4,missing,5,6,7,8}
            \node [every neuron/.try, neuron \m/.try ] (hiddentwo-\m) at (3,4.5-\y) {};
        \foreach \m [count=\y] in {1,2,3,4,missing,5,6,7,8}
            \node [every missneuron/.try, missneuron \m/.try ] (hiddenmid-\m) at (4.5,4.5-\y) {};
        \foreach \m [count=\y] in {1,2,3,4,missing,5,6,7,8}
            \node [every neuron/.try, neuron \m/.try ] (hiddenpone-\m) at (6,4.5-\y) {};
        \foreach \m [count=\y] in {1,2,3,missing,4,5,6}
            \node [every neuron/.try, neuron \m/.try ] (hiddenp-\m) at (7,3.5-\y) {};
        \foreach \m [count=\y] in {1,2,3,missing,4,5,6}
            \node [every neuron/.try, neuron \m/.try ] (sps-\m) at (8,3.5-\y) {};
        \foreach \i in {1,...,4}
            \draw [->] (input-\i) -- (inputnorm-\i);
        \foreach \i in {1,...,4}
            \foreach \j in {1,...,8}
                \draw [->] (inputnorm-\i) -- (hiddenone-\j);
        \foreach \i in {1,...,8}
            \foreach \j in {1,...,8}
                \draw [->] (hiddenone-\i) -- (hiddentwo-\j);
        \foreach \i in {1,...,8}
            \foreach \j in {1,...,8}
                \draw [->] (hiddentwo-\i) -- (hiddenmid-\j);
        \foreach \i in {1,...,8}
            \foreach \j in {1,...,8}
                \draw [->] (hiddenmid-\i) -- (hiddenpone-\j);
        \foreach \i in {1,...,8}
            \foreach \j in {1,...,6}
                \draw [->] (hiddenpone-\i) -- (hiddenp-\j);
        \foreach \i in {1,...,6}
            \draw [->] (hiddenp-\i) -- (sps-\i);
        \foreach \l [count=\x from 0] in {SPS\\Parameters\\$\boldsymbol{\theta}$}
            \node [align=center, above] at (\x,2.2) {\l};
        \foreach \l [count=\x from 0] in {Normalized\\Parameters\\$f_{\textrm{norm}}(\boldsymbol{\theta})$}
            \node [align=center, above] at (1+\x,2.2) {\l};
        \foreach \l [count=\x from 0] in {Hidden\\Layer 1\\$\phi^1(\ldots)$}
            \node [align=center, above] at (2+\x,4.2) {\l};
        \foreach \l [count=\x from 0] in {Hidden\\Layer 2\\$\phi^2(\ldots)$}
            \node [align=center, above] at (3+\x,4.2) {\l};
        \foreach \l [count=\x from 0] in {Hidden\\Layer $L-1$\\$\phi^{L-1}(\ldots)$}
            \node [align=center, above] at (6+\x,4.2) {\l};
        \foreach \l [count=\x from 0] in {Linear\\Activation\\$\phi^{L}(\ldots)$}
            \node [align=center, above] at (7+\x,3.2) {\l};
        \foreach \l [count=\x from 0] in {Filter\\Magnitudes\\$f_\textrm{denorm}(\ldots)$}
            \node [align=center, above] at (8+\x,3.2) {\l};
    \end{tikzpicture}
    \caption{Illustration of the neural network architecture used in this work. At the far left, a vector $\boldsymbol{\theta}$ of physical parameters is taken as input, from which a precomputed mean is subtracted and standard deviation divided to normalize the inputs. Then, a set of $L-1$ hidden layers (where $L = 6$ for the entirety of this work), each of which has $N$ fully-connected nodes using the GELU activation function as described in Section \ref{sec:emul}. The final fully-connected layer, layer $L$, has linear activation and has a number of nodes equal to the number of predicted photometric filters (137 in this work). Finally, the output has a precomputed mean re-added (the precomputed standard deviation is fixed to $1$), which represents the predicted photometry.}
    \label{fig:nn}
\end{figure*}

Choosing the proper emulator architecture for a particular SED-fitting problem is one of the central questions this work is seeking to address. Put simply, our goal is to provide an SPS alternative that produces precise photometry estimates while requiring as little execution time as possible. Emulator architectures that are more complex, which provide for the ability to fit more complex input-output relationships more precisely, necessarily require longer execution times in comparison to simpler architectures. Thus, there is a key balance to be had -- ideally, we would like to choose the \emph{simplest} (i.e. fastest) possible emulator architecture that provides sufficiently accurate results (i.e. posteriors) when used for fitting SEDs. Choosing an overly complex emulator architecture may be counterproductive if that additional complexity is not also adding a notable improvement in the recovery of posteriors, which is an extra layer of nonlinear transformations beyond simply reproducing the spectral flux density accurately.

Following previous work \citep[i.e.][]{han12,han14,alsing20}, the emulation method selected for this work is a multilayer perceptron (MLP) ANN. In the context of emulation, MLP ANNs have a very desirable runtime advantage over other machine learning techniques, such as the $k$-nearest neighbors algorithm and Gaussian processes, in that they are incredibly fast to call and have a small memory requirement (whereas other methods have memory footprints and execution times that can scale with the size of the training set). An MLP is constructed of $L$ trainable layers, where layer $i$ contains $N_i$ nodes which each use an activation function $\phi^i(\mathbf{x})$ (which generally operates element-wise). The MLP requires a set of weight matrices $\mathbf{W}^i$ and bias vectors $\mathbf{b}^i$ for each layer, such that evaluating
\begin{equation}
    \mathbf{x}^i = \phi^i\left(\mathbf{W}^i\mathbf{x}^{i-1} + \mathbf{b}^{i}\right)
\end{equation}
for each layer produces a useful output from the last layer, given an input vector $\mathbf{x}^0$ (which in this case is our 18 parameters). For the purposes of this work, we use a total of $L=6$ layers for our emulator, where the first $5$ layers have $N$ nodes per layer and use a nonlinear activation function. The final layer consists of $137$ nodes to match the $137$ bands of photometry to be predicted (see Section \ref{sec:sps}) and uses linear activation (i.e. $f(\mathbf{x}) = \mathbf{x}$). The full architecture described in this section is illustrated in Figure \ref{fig:nn}.

To find where the balance between ANN architecture and emulator precision occurs, we train a total of $6$ ANN emulators with varying numbers of nodes per layer, $N$:
\begin{equation}
    N = 32, 64, 128, 256, 512, 1024.
\end{equation}
As is demonstrated in Section \ref{sec:apriori}, this selection of $N$ provides a wide range in emulator accuracies (i.e. an emulator's typical bias with respect to the true \texttt{FSPS} predictions) and precisions (i.e. the typical scale of the differences between an emulator's predictions and the true \texttt{FSPS} predictions) to gauge the dependence of posterior quality on emulator precision. All other hyperparameters relating to ANN architecture and training procedure (e.g. the depth of the network $L$, training set size, activation function, loss function) are fixed.

We use the Gaussian Error Linear Unit \citep[GELU,][]{hendrycks16} activation function for the nonlinear layers in our emulator, which we found to produce the best balance between execution time and emulation precision. It is formally defined as
\begin{equation}\label{eq:geluexact}
    \textrm{GELU}(x) \equiv \frac{x}{2}\left[1 + \textrm{erf}\,{\left(\frac{x}{\sqrt{2}}\right)}\right].
\end{equation}
However, as there is no closed-form expression for $\textrm{erf}(x)$, one must approximate it. We elect to use the ``tanh'' approximation,
\begin{equation}\label{eq:geluapprox}
    \textrm{GELU}(x) \approx \frac{x}{2}\left(1 + \tanh\left[\sqrt{\frac{2}{\pi}}\left(x + 0.044715x^3\right)\right]\right),
\end{equation}
which can be accessed using the \texttt{jax.nn.gelu(x, approximate=True)} function within the \texttt{JAX} \citep{bradbury18} numerical library in Python or the \texttt{Flux.gelu(x)} function in the \texttt{Flux} numerical library in Julia.

In addition to this ordinary ANN architecture, special layers are added to the input and output layers of the emulator to normalize and denormalize the inputs and outputs using precomputed means and standard deviations of the training set parameters and photometry. Specifically, a normalization layer is added to the input layer, which subtracts the precomputed mean to center the data around zero (where floating-point precision is highest) and divides by the precomputed standard deviation to force each parameter to have approximately the same scale. Finally, a denormalization layer is added to the output which performs the opposite operations on the output photometry with the exception being that the precomputed standard deviations are all set to one to ensure each filter is weighted equally in the loss function.

In order to determine the optimal set of weight matrices $\textbf{W}^i$ and bias vectors $\textbf{b}^i$ for a given emulator, one must provide the emulator with a large set of input/output pairs so it can learn how the pairs are related. In this work, we generate a set of $10^7$ parameter-SED pairs for training. For comparison, \citet{alsing20} chose to use $10^6$ training samples; we chose to increase the number of training samples due to the additional parameters that exist in our model. While not done in this work, a useful avenue of inquiry would be to determine the ideal training set size for these types of emulators -- larger training set sizes grant an emulator more examples to hone their accuracy and precision on, but at some point face diminishing returns. Parameter vectors are sampled from the distributions listed in Table \ref{tab:physmod}. Similar to \citet{alsing20}, we base these distributions on the distributions used as priors in our SED fits. However, all non-uniform distributions are altered to add more probability density in their tails to increase the number of samples the emulator is trained on in regions of parameter space that are rarely occupied (see Section \ref{sec:dists} for further discussion on the choice of training set distributions). These parameter vectors are then fed into the SPS portion of our \texttt{Prospector} workflow to generate \texttt{FSPS} spectra, which are then fed to \texttt{sedpy} to convert the spectra into photometry (expressed in $\textrm{arsinh}$ magnitudes as per Section \ref{sec:sps}). When training an emulator, we use the root mean square error (RMSE, i.e. the square root of MSE) loss function, which is the same as used by \citet{alsing20}.\footnote{It should be noted that while \citet{alsing20} describes a mean square error (MSE) loss function, the \texttt{Speculator} source code shows that the RMSE loss function is used. We make this distinction here as we found trained emulators' accuracy and precision improve significantly when the RMSE loss function is used in place of MSE.} We optimize each emulator's weights and biases in each step using the NADAM optimizer \citep{dozat16}, a variation of the ADAM optimizer \citep{kingma14} that incorporates Nesterov momentum \citep{nesterov83}.\footnote{In short, gradient descent with Nesterov momentum delays the computation of the gradient until \emph{after} the momentum has been applied, as opposed to the traditional method where it is applied after the gradient has been computed.} The decay of momentum parameters are set to $\beta_1 = 0.9$ and $\beta_2 = 0.999$.

The network is trained in three steps following a learning rate $\eta$ and batch size $N_\textrm{batch}$ schedule:
\begin{eqnarray}
    \eta = 10^{-3}, 10^{-4}, 10^{-5} \\
    N_\textrm{batch} = 10^3, 10^3, 10^3
\end{eqnarray}
In each step of the schedule, the full training set is randomly split into a training set and a validation set, using a $1\!:\!19$ validation-training split (i.e. $5\%$ of available training data is devoted to validation). Training continues until either the loss over the validation set has not improved over the past $20$ epochs or a total of $1000$ epochs have been reached in that step. At the end of a given step, the weights and biases that resulted in the lowest validation loss are saved and training continues until the final step has completed. These emulators typically take of order $\sim10\ \textrm{hours}$ to train on a Nvidia Tesla K80 GPU.

\subsection{Fitting Photometry SEDs}\label{sec:photdata}

The most stringent test of an SPS emulator is to produce accurate posterior distributions when conditioned on galaxy SEDs. In order to test each emulator's ability to meet this goal, we use each emulator to fit a selection of photometric galaxy SEDs to test the emulator across a broad range of galaxy SED types. Specifically, we fit a selection of galaxies in the 3D-HST photometric catalog \citep{brammer12,skelton14}, fitting all $12319$ galaxies fit by \citet{leja19} in the COSMOS field. This sample size was selected to provide a robust measurement of the rate of $3\sigma$ posterior discrepancies, which theoretically occur at a rate of about 1 in 370. In addition to the observed photometric spectral flux densities and their measurement uncertainties, the $z_\textrm{best}$ redshift from \citet{momcheva16} is extracted from the catalog, and redshift is then fixed to that value in our fits to avoid significant posterior discrepancies solely arising from multimodal photometric redshift solutions. All of the 3D-HST data used in this work are available on MAST: \dataset[10.17909/T9JW9Z]{\doi{10.17909/T9JW9Z}}.

We fit these SEDs using the \texttt{Prospector}\footnote{The version of \texttt{Prospector} used in this work was v1.1.0.} \citep{leja17,johnson21b} SED fitting code, with our priors set to the distributions listed in the rightmost column of Table \ref{tab:physmod}. These priors are largely the same as the priors used by \citet{leja19}, with modifications including the aforementioned lower limit for total stellar mass formed in addition to distributions for the added dust emission parameters that approximate their stacked posteriors from the \citet{leja17} work.\footnote{See \citet{leja17}, Figure 29.} For sampling, we use the \texttt{dynesty}\footnote{The version of \texttt{dynesty} used in this work was v1.2.3.} \citep{skilling04,skilling06,speagle20,speagle22} nested sampling routine. Fitting a single SED with \texttt{Prospector} with the settings used in this work typically takes approximately $\sim\!1.8\times 10^6$ likelihood function calls, which each emulator requires $\sim\!30\ \textrm{CPU min}$ to execute and \texttt{FSPS} requires $\sim20\ \textrm{CPU hr}$ to do the same.\footnote{It should be noted that emulator-based fits are currently limited to a factor of $\sim\! 40$ speed-up in comparison to non-emulator-based fits due to bottlenecks unrelated to SPS in \texttt{Prospector} and \texttt{dynesty}.} Following the procedure of \citet{leja19}, we assume an error floor in the observed spectral flux densities of $5\%$ in each band to account for systematic uncertainties in \texttt{FSPS} predictions; however, to further account for the known emulation error, we convolve the observational errors with emulator precision in that band (i.e. the standard deviation of the test set residuals, see Section \ref{sec:apriori}). That is, for a given observational uncertainty $\sigma_{\textrm{obs}}$, estimated emulation precision $\sigma_{\textrm{emul}}$, and observed flux $E_{\nu}$, we assume an overall uncertainty $\sigma_\textrm{assume}$ of
\begin{equation}
        \sigma_\textrm{assume} = \sqrt{\sigma_\textrm{obs}^2 + \sigma_\textrm{emul}^2 + (0.05E_\nu)^2}.
\end{equation}

Another aspect of the fitting routine we are particularly concerned with is the settings used for the \texttt{dynesty} sampler, especially as they pertain to sampling variance. As the purpose of this work is to compare different emulation architectures against each other, we want to ensure that the differences between the emulators are the primary cause for the differences between fits. The effect of emulation errors could be hidden if sampling variance is too large, such that an emulator's fit could be found to be discrepant with respect to a reference fit even if the emulator is far more precise than the observations. Indeed, we found that independent realizations of the same fit would occasionally result in highly discrepant posteriors when using default \texttt{dynesty} settings. However, the combination of \texttt{nlive\_init=2000} (default \texttt{100}) and \texttt{target\_n\_effective=200000} (default \texttt{10000}) sufficiently decreases the sampling noise to make the emulation errors more evident. Increasing the value of \texttt{nlive\_init}, the number of live points used by \texttt{dynesty}, allows \texttt{dynesty} to explore more areas of parameter space, reducing the risk of ``missing'' solutions. Meanwhile, increasing the value of \texttt{target\_n\_effective}, the target effective sample size \citep{kish65} of posterior samples defined as $n_\textrm{eff} = (\sum_{i} w_i)^2/\sum_{i} w_i^2$ where $w_i$ are the sample weights, allows for more robust measurements of the extreme percentiles of the posterior (e.g. percentiles corresponding to $\pm 3\sigma$). Both of these qualities are relevant, as we will ultimately be wanting to compare how often a solution proposed by one emulator is statistically excluded by a different emulator's fit, and in doing so we want to reduce the number of these exclusions that occur solely due to the sampler.\footnote{We would like to emphasize that we are not trying to claim that these settings are to be used in general use -- we are selecting these settings solely to make the small differences between emulators more evident at the $\sim\!3\sigma$ level.}

Each of the 12319 galaxies is fit once with each emulator, except for the 1024 node-per-layer emulator, which each galaxy is fit with twice to provide an estimate on the discrepancies between posteriors solely due to sampling uncertainties. Furthermore, in addition to fitting the galaxy SEDs with the emulators, we also fit a random sub-sample of $10^3$ galaxies with \texttt{FSPS} to isolate any errors that may be caused by emulation alone. Finally, we generate a mock catalog of $10^4$ galaxy SEDs and fit these with the emulator as well to verify that the posteriors are representative of galaxies' true properties. These are generated in the same way as the training and test sets, with the exceptions that the parameters are sampled from the prior distributions rather than the training/test set distributions (although redshift is still sampled from a $\textrm{Uniform}(0.5,3.0)$ distribution), a random error is added to each filter flux, with the error being randomly sampled from a normal distribution with $\mu = 0\ \textrm{mag}$ and $\sigma = 0.1\ \textrm{mag}$, and the filter set is restricted to only those filters provided by the 3D-HST COSMOS catalog.

\section{Results}\label{sec:results}

Our results are organized into two separate sets of tests: Section \ref{sec:apriori} evaluates the performance of the emulators in flux space, while Section \ref{sec:aposteriori} evaluates the performance of the emulators in derived parameter space, which is a more expensive computational test since it requires an estimate of the ``truth.''

\subsection{A Priori Emulation Performance}\label{sec:apriori}

\begin{figure}
    \plotone{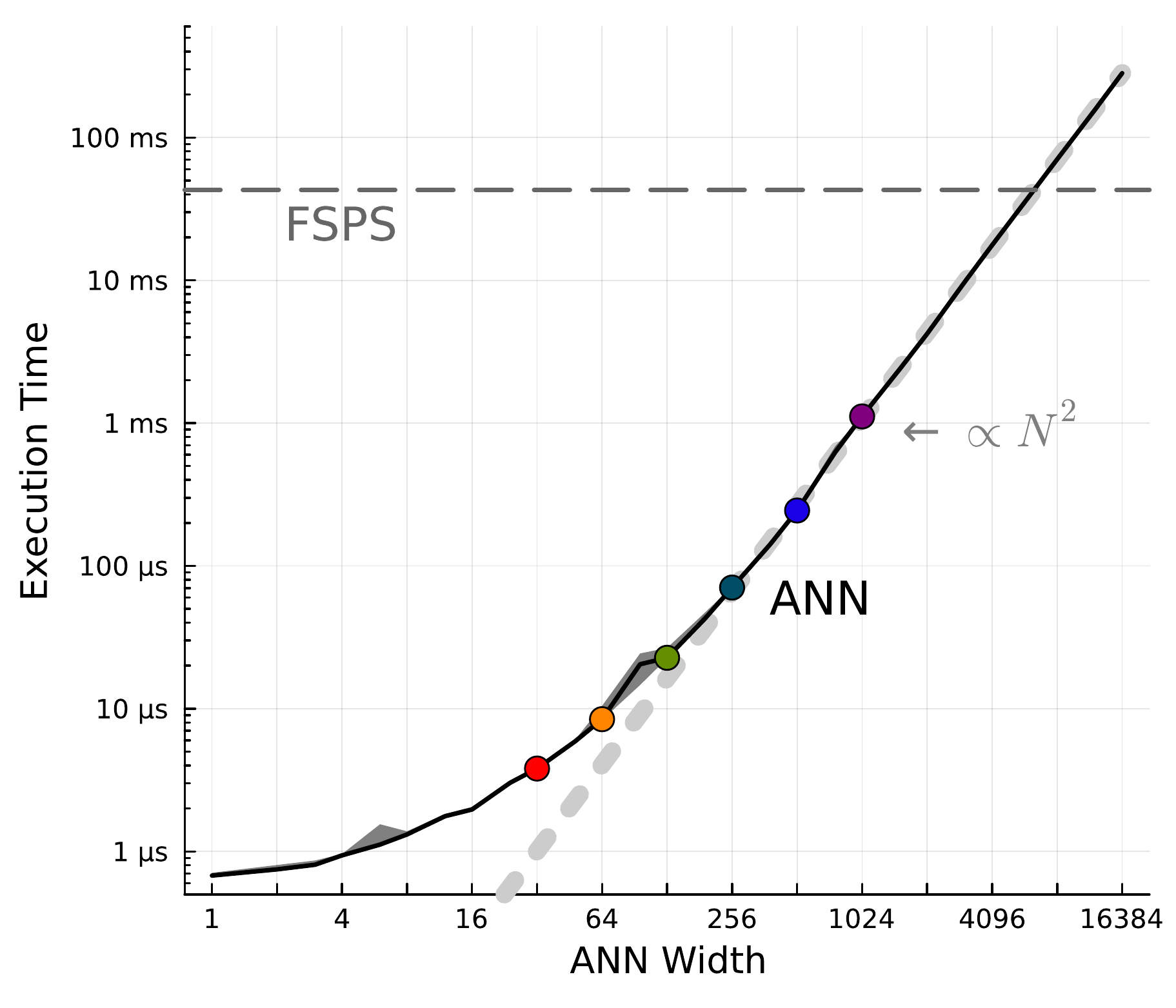}
    \caption{Execution time of ANN emulators (on a CPU) as a function of the width of the network (i.e. the number of nodes per layer). The solid black line denotes the median execution time among all benchmark trials evaluated with a given architecture, while the shaded region encompasses the $\pm 1\sigma$ quantiles of those trials. The dashed gray line near the top of the plot denotes the execution time of \texttt{FSPS} with the settings used for our physical model. The colored points denote the network widths used for the emulators trained in this work. For network widths $N \geq 256$, the execution time scales approximately as $N^{2.01\pm0.01}$.}
    \label{fig:emulspeed}
\end{figure}

Here we demonstrate how the execution time of an emulator scales with the architecture. The dashed gray line shown in Figure \ref{fig:emulspeed} shows the measured $\sim\!40\ \textrm{ms}$ execution time for \texttt{FSPS} (with the proper settings for use with our physical model), which is the benchmark that all emulators must exceed in order to be useful. Meanwhile, the solid black curve denotes the typical execution times for ANN emulators of various different widths on a CPU.\footnote{ANNs typically perform much faster on GPUs than CPUs, but we restrict this study to timings on a CPU due to the fitting software used in this work not being able to make effective use of a GPU's capabilities. Loosely speaking, one may expect about a factor of $\sim\! 10$ reduction in execution time when using a GPU, as per \citet{alsing20}.} As expected, emulators that have smaller architectures require less execution time. For example, an emulator with 1 node per layer (which would only feasibly be able to emulate the normalization of an SED) is approximately $10^5$ times faster than \texttt{FSPS}, while an emulator with 4096 nodes per layer requires roughly as much execution time as \texttt{FSPS} (and is thus the absolute upper limit in ANN width for an SPS emulator on a CPU). For emulators with more than $\sim\! 256$ nodes per layer, we find that the execution time of the emulator tends to scale quadratically with respect to the number of nodes per layer (i.e. doubling the width of the emulator quadruples the execution time), with the execution time scaling as $\propto N^{2.01\pm0.01}$ via a least squares fit, directly attributable to the matrix multiplication involving $N\times N$ matrices in the hidden layers. For emulators with less than $\sim\! 256$ nodes per layer, the curve flattens due to the linear output layer (which has a fixed width of 137 nodes due to the 137 emulated filter magnitudes) becoming the most computationally expensive step in the execution, asymptotically approaching an execution time of $\sim\! 800\ \textrm{ns}$ for extremely small widths. Thus, for large architectures, adding additional filters to the emulator will not significantly change the overall execution time of the emulator, but these additional filters may change the emulator's precision after training due to the additional information the emulator needs to learn.

\begin{figure*}
    \plottwo{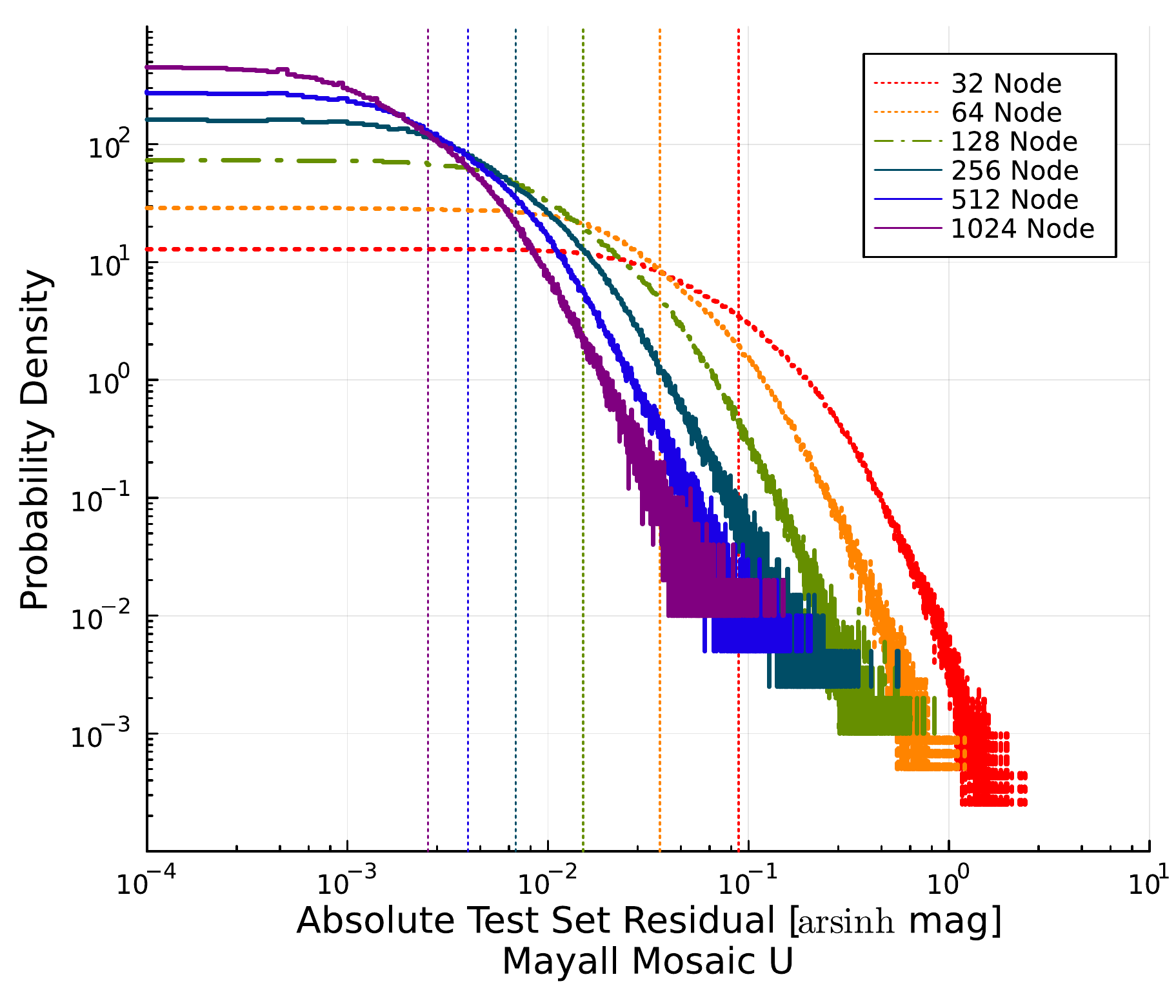}{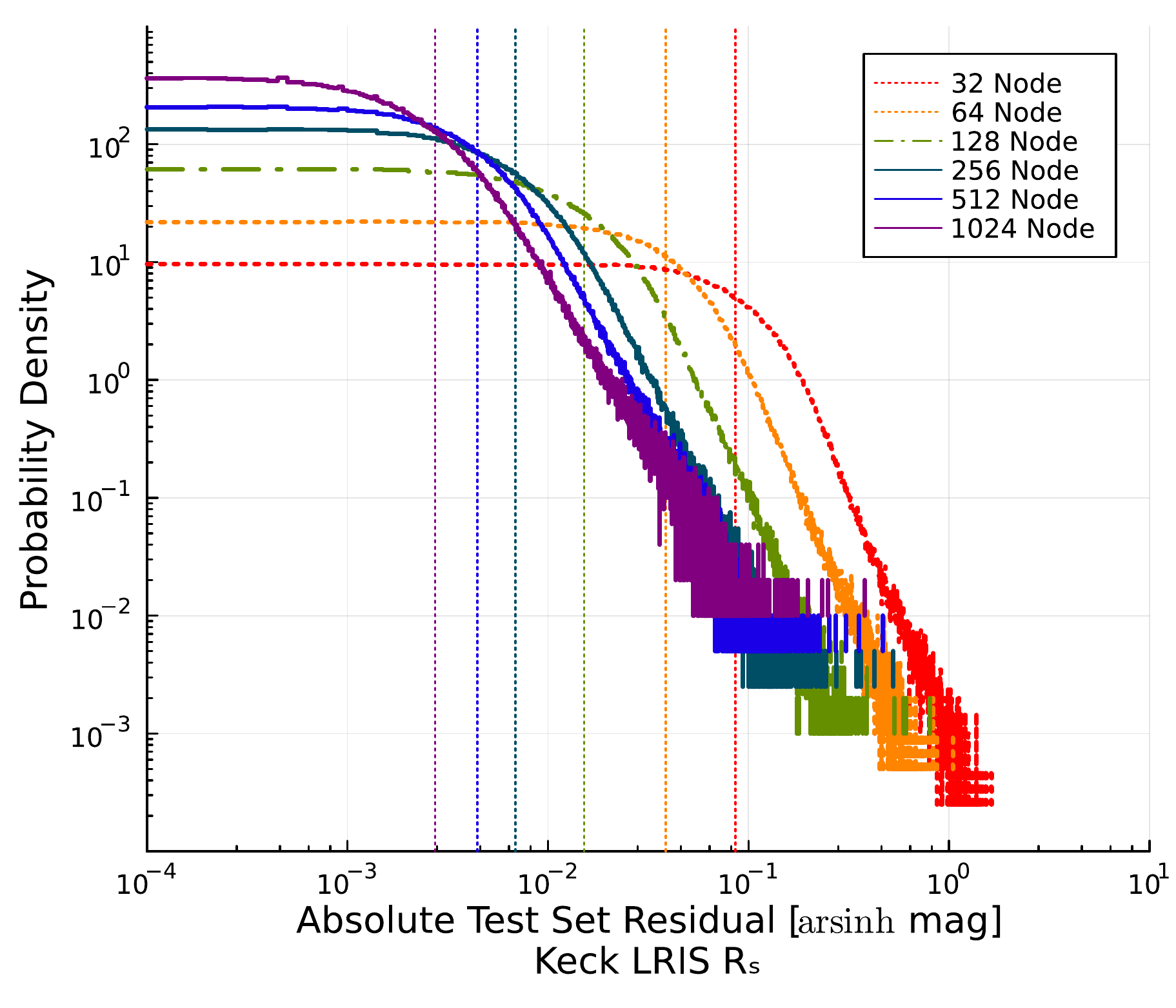}
    \plottwo{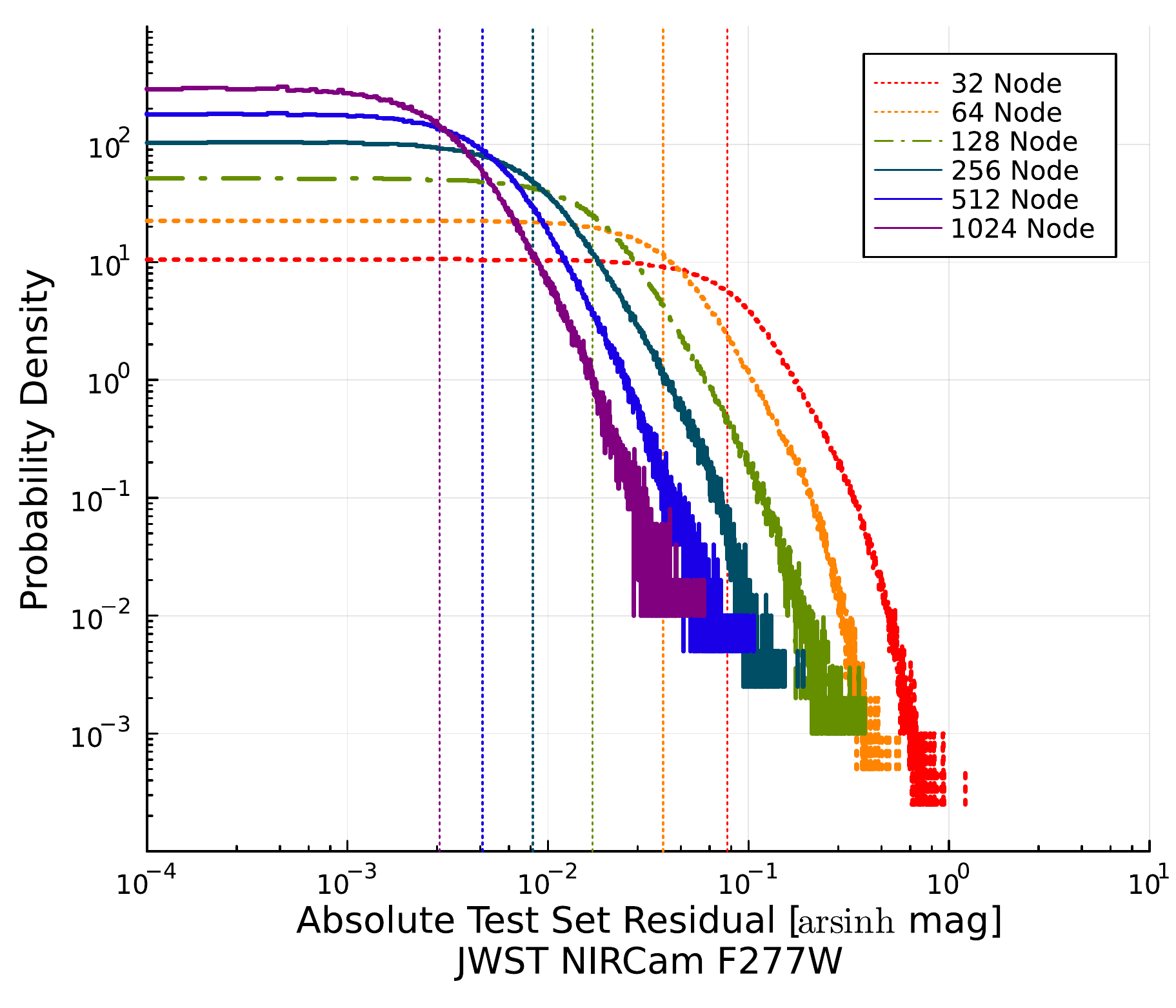}{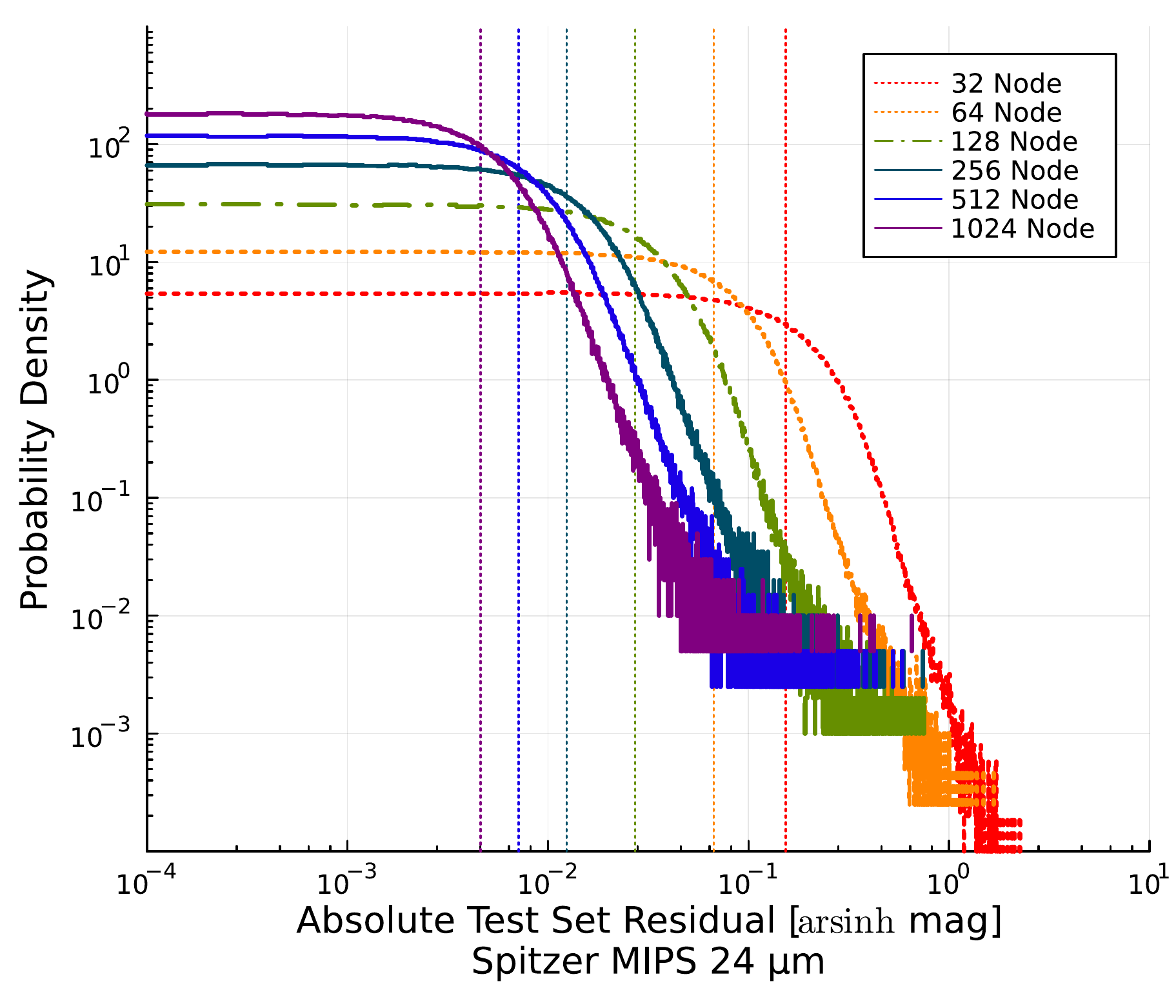}
    \caption{The distribution of absolute test set residuals (i.e. $\lvert \textrm{emulated} - \textrm{actual}\rvert$) for each emulator, evaluated for the Mayall Mosaic $U$, Keck LRIS $R_s$, JWST NIRCam F277W, and Spitzer MIPS $24\ \mu\textrm{m}$ filters. The thin vertical dotted lines denote the residual corresponding to the distribution's $\sim\!68\%\textrm{ile}$.}
    \label{fig:mydists}
\end{figure*}

Next we assess how the precision of an emulator in flux space scales with network width. For each filter $i$, we estimate emulator errors from the test set (a dataset of $10^6$ parameter-SED pairs generated using same distributions as the training set) by computing the mean $\mu_i$ (a measure of the emulator's accuracy) and standard deviation $\sigma_i$ (a measure of the emulator's precision) of the emulation errors in that filter, which are the differences between the flux predicted by the emulator and the flux predicted by \texttt{FSPS}. Overall, the accuracy of all emulators trained in this work is very good, with the maximum $\mu_i$ of any filter/emulator combination of about $4\times 10^{-3}\ \textrm{mag}$.\footnote{For brevity, the unit ``$\textrm{mag}$'' used for the remainder of this work is understood to be referring to the $\textrm{arsinh}$ magnitude system as defined in Section \ref{sec:sps}. This is distinction has no meaningful impact for fluxes brighter than $m_\textrm{AB} \approx 32$.} The full distributions for each emulator for a selection of 4 filters can be seen in Figure \ref{fig:mydists}.

\begin{figure*}
    \plotone{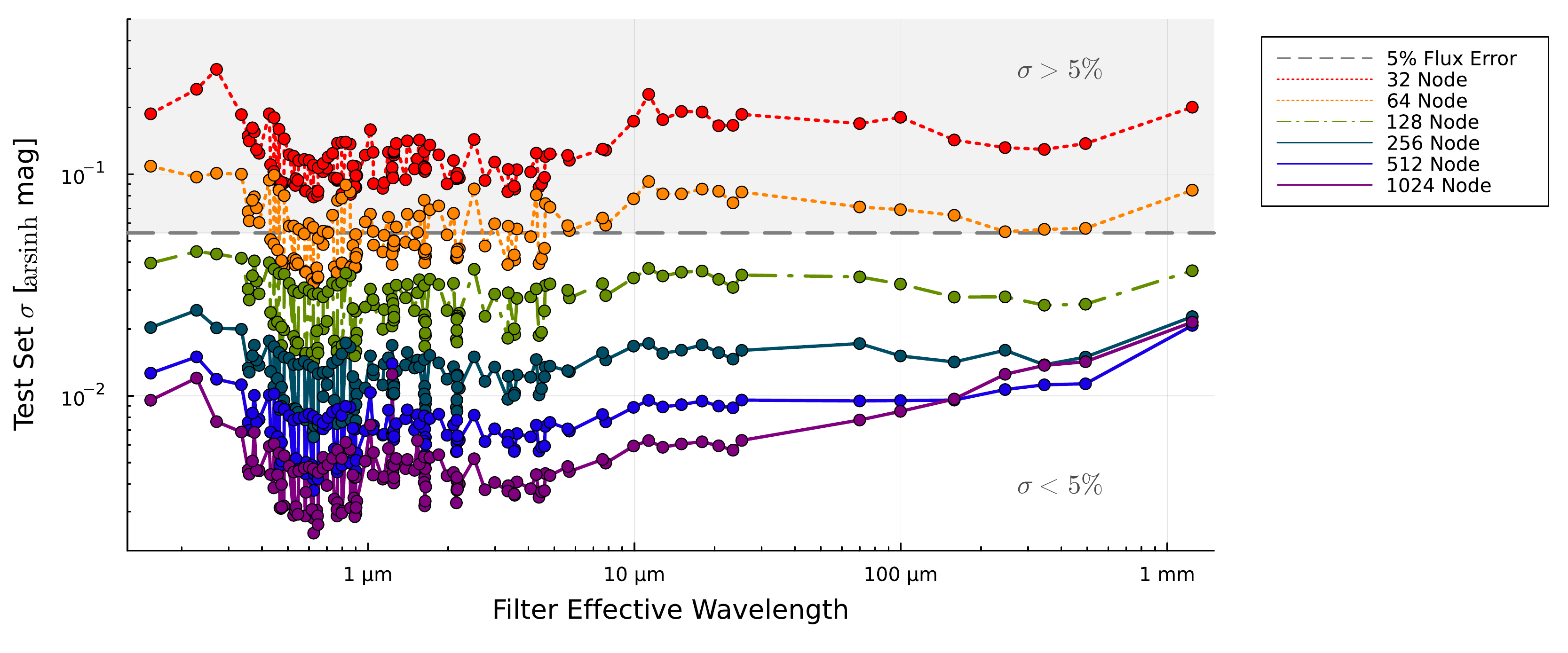}
    \caption{The emulation precision of each emulator trained in this work as a function of filter effective wavelength, where the precision is expressed in terms of $\textrm{arsinh}$ magnitudes (the native outputs of the emulator, as defined in Section \ref{sec:sps}). Emulation precision generally doubles (i.e. $\sigma$ decreases by a factor of $2$) as network width doubles, except for filters in the far-IR and for extremely wide networks, which see diminishing returns. Solid lines denote architectures deemed sufficiently accurate and precise for SED fitting in this work, while dash-dot lines denote emulators deemed marginally sufficient and dotted lines denote emulators deemed insufficient.}
    \label{fig:prec}
\end{figure*}

However, the precision is a more relevant factor in parameter inference. In Figure \ref{fig:prec}, we show the emulation precision $\sigma_i$ for each filter in each emulator. We first notice that the precision of a given emulator scales with the number of nodes per layer -- specifically, we generally find that doubling the width of an emulator doubles the precision (i.e. $\sigma_i$ decreases by a factor of $2$). Given a fiducial emulator architecture and its precision after being trained, this provides a useful method of estimating the architecture that will be required in order to meet a given precision target. However, it appears that this trend breaks down for extremely complex architectures, with the 1024 node-per-layer emulator displaying diminishing returns in this regard. Additionally, we find that the 64, 128, 256, 512, and 1024 node-per-layer emulators have precisions that exceed the precision of the observations we fit in this work (which have an error floor of $5\%$) for at least a subsample of the emulated bands, with the 128 node-per-layer network being the simplest emulator to exceed this precision in all 137 of its filters. All emulators exhibit very little bias, with the 32 node-per-layer emulator showing biases no greater than $5\times 10^{-3}\ \textrm{mag}$ and the 1024 node-per-layer emulator showing biases no greater than $5\times 10^{-4}\ \textrm{mag}$.

Additionally, we note that there is a spectral trend in the precision of a given emulator -- with this filter set, we find that the emulators tend to be the most precise in the optical and NIR, and have diminished precision near the extreme ends of the modeled SEDs. This is likely partly due to the fact these locations are where the mass-to-light ratio ($M/L$) is most variable \citep{conroy13}, in addition to the fact that the selected filters are most densely concentrated in the optical and NIR and all filters are weighted equally in the loss function. Finally, we can also see that there are instances where the precision of a given emulator can change rapidly from one filter to the next, particularly in the optical and NIR. The cause for this is likely two-fold, both of which are sensitive to redshift. Firstly, as a rest-frame SED is redshifted, sharp spectral features present in the SED (e.g. Lyman and Balmer breaks, emission lines) will rapidly change filter magnitudes as these features pass through each filter's transmission curve, which means that filters whose transmission curves contain these spectral features (when subject to the redshifts present in our test set) will be affected more than filters that do not. Secondly, while our filter selection primarily contains wideband filters, some of the filters have narrower transmission curves, including 18 Subaru Suprime-Cam intermediate bands between $400\ \textrm{nm}$ and $900\ \textrm{nm}$ and 12 JWST NIRCam medium band filters between $1\ \mu\textrm{m}$ and $5\ \mu\textrm{m}$, which are affected even more strongly as spectral features pass through their transmission curves.

\begin{figure*}
    \plottwo{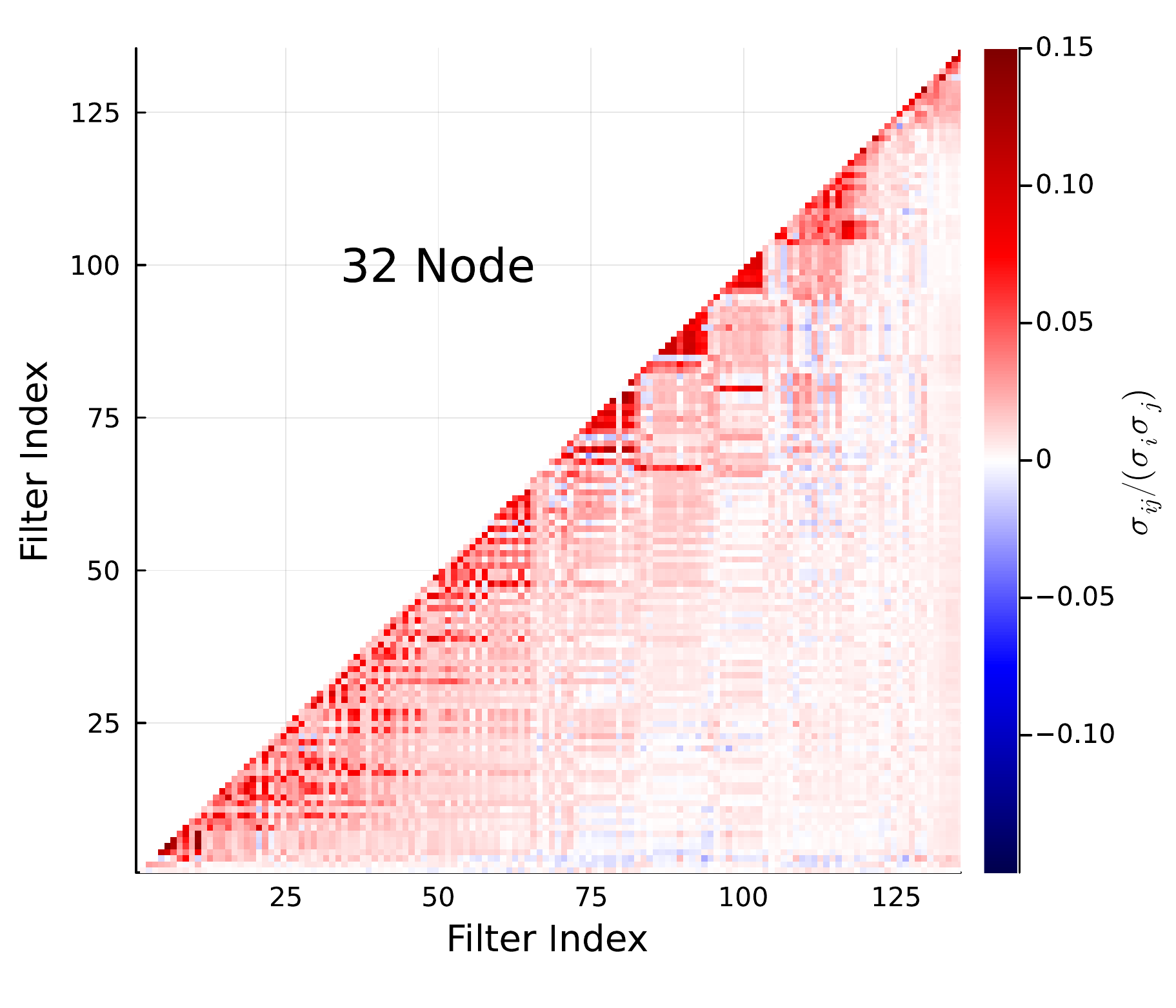}{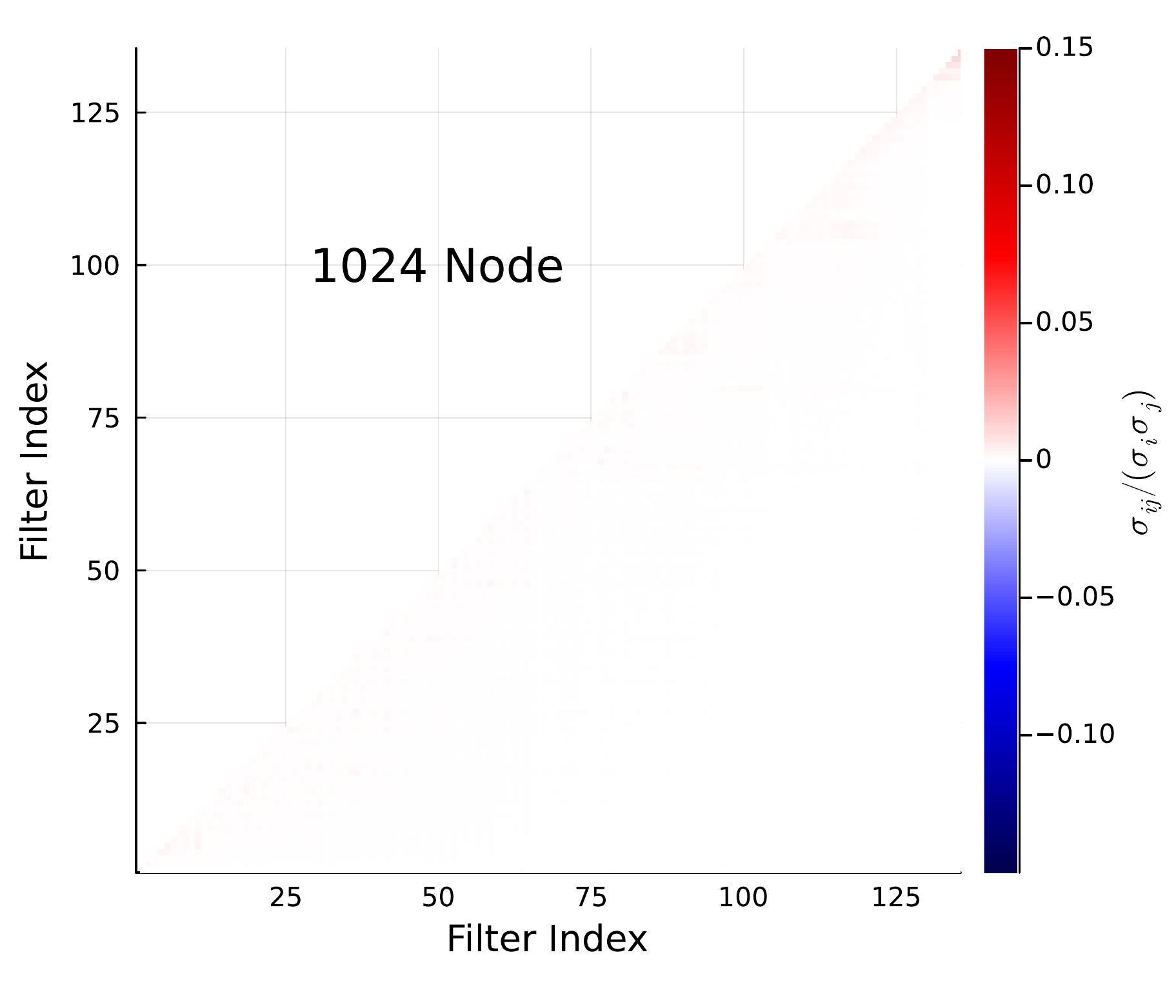}
    \caption{Covariance matrices for the 32 node-per-layer emulator (left) and 1024 node-per-layer emulator (right), normalized by the emulation precision in the relevant filters (i.e. plotted is $\sigma_{ij}/(\sigma_i \sigma_j)$). Note the differing colorbar scales between the two plots. Compared to the 32 node-per-layer emulator, the 1024 node-per-layer emulator provides covariances that are both smaller in scale and less spectrally widespread (i.e. they are limited to filters with similar effective wavelengths).}
    \label{fig:cov}
\end{figure*}

Now we assess to what extent the emulator errors are correlated. This is important -- if an emulator typically provides an incorrect prediction of the observed spectral flux density for a single filter while simultaneously being correct in the remaining filters, then the errors in the emulator's predictions can be regarded as independent with regard to filter and the results of the filter-by-filter precision analysis from Figure \ref{fig:prec} can be treated as independent noise (and thus, the sampler should be able to properly account for an emulator's errors). However, if an emulator residuals are typically correlated for a \emph{group} of filters (i.e. if emulation errors are correlated between some number of filters), it may lead to systematic color offsets that yield incorrect results. For example, if an otherwise precise emulator typically predicts a Balmer break that is too small (i.e. predicting too much flux for filters on the blueward side of the jump while simultaneously predicting too little flux for filters on the redward side), it would likely lead to a stellar age posterior that is artificially biased towards younger ages, especially since these correlations would not be accounted for if the emulator's errors are treated as independent with regard to filter.

A simple way to investigate this effect is by computing the covariance matrix of the emulator's test set residuals, where $\sigma_{ij}$ is the covariance between errors in filter $i$ and filter $j$ (and with this notation $\sigma_{ii} = \sigma_{i}^2$). We display these covariance matrices for the 32 and 1024 node-per-layer emulators in Figure \ref{fig:cov}, where we normalize each of the covariances by the corresponding filter precisions (i.e. $\sigma_{ij}/(\sigma_i \sigma_j)$) to compare how strong their covariances are in comparison to the individual precisions. We find that in the 32 node-per-layer emulator, the covariances are both large in comparison to the filter precision (i.e. maximal $\lvert\sigma_{ij}/(\sigma_i \sigma_j)\rvert$ of order $10^{-1}$) and spectrally widespread, indicating that this emulator is very likely to be significantly wrong (and wrong in the same direction) in many unrelated filters when it is wrong in a given filter (i.e. its errors are highly correlated). On the contrary, we find that emulators with more nodes per layer have increasingly less correlation in their flux errors -- on the extreme end, the 1024 node-per-layer emulator contains much smaller covariances (i.e. maximal $\lvert\sigma_{ij}/(\sigma_i \sigma_j)\rvert$ of order $10^{-2}$), and these covariances are mostly concentrated to filters that are spectrally nearby (and in some cases are nearly identical filters, e.g. the $V$-band on different telescopes/instruments). Thus, the standard assumption of uncorrelated Gaussian errors is a much safer assumption for a $1024$ node-per-layer emulator than it is for the $32$ node-per-layer emulator.

\subsection{A Posteriori Emulation Performance}\label{sec:aposteriori}

\begin{figure*}
    \plottwo{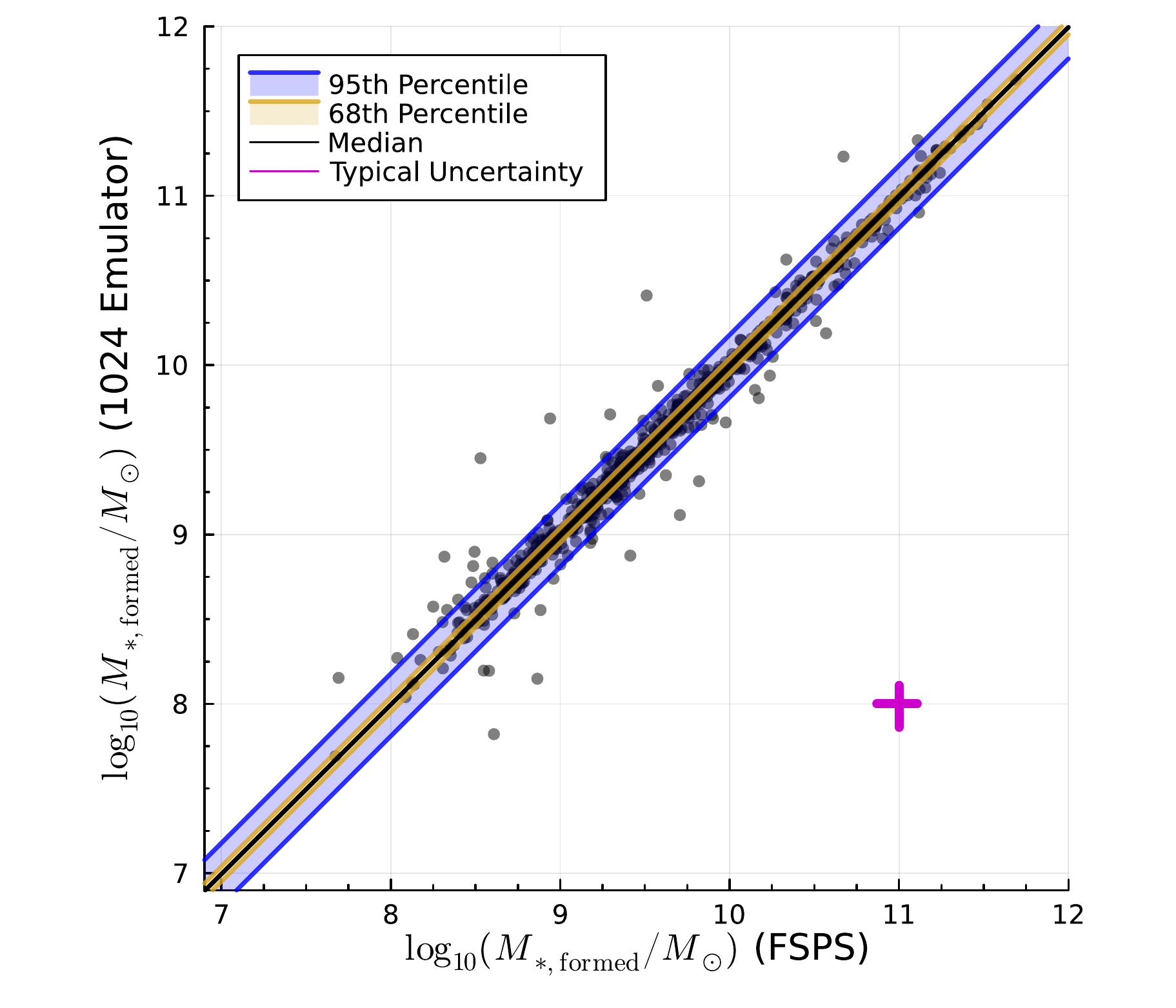}{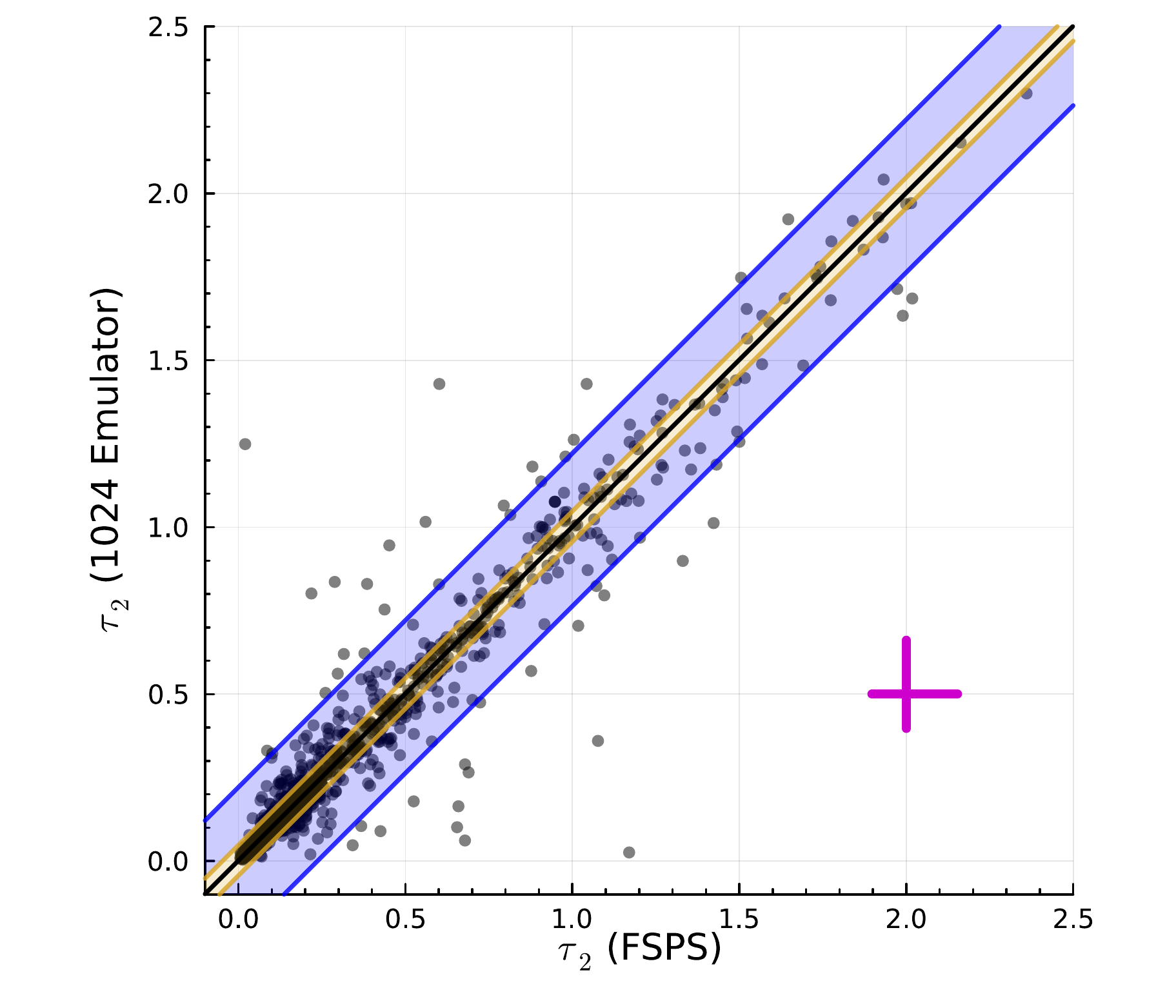}
    \plottwo{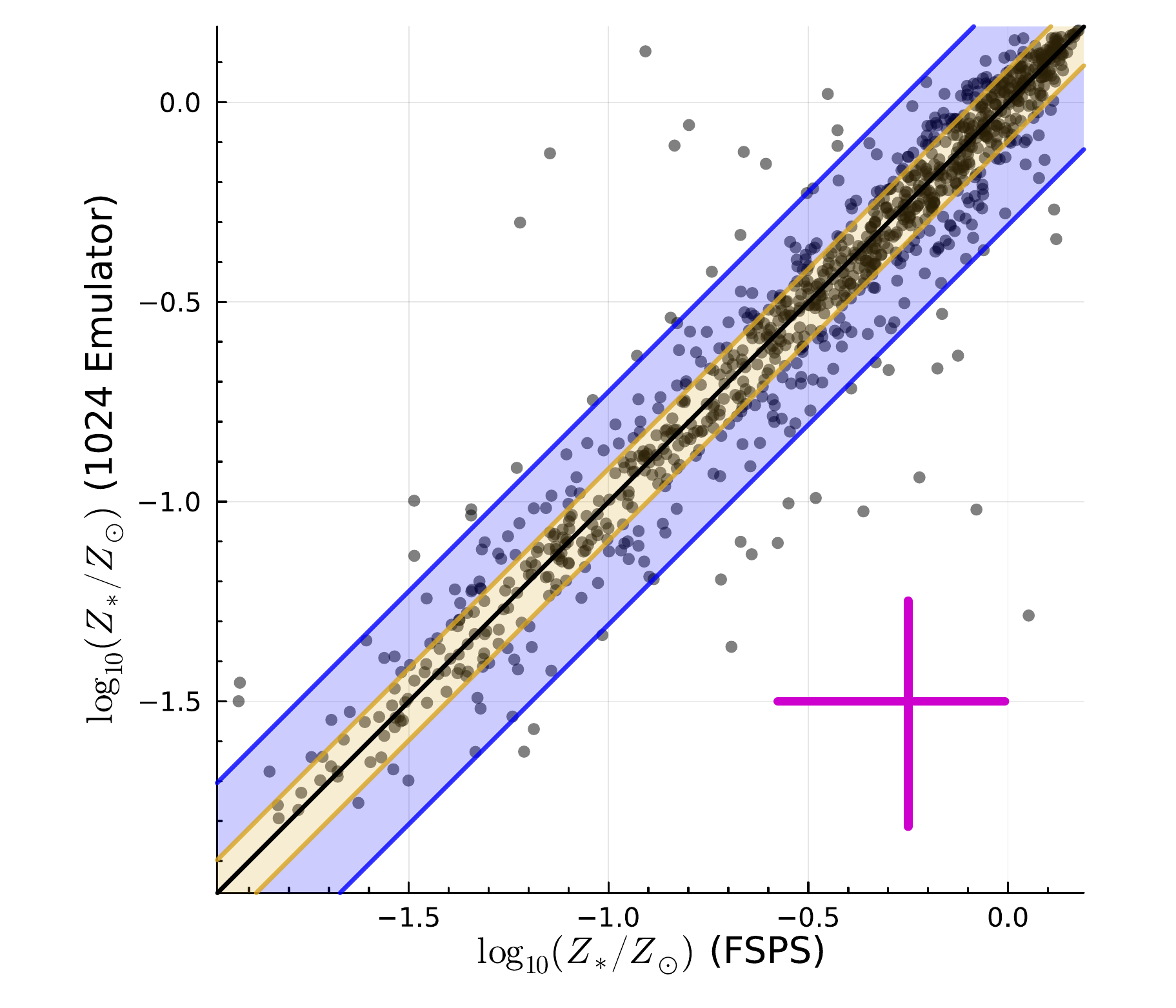}{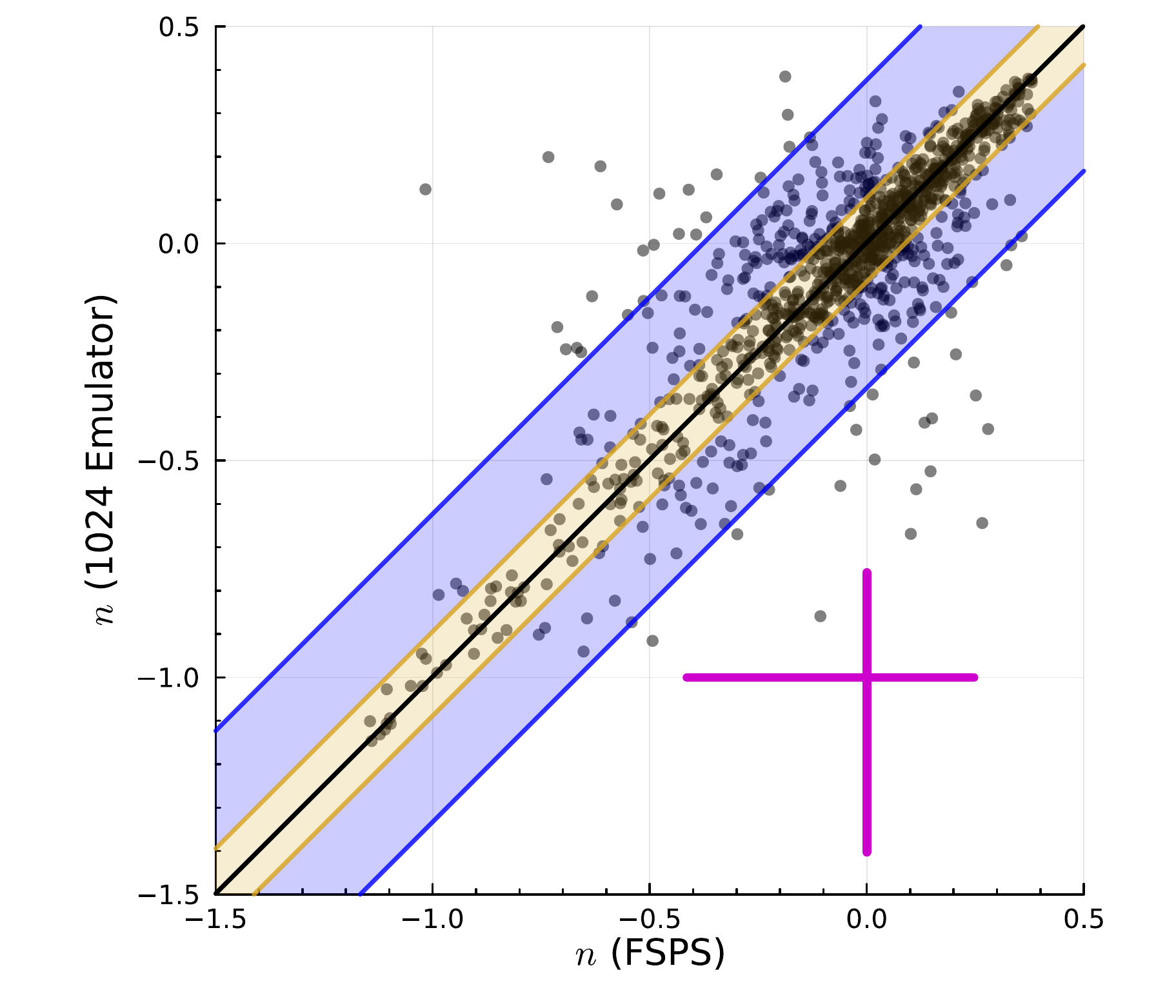}
    \plottwo{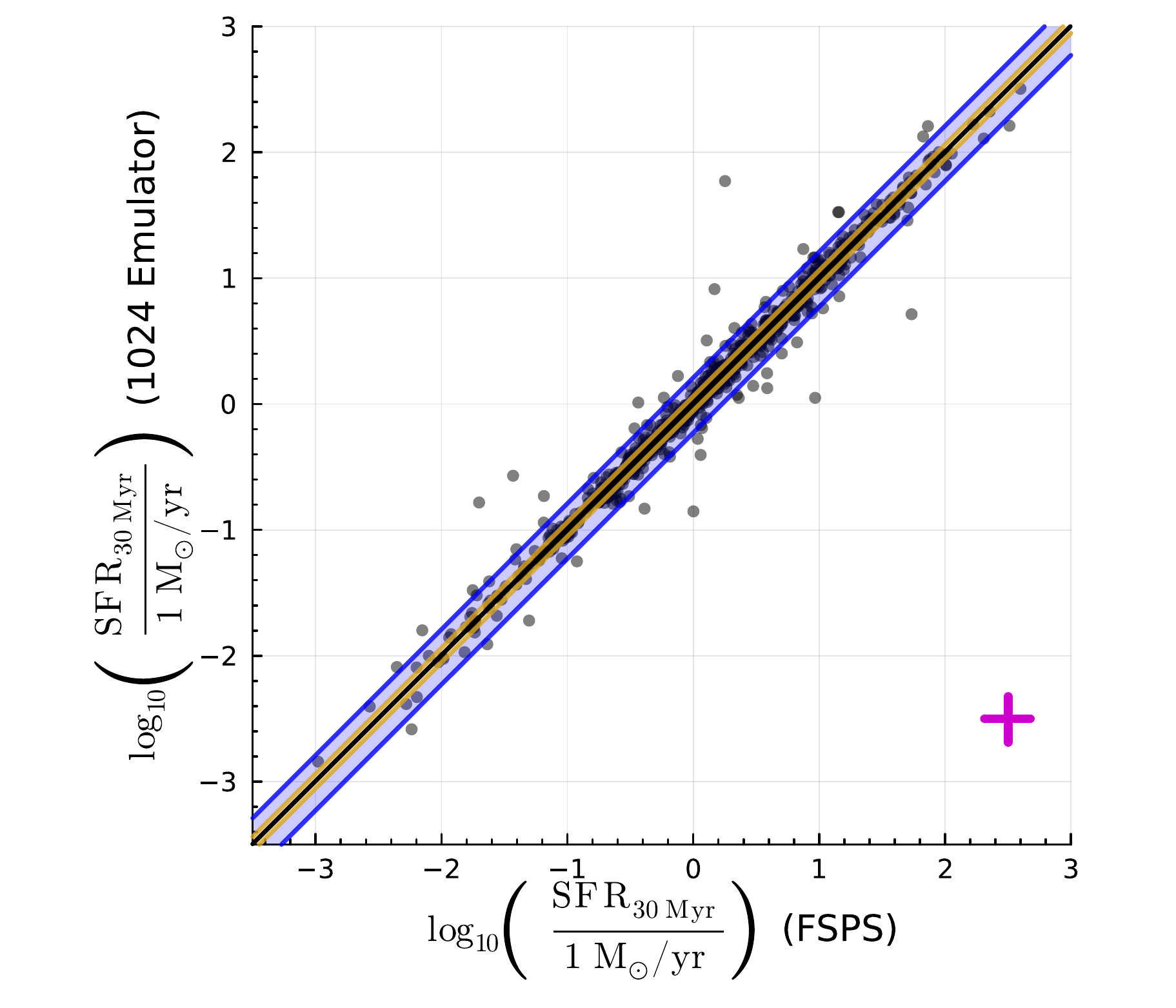}{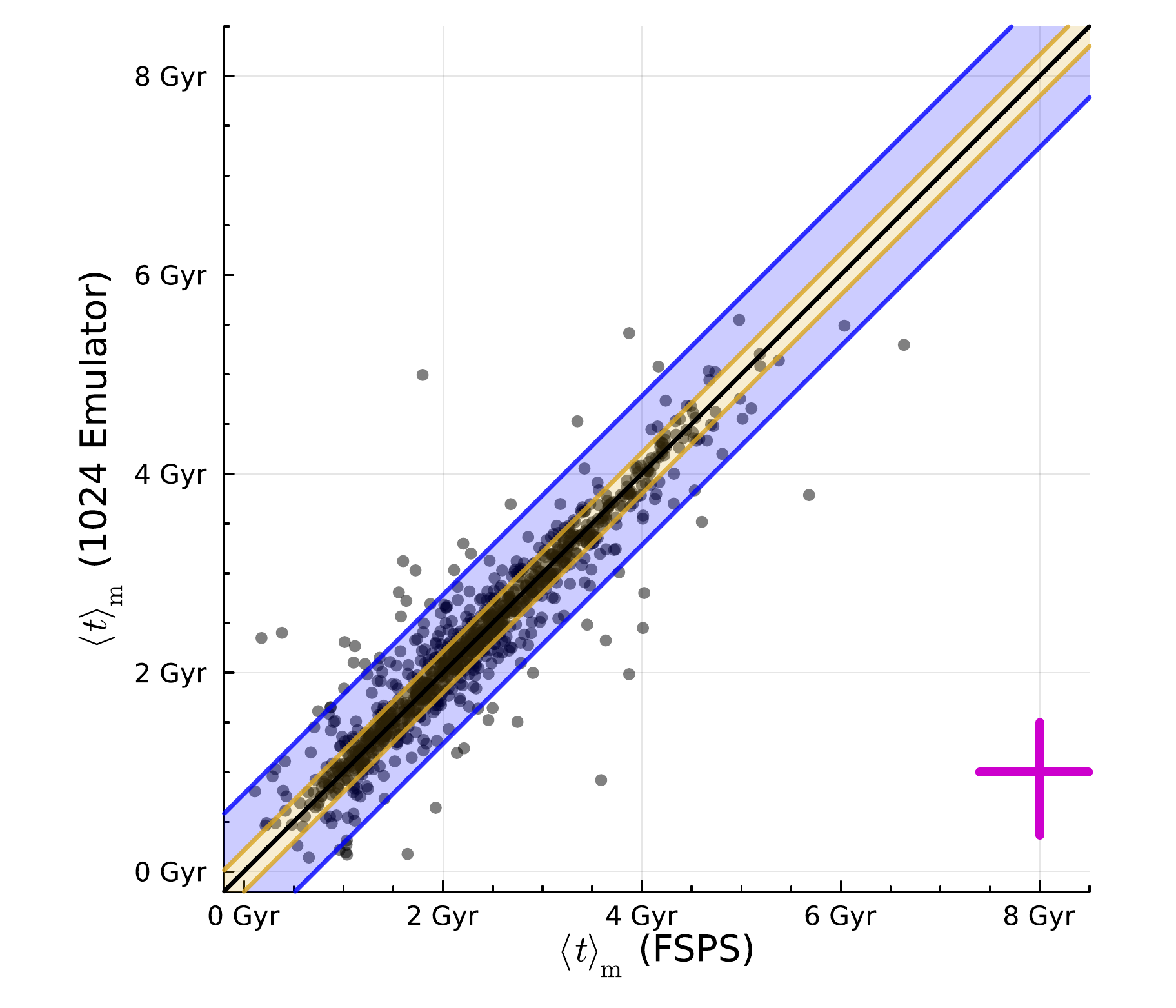}
    \caption{A comparison between posteriors recovered from 1000 \texttt{Prospector} fits to 3D-HST photometry, run with a 1024 node-per-layer emulator and with \texttt{FSPS}, to verify the ability of emulators to recover posteriors consistent with \texttt{FSPS}. Plots are shown for the stellar mass formed $M_{*,\textrm{formed}}$, diffuse dust optical depth $\tau_2$, stellar metallicity $Z_*$, and dust power law index $n$ (top four panels), in addition to the star formation rate $\textrm{SFR}$ for the most recent $30\ \textrm{Myr}$ and the mass-weighted age $\langle t \rangle_m$, which are derived parameters (bottom two panels). The posterior median from the \texttt{FSPS}-based fit is shown on the $x$-axis and the posterior median from the emulator-based fit is shown on the $y$-axis. The gold and blue shaded regions enclose the middle 68th and 95th percentiles of the differences in posterior medians, while the black line denotes the median difference. Shown in pink is a typical posterior width for the plotted parameter.}
    \label{fig:fspscompare}
\end{figure*}

As the analysis that follows in this section is predicated on the assumption that our most accurate/precise emulator (the 1024 node-per-layer emulator) is a suitable replacement for \texttt{FSPS}, we first need to establish that this is actually the case. In Figure \ref{fig:fspscompare}, we compare the posterior medians recovered by the emulator-based fits (using the 1024 node-per-layer emulator, the most accurate and precise emulator) to the posterior medians recovered for the random sample of $10^3$ 3D-HST galaxies using traditional \texttt{FSPS}-based fits for a selection of parameters. In this test, there is strong agreement in posterior medians between the two SPS methods. By measuring the 68\%-ile of the absolute differences between the posterior medians, we find typical differences of order $0.10\ \textrm{dex}$ for the total stellar mass formed, $0.14\ \textrm{dex}$ for the stellar metallicity, $0.12\ \textrm{dex}$ for the star formation rate over the most recent $30\ \textrm{Myr}$, and $0.11\ \textrm{dex}$ for the mass-weighted age. There is no apparent bias between the two SPS methods and the typical scale of the differences between the recovered posterior medians (shown by the gold shaded region on the plots) is of a similar scale to the typical widths of the posteriors in these parameters (shown by the pink crosshairs). This demonstrates that the 1024 node-per-layer emulator produces fits that have posteriors equivalent to the posteriors produced by \texttt{FSPS}-based fits.

\begin{figure*}
    \plottwo{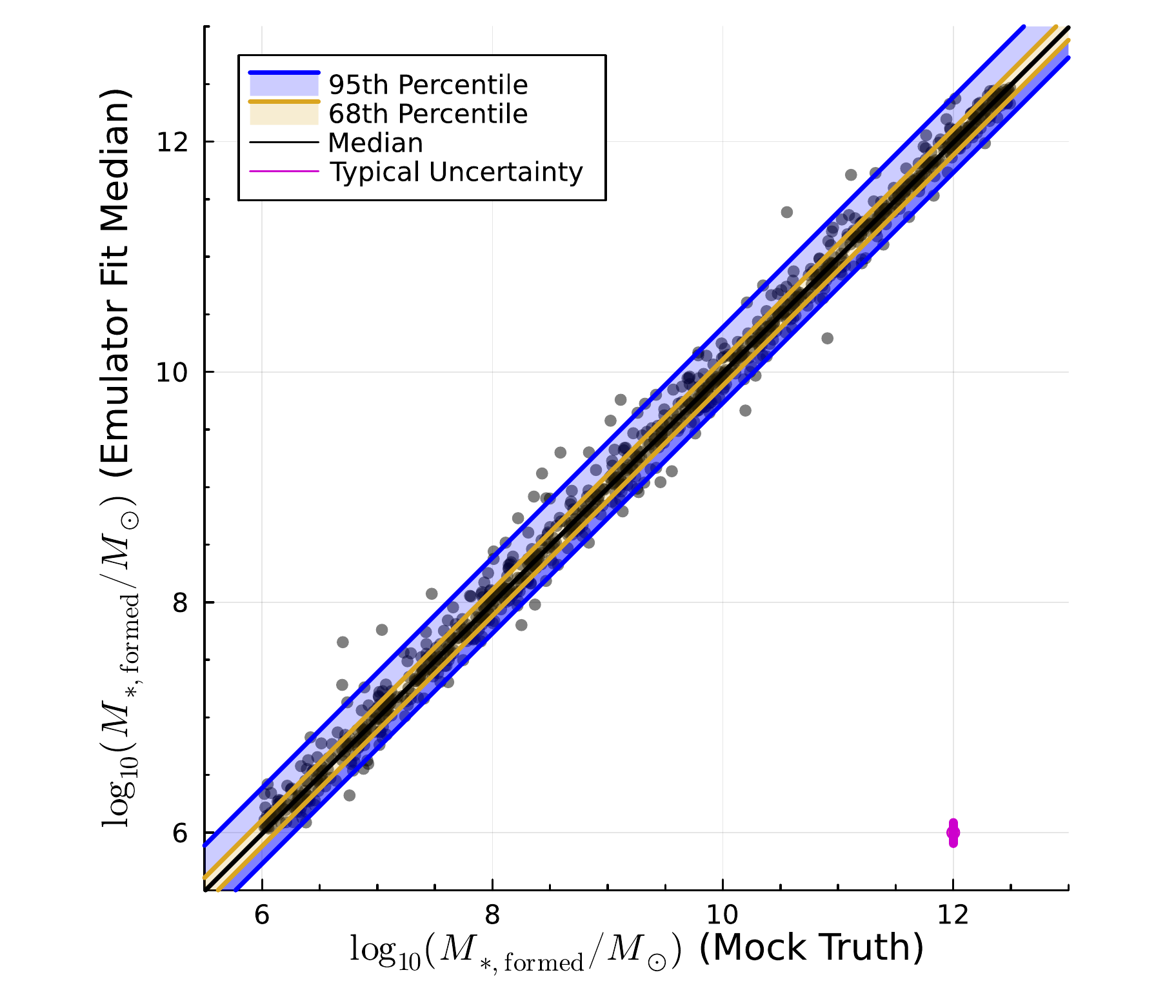}{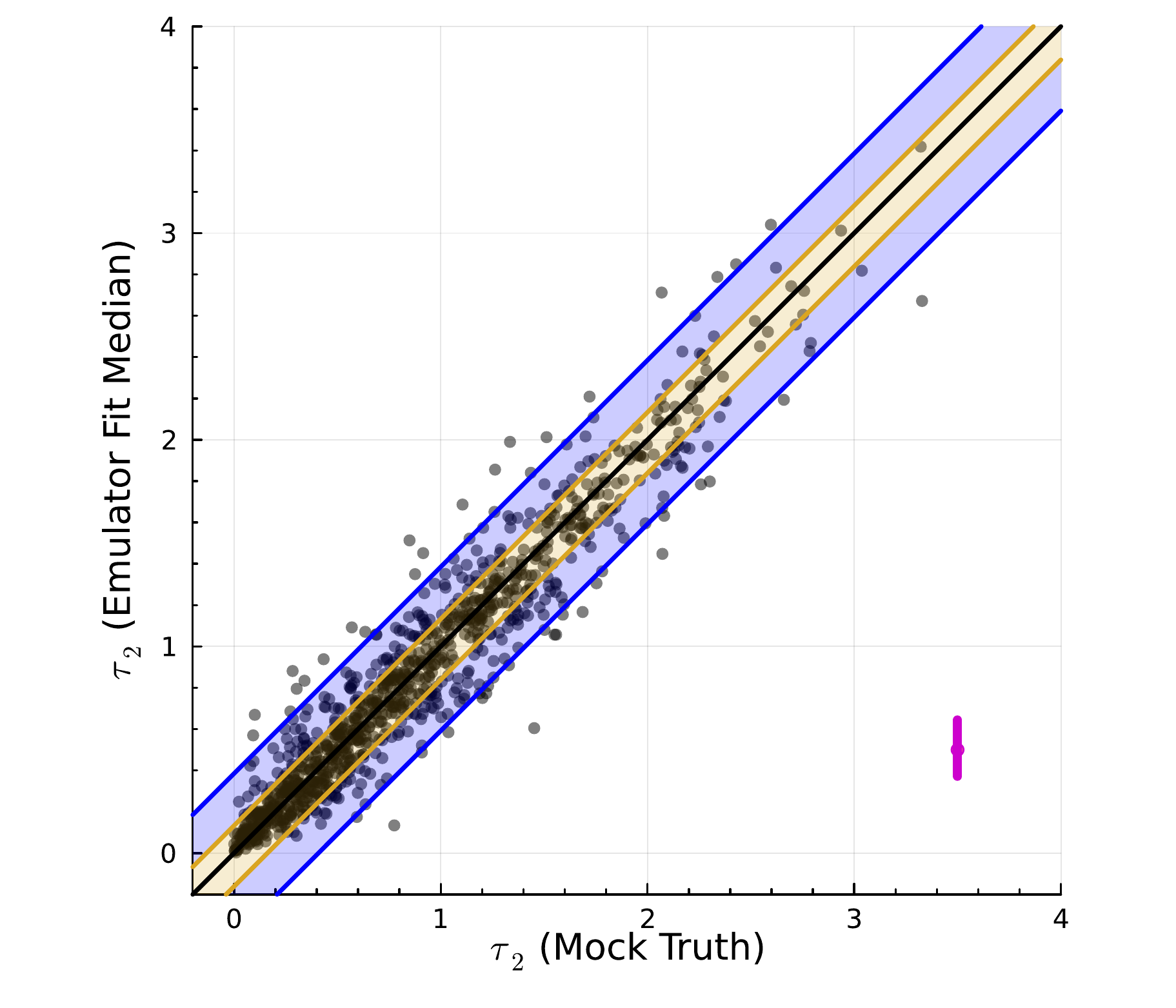}
    \plottwo{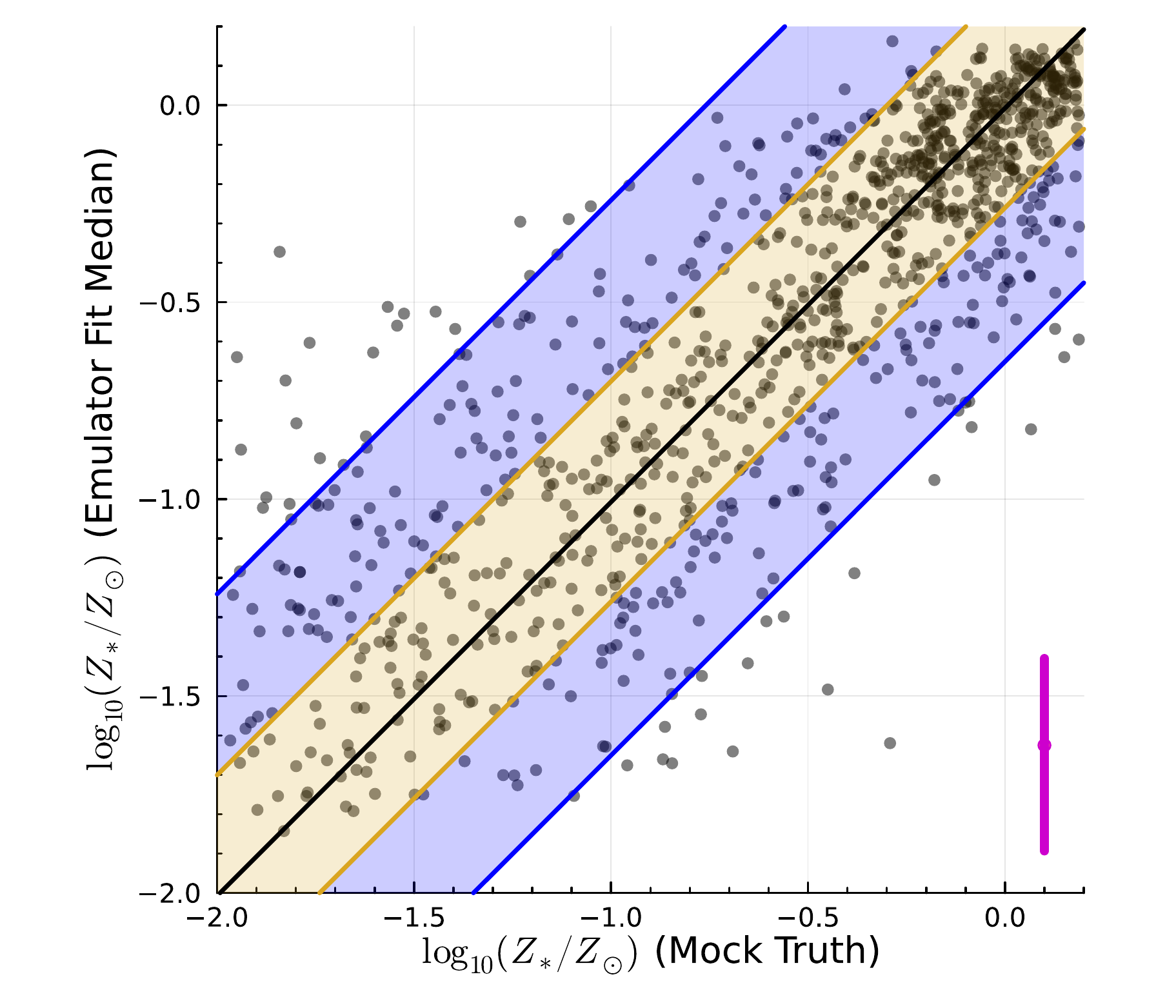}{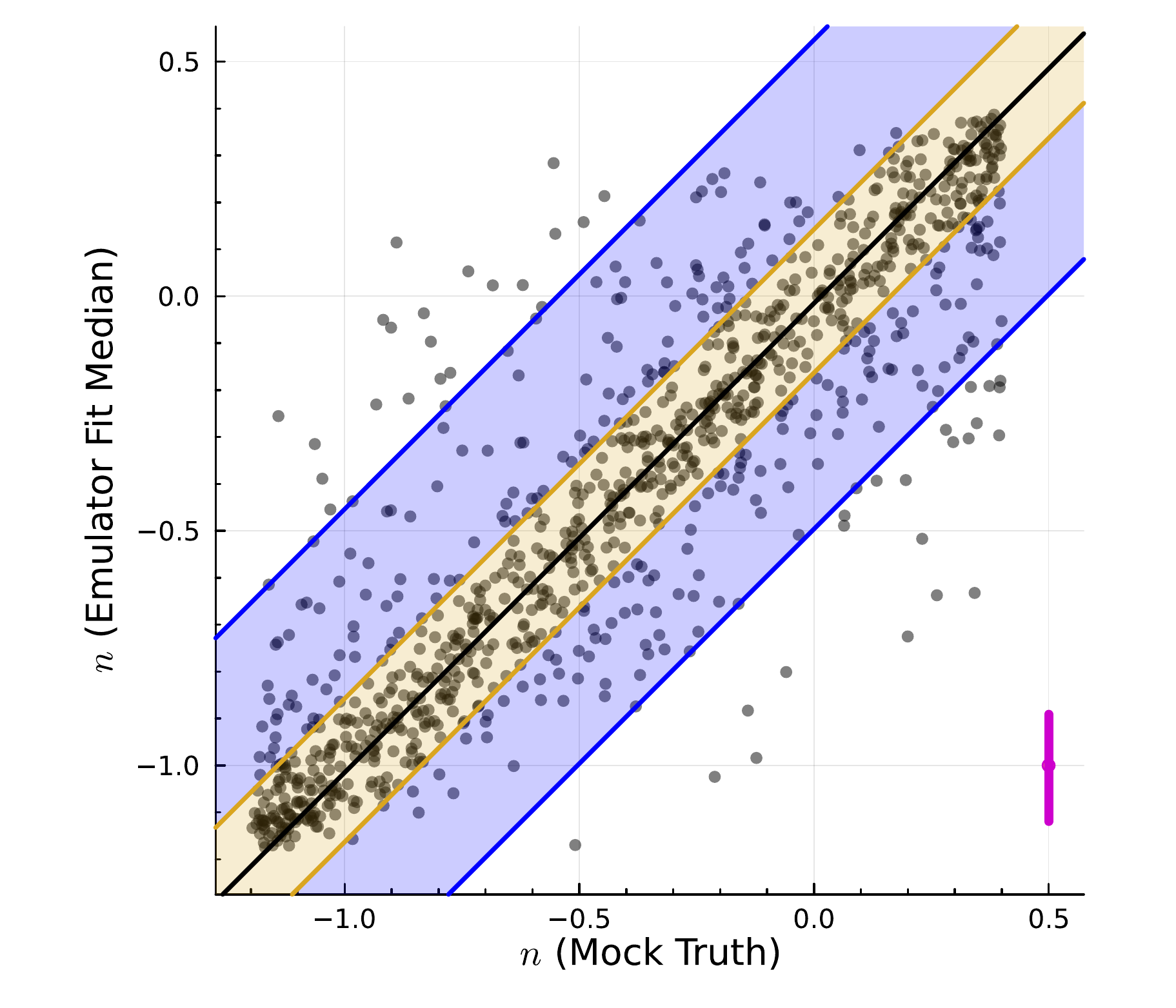}
    \plottwo{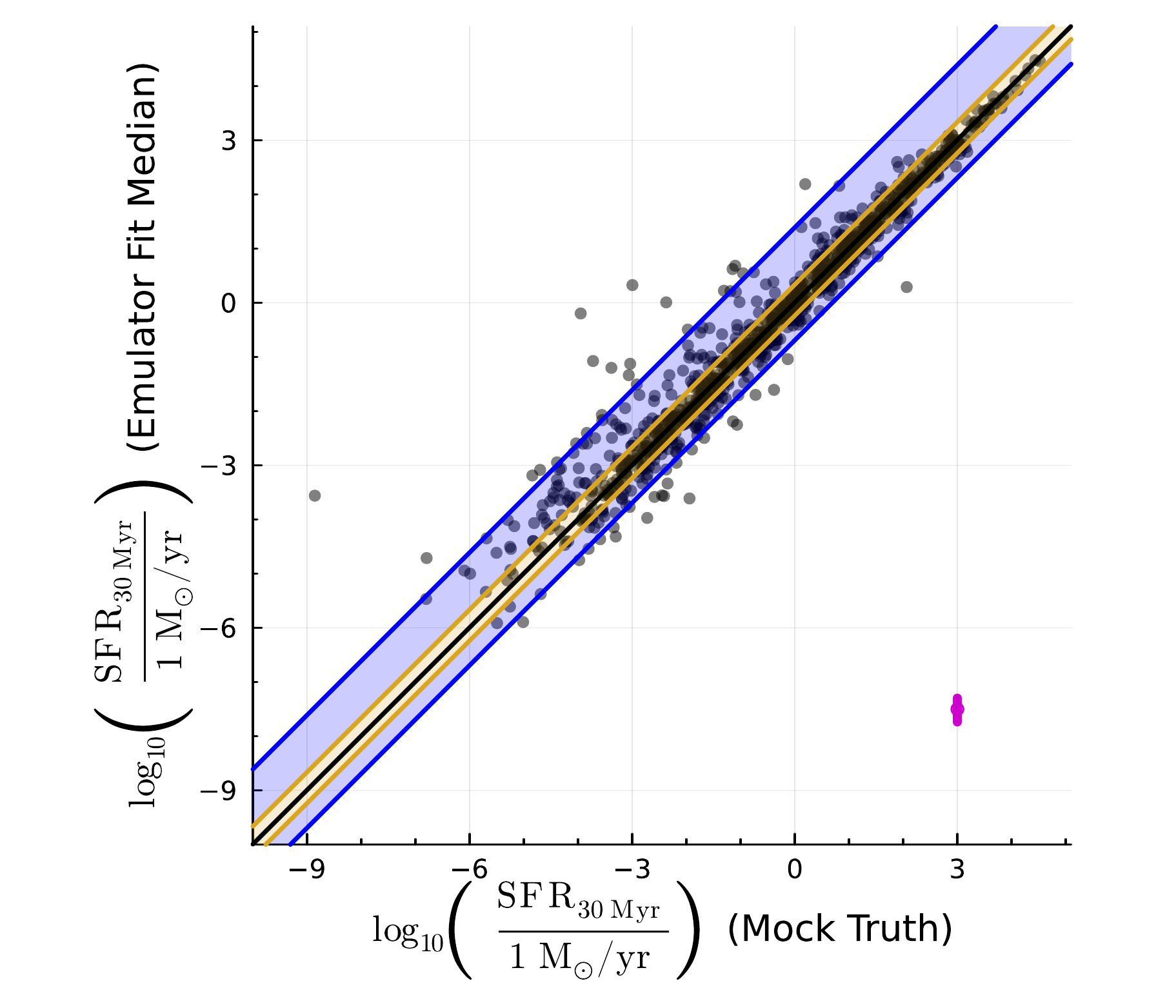}{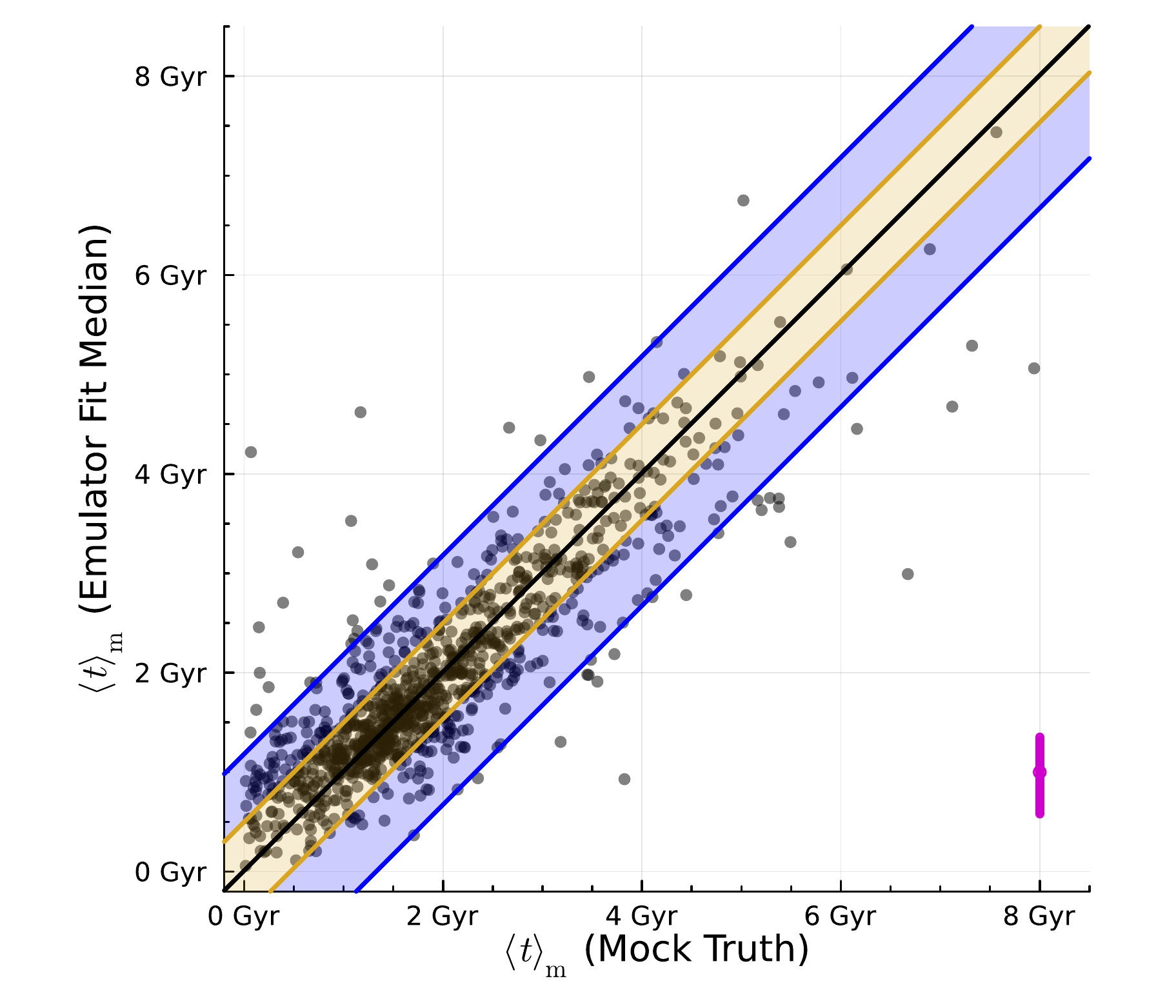}
    \caption{A test of \texttt{Prospector}'s ability to recover accurate estimates of various parameters from mock observations when paired with the 1024 node-per-layer emulator. Results are displayed similar to the \texttt{FSPS} comparisons shown in Figure \ref{fig:fspscompare}, with spreads somewhat wider than found in that test.}
    \label{fig:mockcompare}
\end{figure*}

\begin{figure*}
    \plottwo{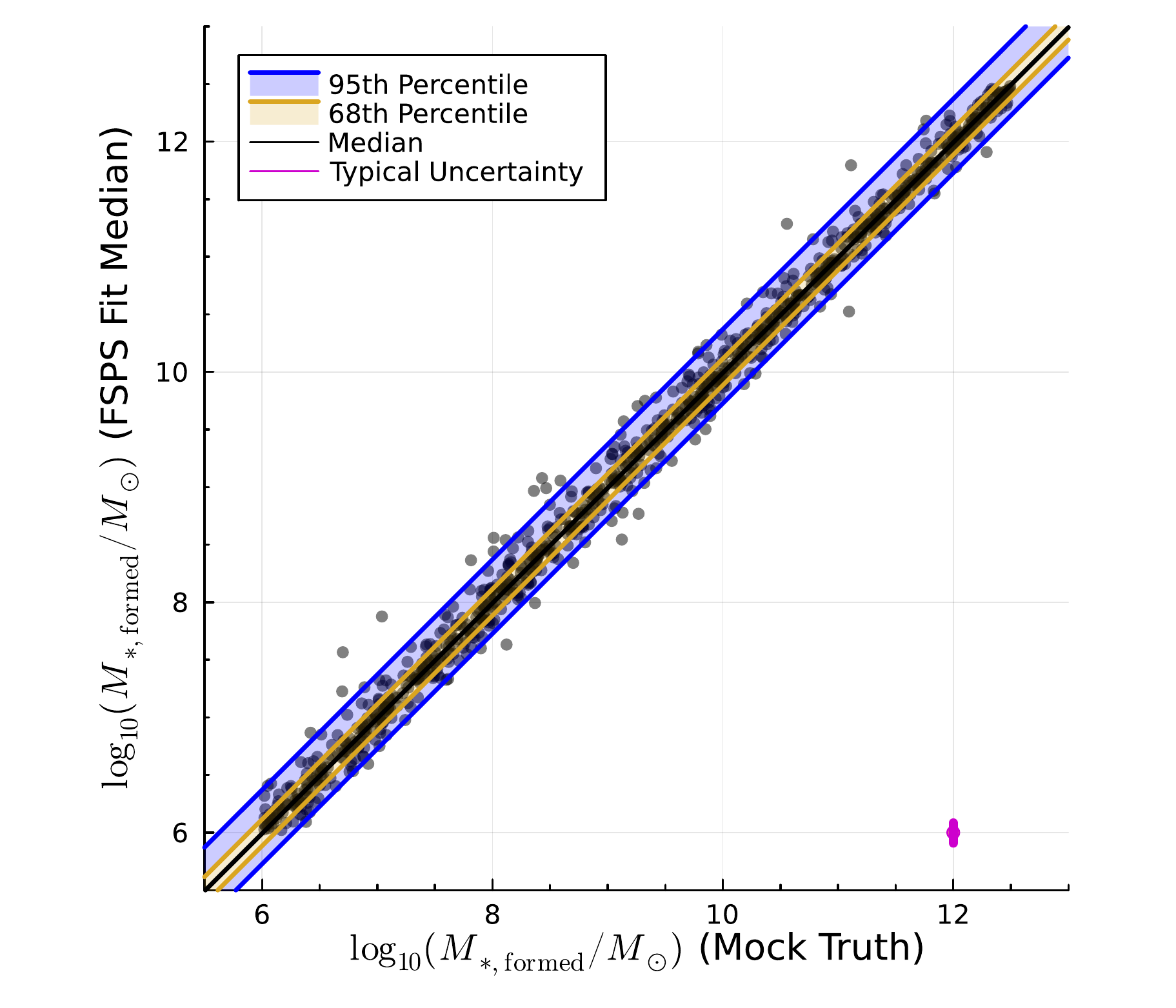}{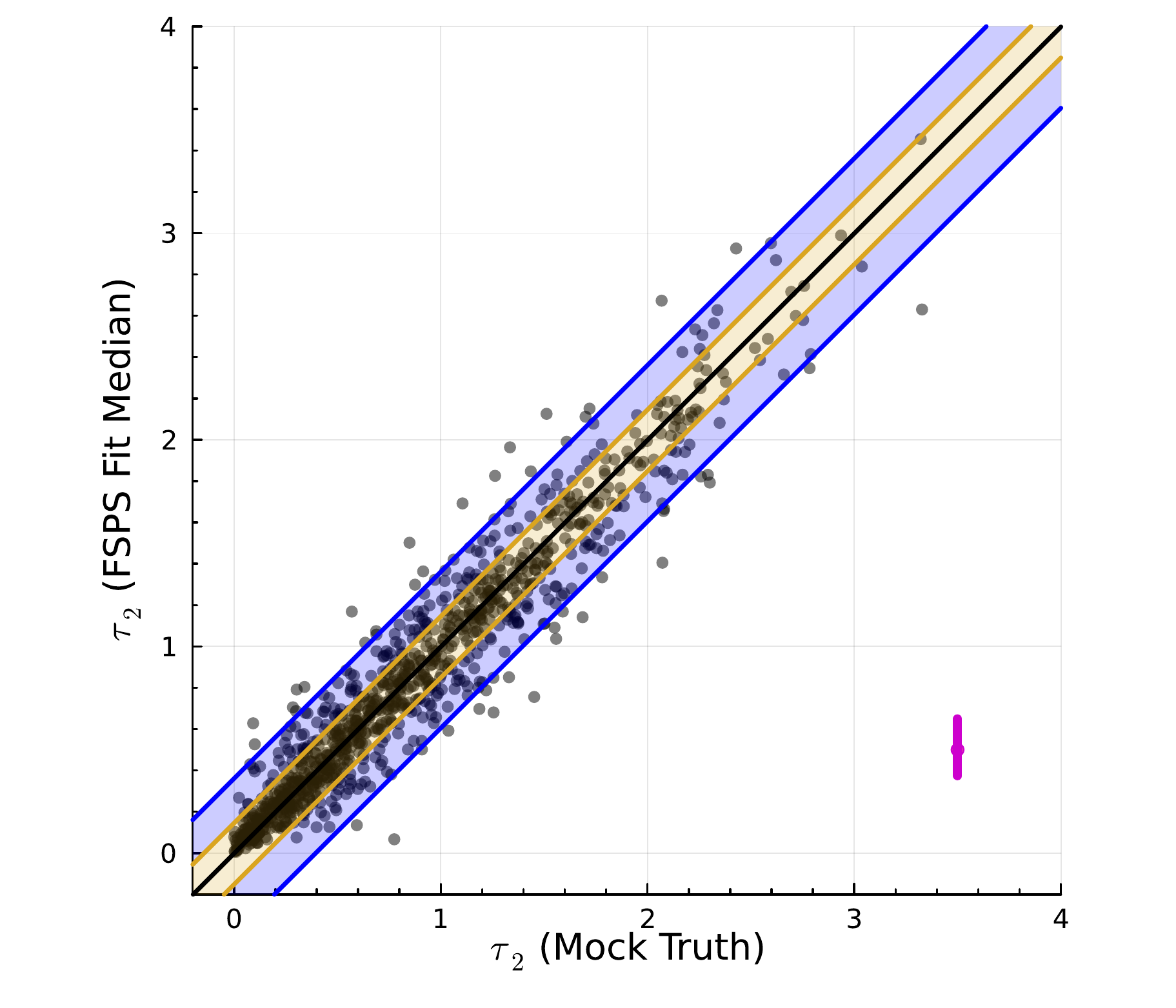}
    \plottwo{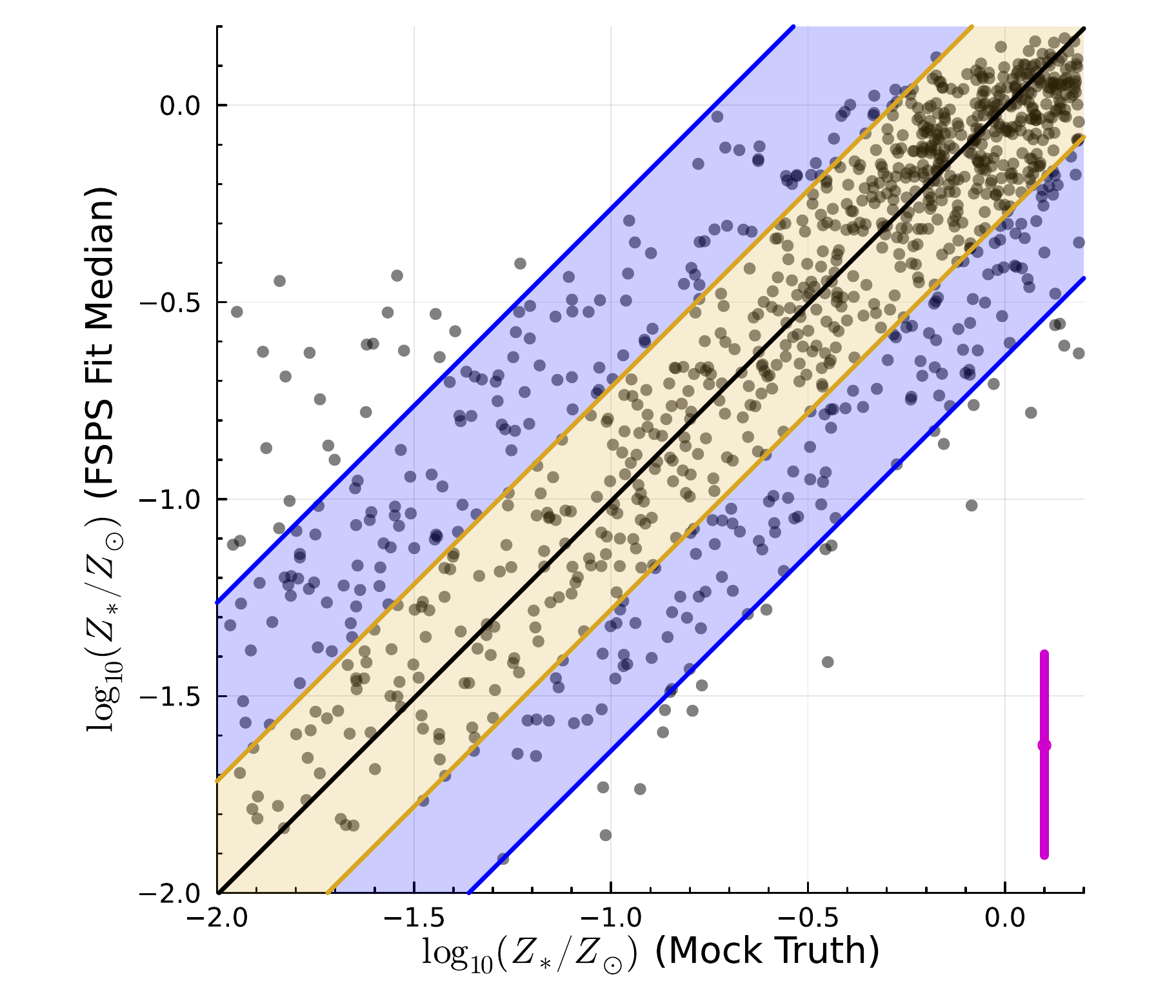}{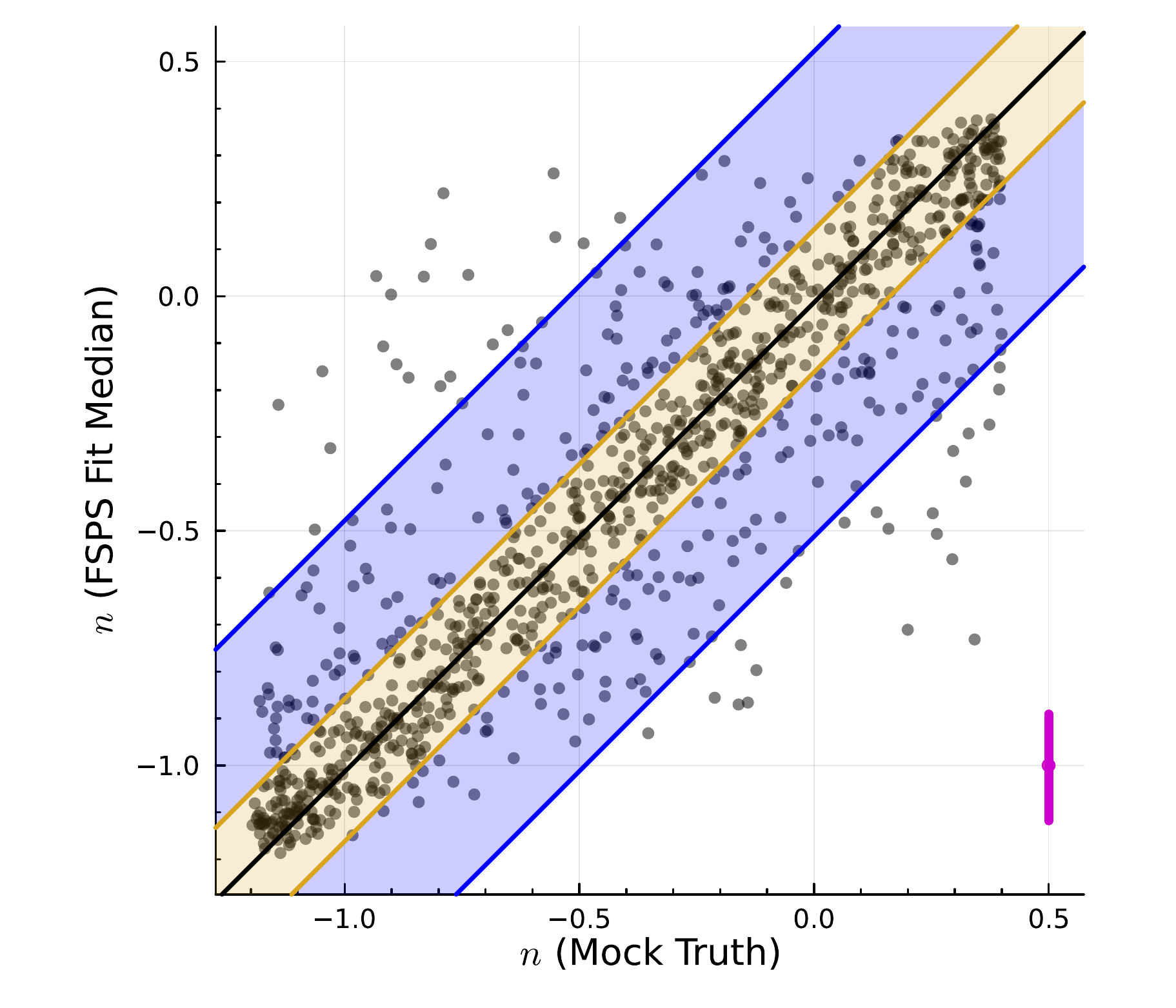}
    \plottwo{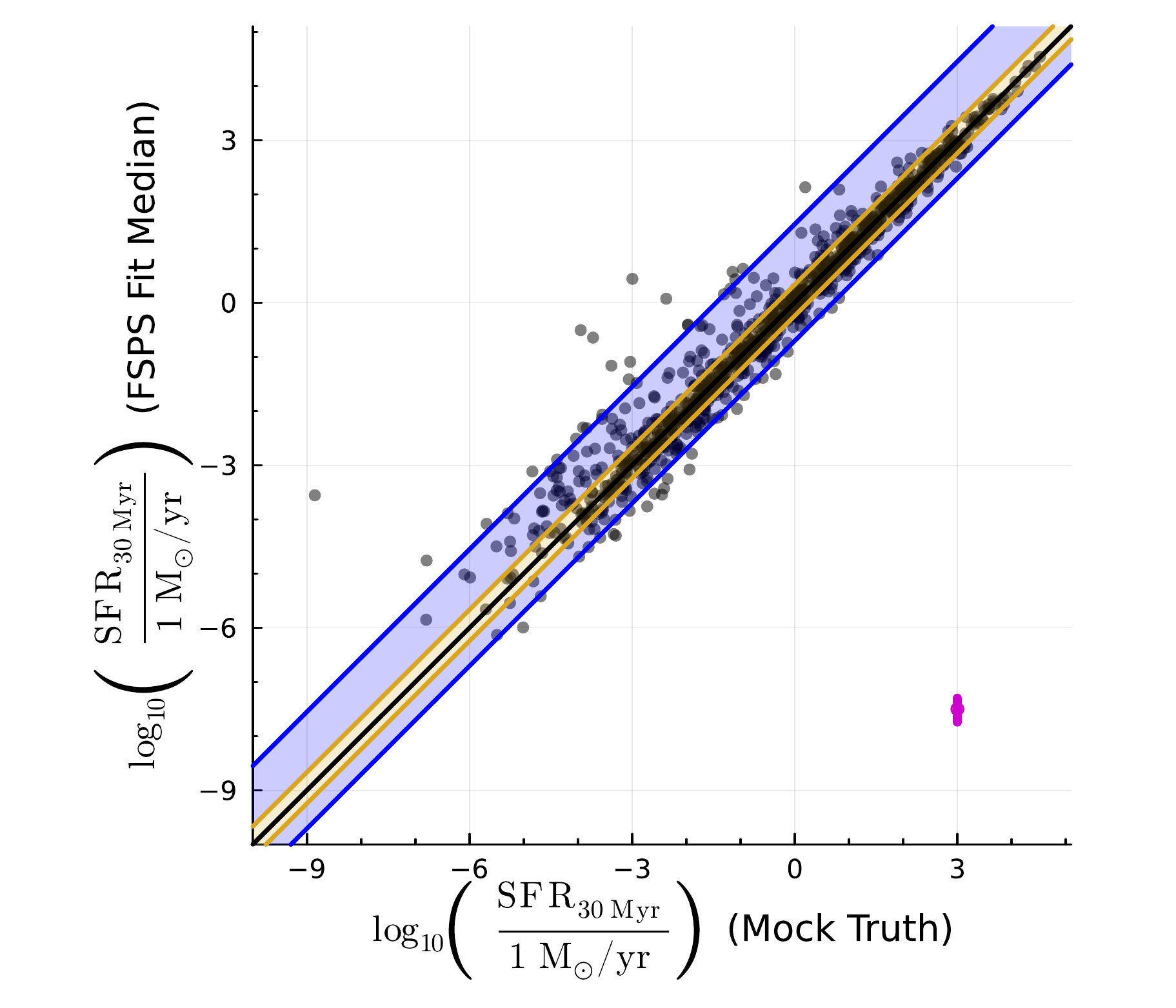}{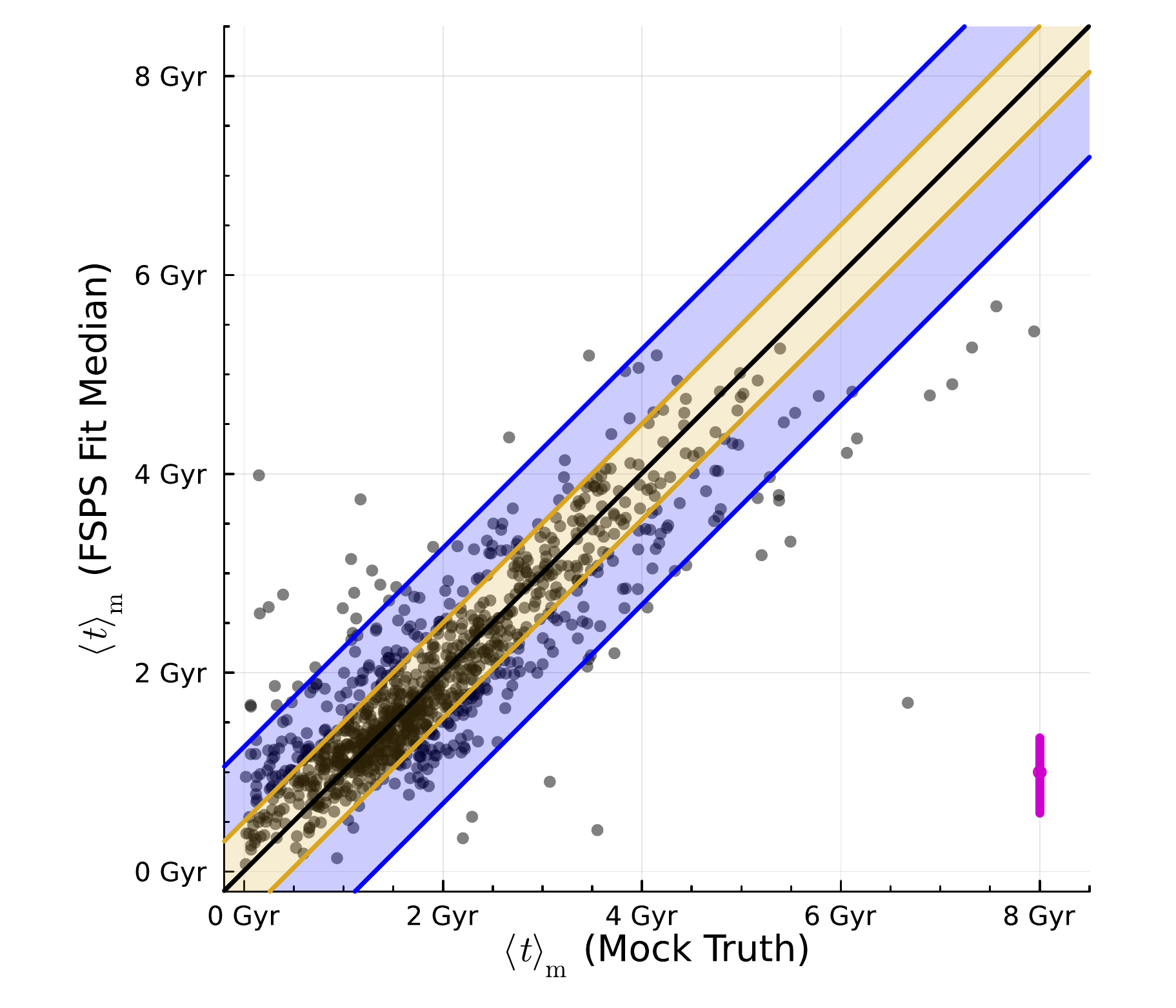}
    \caption{A test of \texttt{Prospector}'s ability to recover accurate estimates of various parameters from mock observations when paired with \texttt{FSPS}, similar to Figure \ref{fig:mockcompare}.}
    \label{fig:fspsmockcompare}
\end{figure*}

Additionally, while we have demonstrated that the emulator can recover posteriors that are consistent with the code it emulates (\texttt{FSPS}), we must also verify that these posteriors are actually representative of a galaxy's true physical properties, which can be evaluated using the fits to the sample of mock galaxies. The results of these fits are shown in Figures \ref{fig:mockcompare} and \ref{fig:fspsmockcompare}, the former for a comparison between 1024 node-per-layer emulator fits and the mock truth and the latter for a comparison between \texttt{FSPS} fits and the mock truth. Here we see that the posteriors recovered by both the emulator and \texttt{FSPS} are typically consistent with the true parameter values.

Using the same metric of measuring the 68\%-ile between the posterior median and the mock truth, we find that mock recovery performance is nearly identical between the two SPS methods, with typical differences of $0.15\ \textrm{dex}$ for the stellar mass formed, $0.32\ \textrm{dex}$ for the stellar metallicity, $0.50\ \textrm{dex}$ for the star formation rate over the most recent $30\ \textrm{Myr}$, and $0.23\ \textrm{dex}$ for the mass-weighted age. While the differences found here are larger than the scatter found between emulator and \texttt{FSPS} posterior medians in the previous 3D-HST test, and thus could potentially be explained by the constant $0.1\ \textrm{mag}$ random error built into the mock photometry, we exclude this possibility given that the differences between the emulator and \texttt{FSPS} posteriors medians on these mock observation fits are nearly identical to the 3D-HST results. Thus, we conclude that the 1024 node-per-layer emulator posterior medians are of the same high quality as the \texttt{FSPS} posterior medians, and the emulator does not limit \texttt{Prospector}'s ability to recover the true parameter values for a given SED.

\begin{figure}
    \plotone{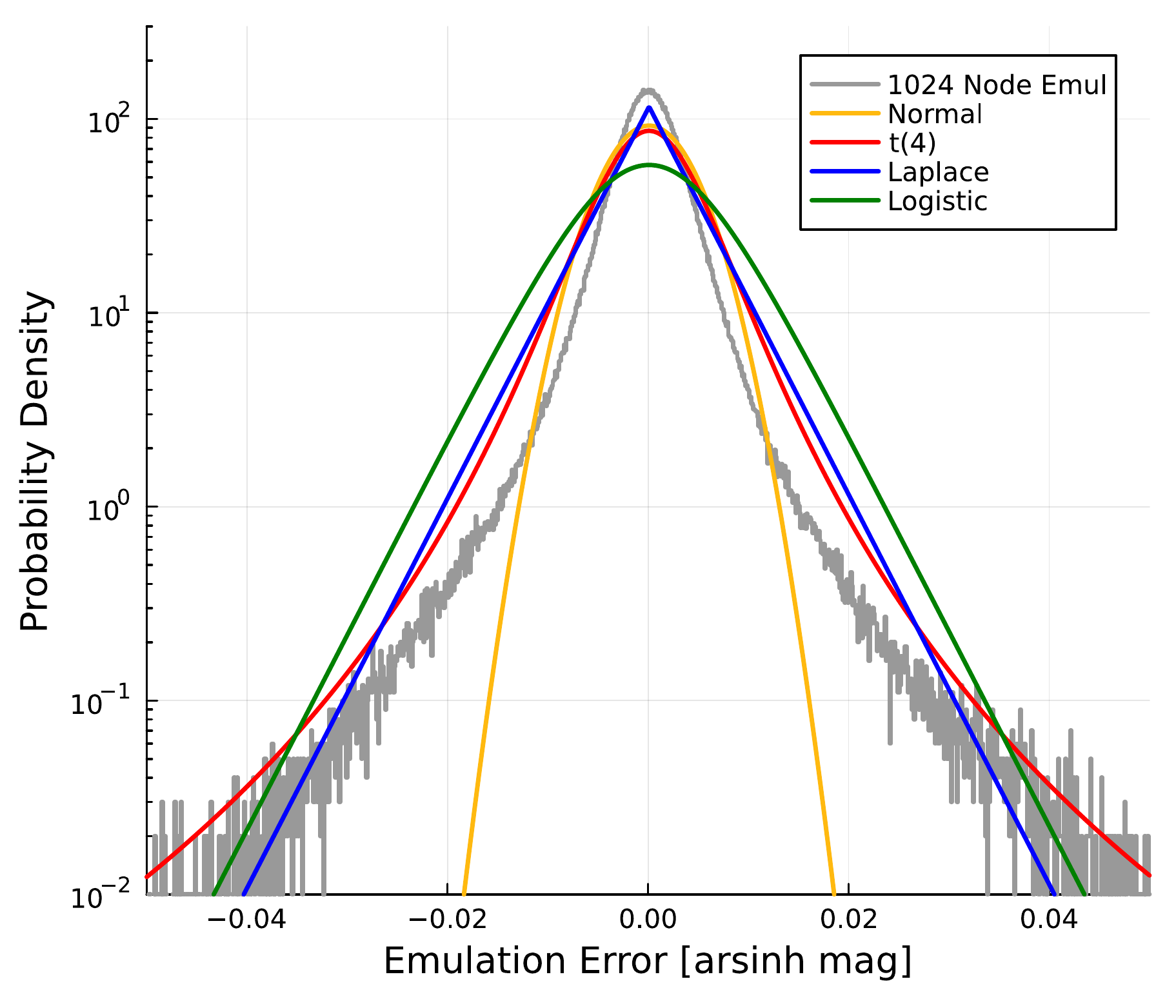}
    \caption{A demonstration of the non-Gaussianity of the emulation errors. Plotted in gray is a histogram of the test set emulation errors for the 1024 node-per-layer emulator's JWST NIRCam F115W filter, while plotted in gold, red, blue, and green are the probability density functions for normal, Student's $t(\nu=4)$, Laplace, and logistic distributions for visual comparison, whose location and scales are set to the mean and standard deviation of emulation errors for this filter.}
    \label{fig:tails}
\end{figure}

\begin{figure*}
    \plottwo{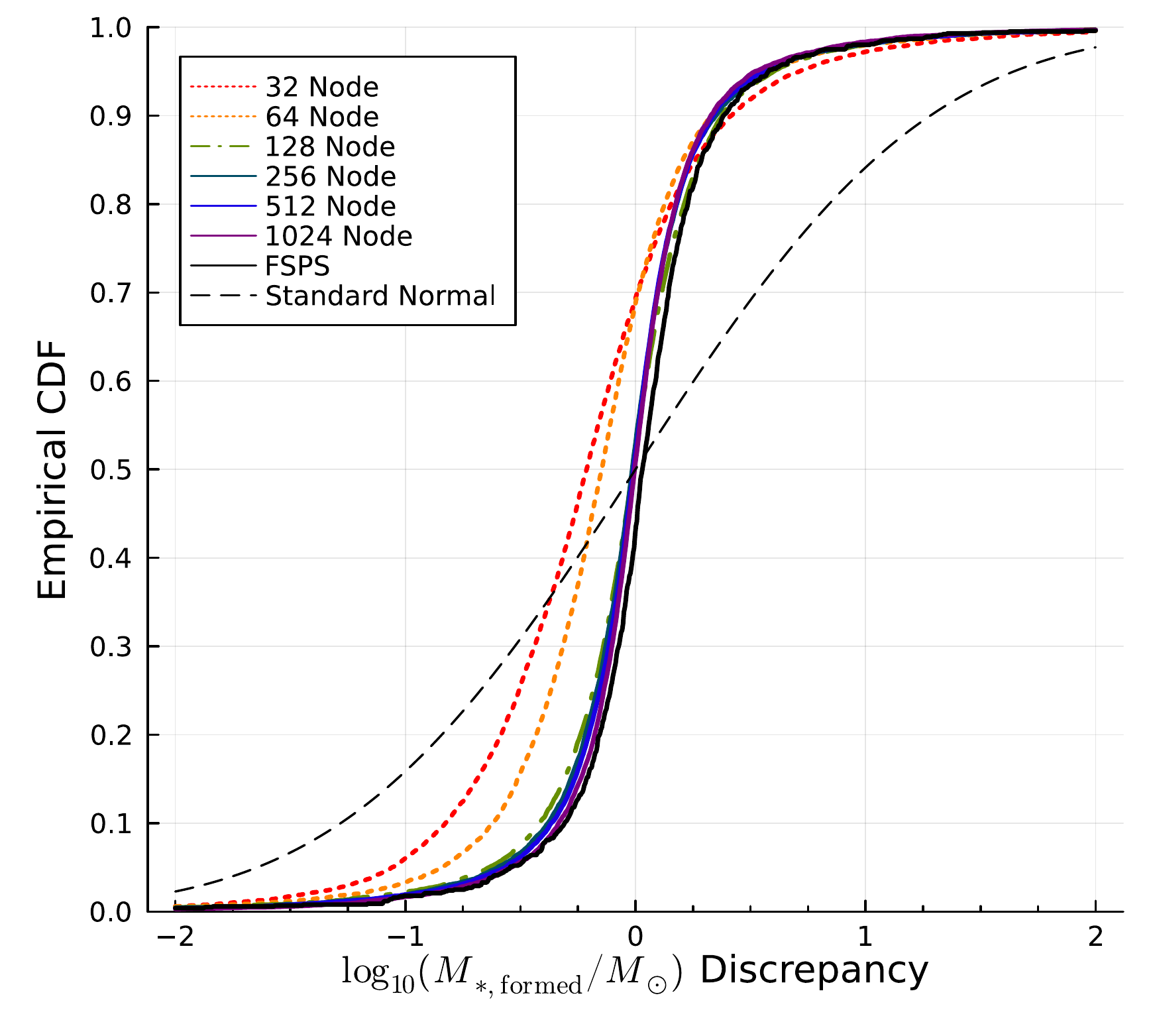}{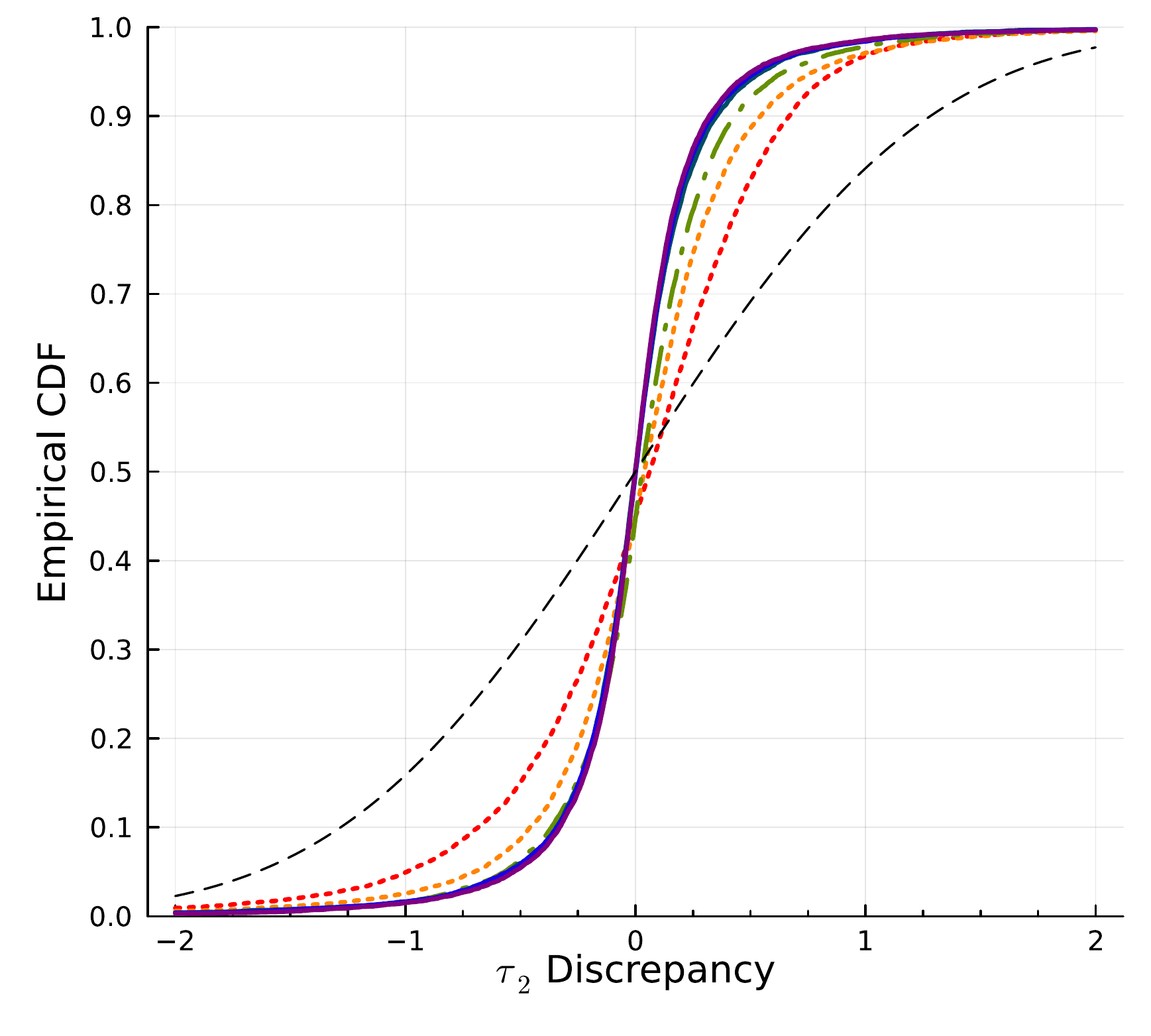}
    \plottwo{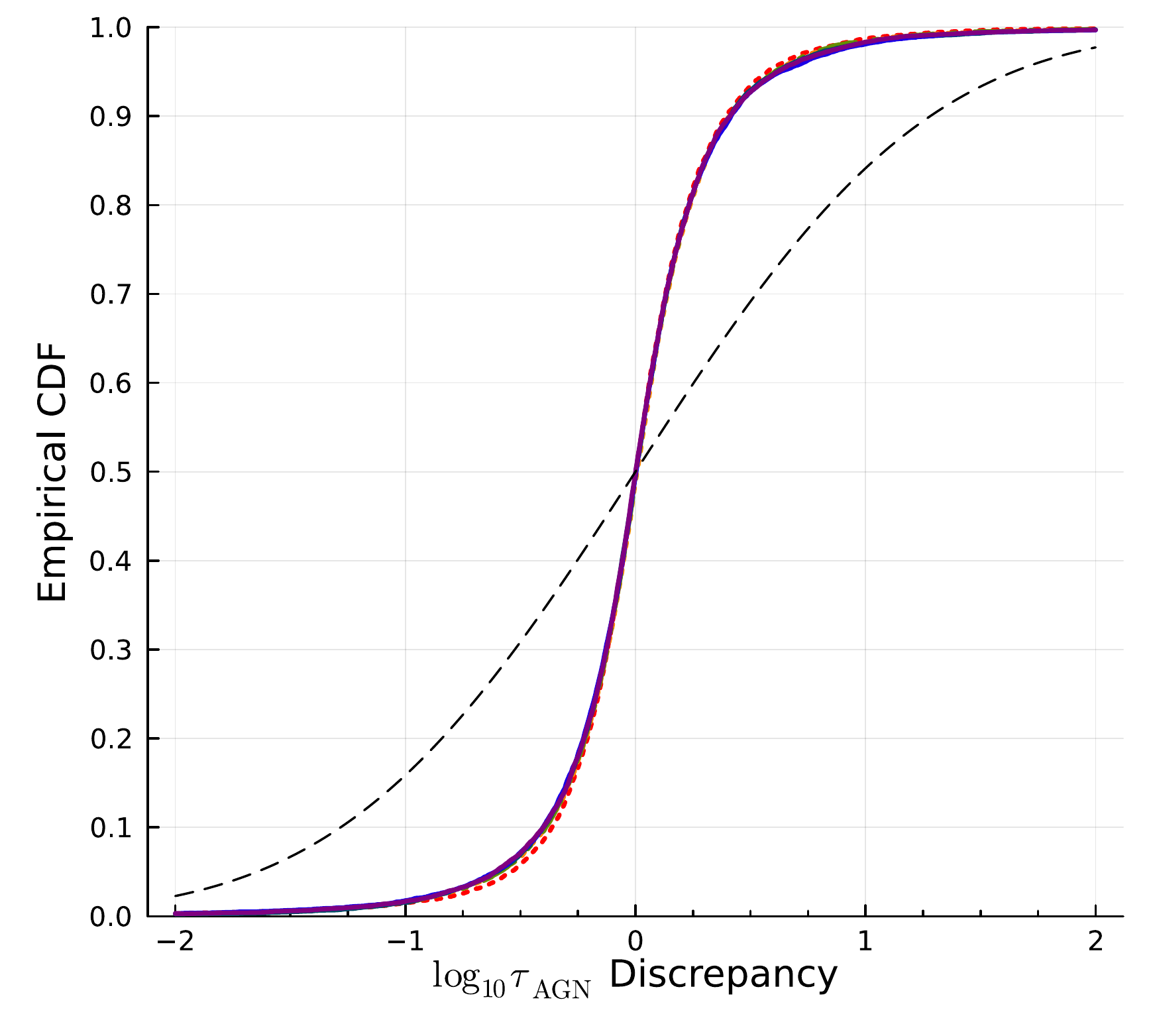}{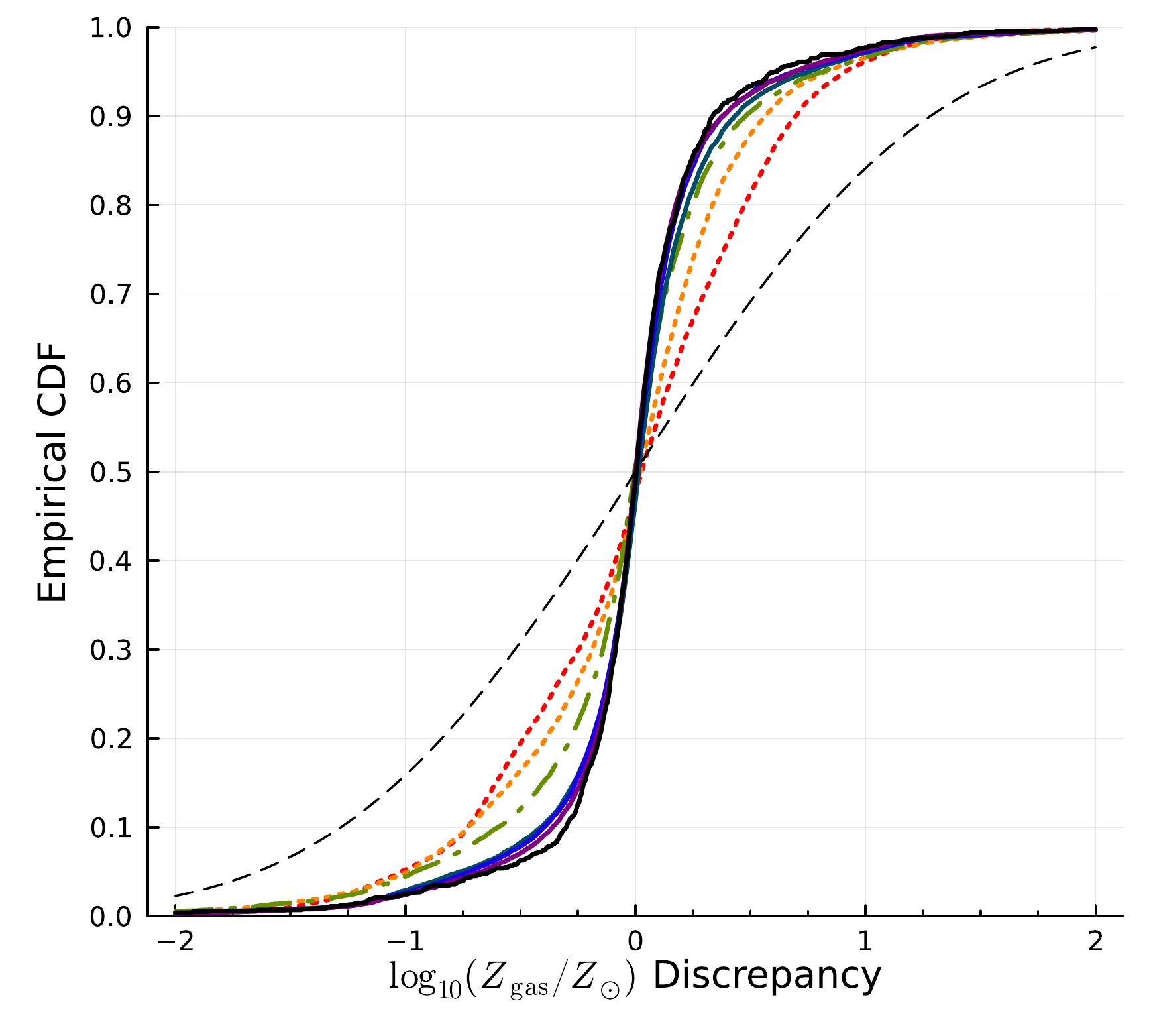}
    \caption{Empirical cumulative distribution functions (ECDFs) for the discrepancy statistic for the cumulative stellar mass formed $M_{*,\textrm{formed}}$, diffuse dust optical depth $\tau_2$, AGN dust optical depth $\tau_\textrm{AGN}$, and gas-phase metallicity $Z_\textrm{gas}$. This is evaluated between the 1024 node-per-layer emulator and another emulator. As ANN width increases, the ECDF tends to asymptotically approach the ECDF reached by the 1024 node-per-layer fit, which denotes the distribution of discrepancies encountered from sampling variance alone. The CDF of a standard normal is shown for reference and to demonstrate the non-Gaussianity of the distributions of discrepancies. Solid lines denote architectures deemed sufficiently accurate and precise for SED fitting in this work, while dash-dot lines denote emulators deemed marginally sufficient and dotted lines denoted emulators deemed insufficient.}
    \label{fig:ecdf}
\end{figure*}

Next we investigate catastrophic outliers, which represent a serious mode of failure in this test. In this context, a catastrophic outlier is obtaining a posterior distribution from an emulator-based fit that would \emph{exclude} (with high probability) a solution obtained from a traditional \texttt{FSPS} fit. One potential source of these errors is emulation errors not accounted for in the estimation of emulation precision. In Figure \ref{fig:tails}, we observe that emulators' errors tend to have tails far longer than that of a Gaussian distribution, making extreme errors more common than one would expect by assuming a normal distribution whose $\sigma$ is set to the estimated precision of an emulator.

To assess the frequency at which these catastrophic emulation errors yield inaccurate posteriors, we define a discrepancy statistic
\begin{equation}
    Q_{i;j,k} = \frac{m_{i,j} - m_{i,k}}{\sqrt{\sigma_{i,j}^2 + \sigma_{i,k}^2}},
\end{equation}
where $m_{i,j}$ is the posterior median value for parameter $i$ from a fit using SPS method $j$ and $\sigma_{i,j}$ is the posterior width (defined as half the width between the marginalized posterior's $15.87\%$ and $84.13\%$ percentiles) for parameter $i$ using SPS method $j$. This is largely analogous to a $z$-score. Given that sampling is an intrinsically random process, it would not necessarily be unexpected for one to obtain different results with independent realizations of the same fit, even if the fluxes were reproduced. Accordingly, we run fits using the same 1024 node-per-layer emulator twice in order to assess the standard rate of catastrophic outliers caused by sampling uncertainty rather than emulator inaccuracy.
\begin{figure*}
    \plottwo{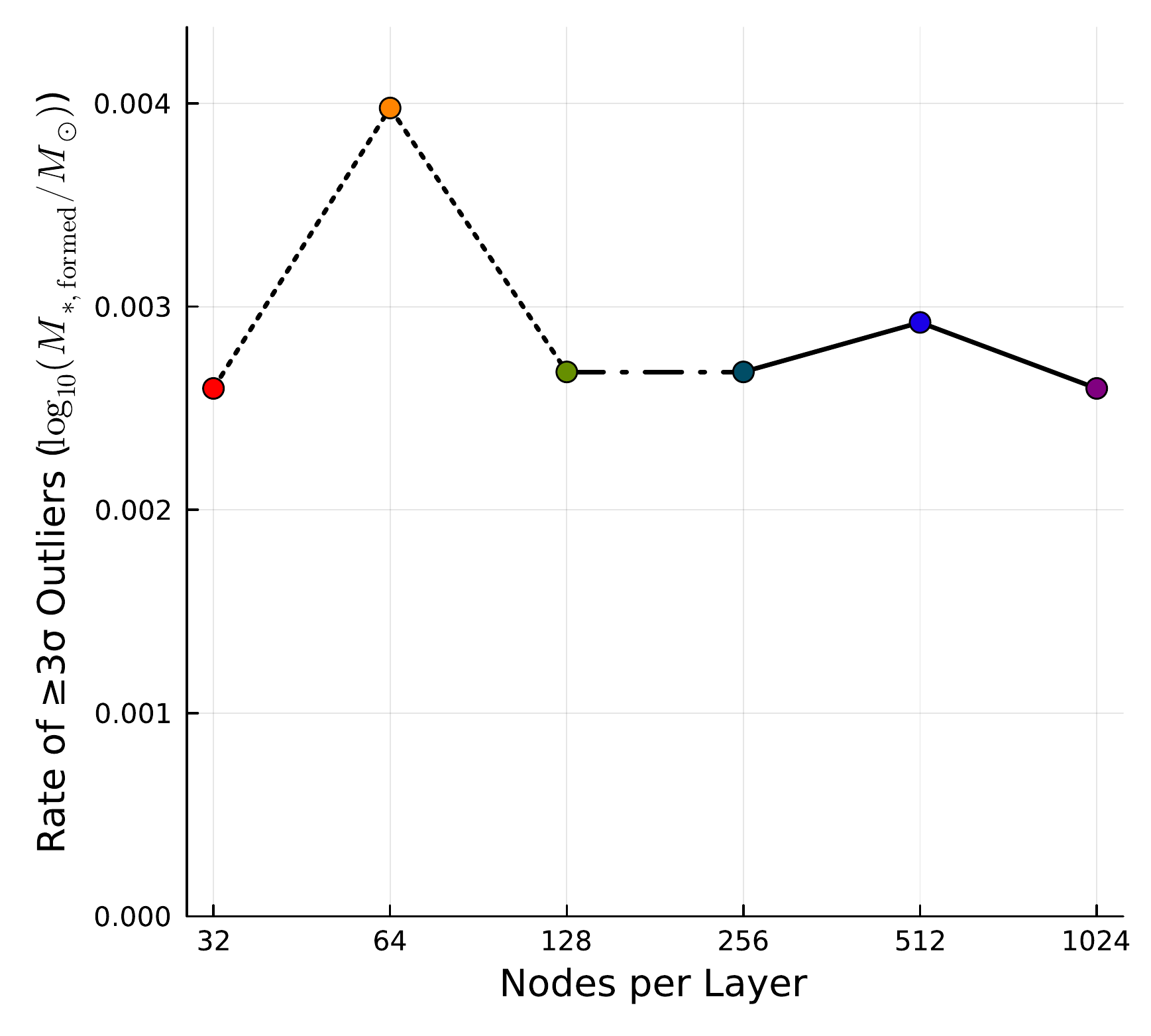}{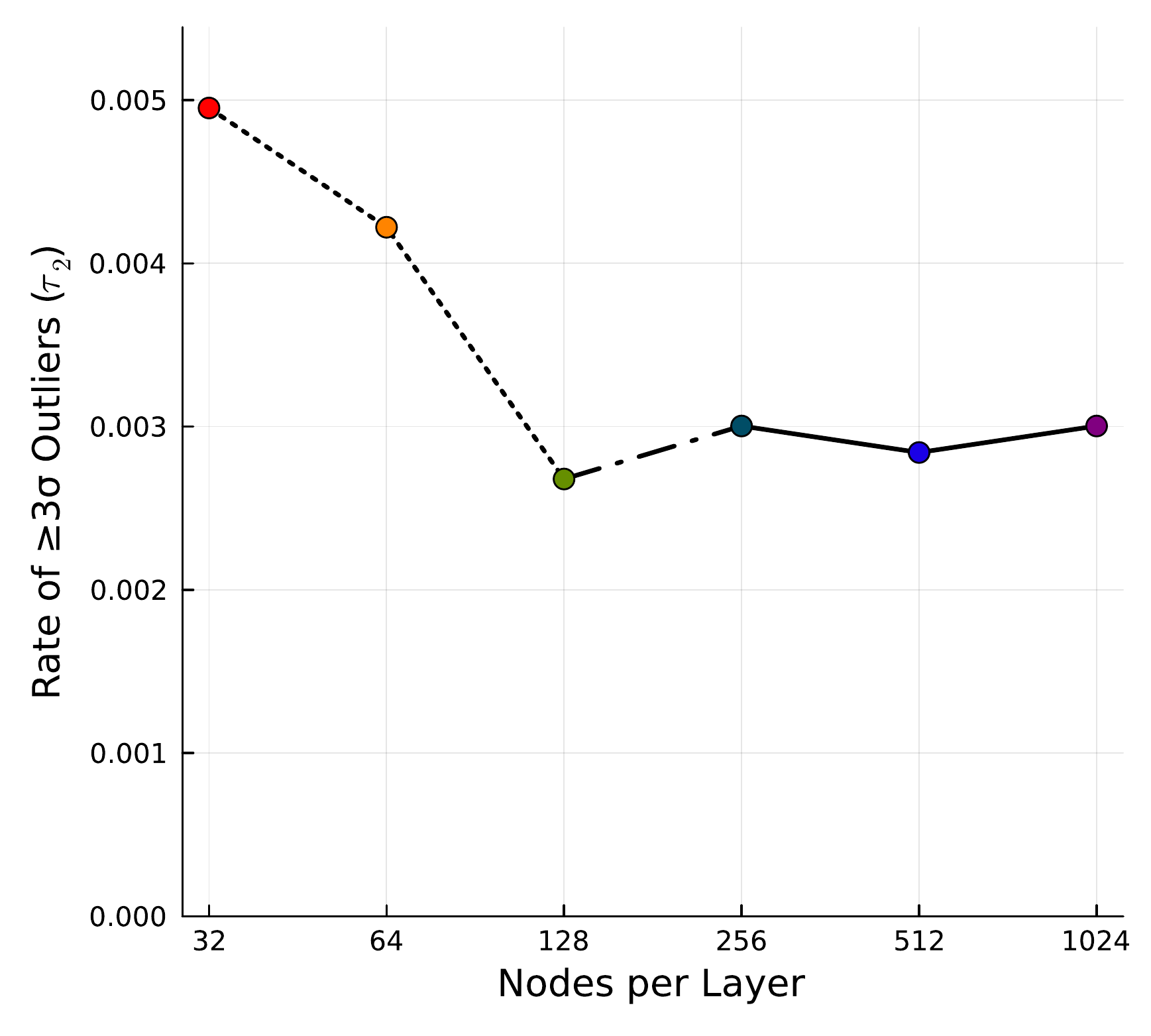}
    \plottwo{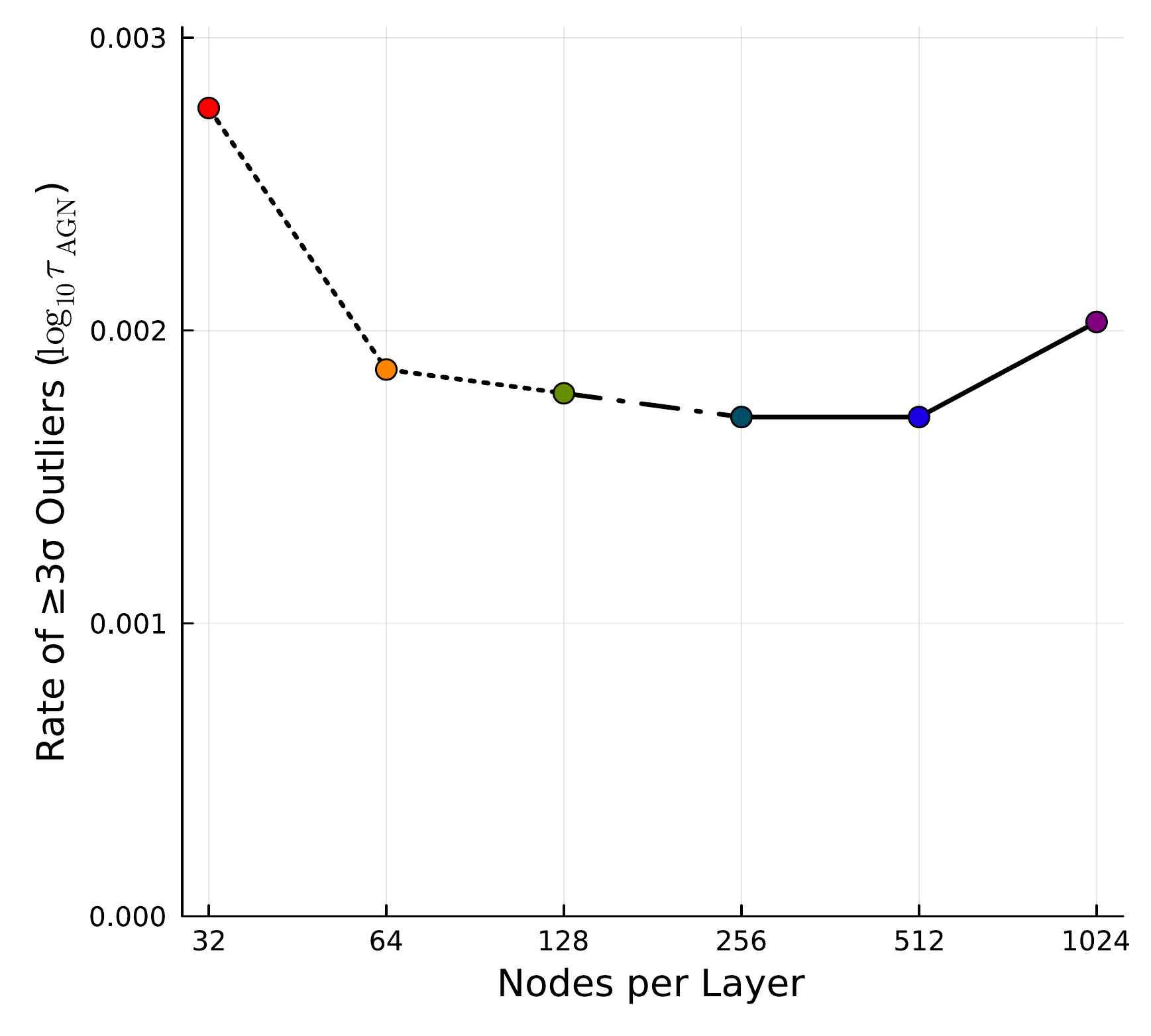}{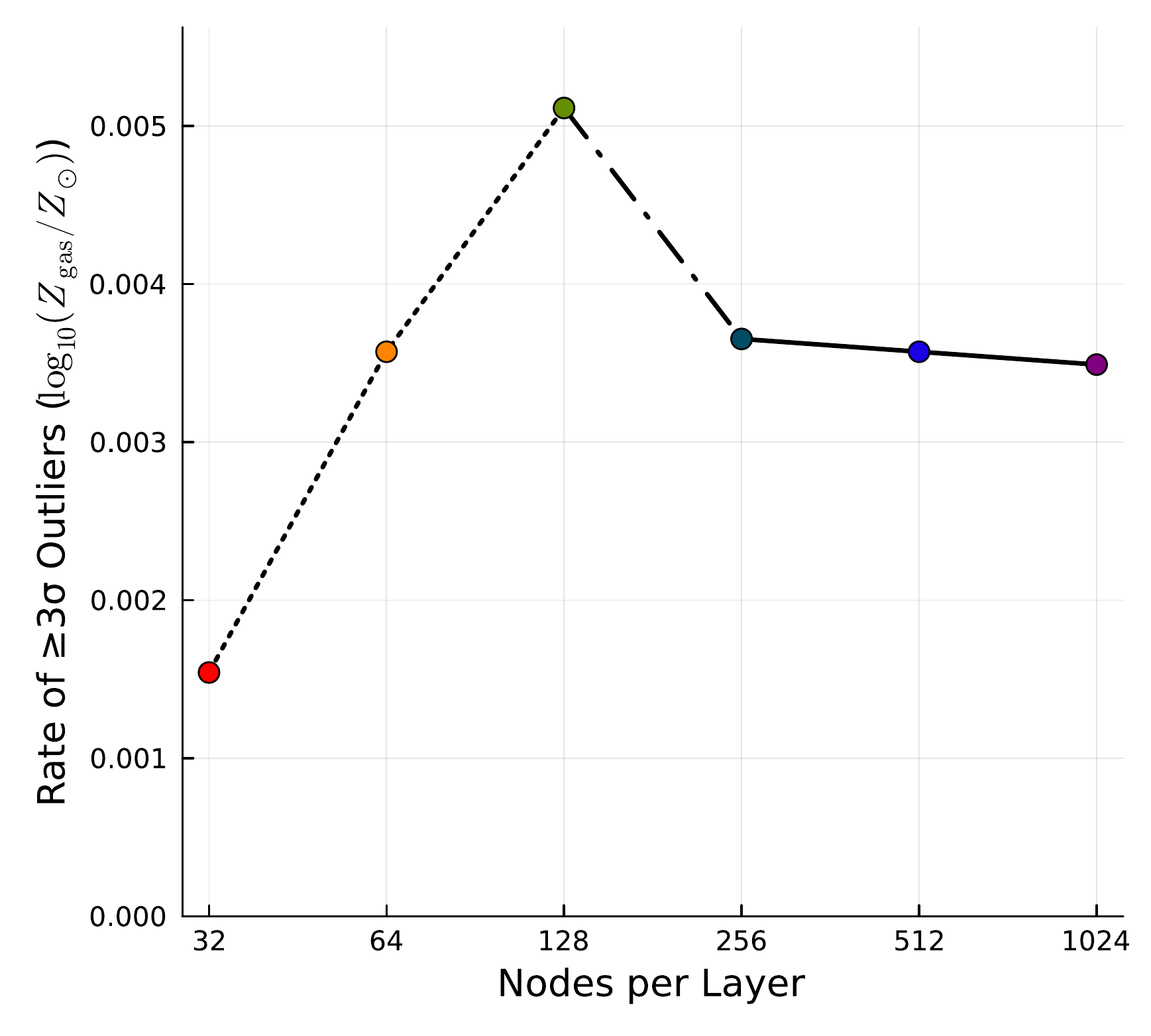}
    \caption{Rate of $3\sigma$ outliers (i.e. rate of $\lvert Q_{i;j,k} \rvert \geq 3$ discrepancies) for 4 of the fitted parameters. We see the $3\sigma$ outlier rates generally approach the rate found in the 1024 node-per-layer emulator, which is caused only by sampling variance, as network width increases.}
    \label{fig:threesigma}
\end{figure*}

In Figure \ref{fig:ecdf}, we view the empirical cumulative distribution functions (ECDFs) of this discrepancy statistic for the cumulative stellar mass formed $M_{*,\textrm{formed}}$ and diffuse dust attenuation optical depth $\tau_2$, both of which are parameters that should be well-constrained by panchromatic SED-fitting \citep[e.g.][]{burgarella05}, in addition to AGN dust optical depth $\tau_{\textrm{AGN}}$ and gas-phase metallicity $Z_{\textrm{gas}}$, which should be less well-constrained. We note that the distributions asymptotically approach the distribution reached by the 1024 node-per-layer emulator, to the extent that the 256 and 512 node-per-layer emulators produce nearly identical distributions of discrepancies to repeat runs of 1024 node-per-layer fits. This indicates that emulators with at least 256 nodes per layer are sufficiently accurate for fitting these 3D-HST SEDs.

We also investigate the rate at which catastrophic outliers occur, which we define as $\lvert Q_{i;j,k} \rvert \geq 3$ (analogous to a $3\sigma$ outlier). Similarly, we see in Figure \ref{fig:threesigma} that these tend to asymptotically approach the 1024 node-per-layer rates as ANN width increases. But, it should be noted that these are not strictly \emph{decreasing} with increasing ANN width for all parameters as one may expect -- in fact, for some parameters (e.g. $Z_{\textrm{gas}}$), the rate actually \emph{increases} with increasing ANN width, while others exhibit noisy behavior (e.g. $M_{*,\textrm{formed}}$). In the case of $Z_{\textrm{gas}}$, this is likely due to the fact that the parameter is rarely well-constrained -- it plays a minor role in a galaxy's rest-frame optical spectrum and is difficult to detect in photometry. Since larger observational errors are assumed with smaller ANN emulators, it is likely that the sampler is correspondingly returning the prior as a result of the larger uncertainties. These wide posteriors create scenarios where it is exceedingly difficult to produce a catastrophic error in some parameters, resulting in the noted performance. All told, the asymptotic behavior is the primary result -- increasing ANN width provides increasingly similar catastrophic outlier rates, with all emulators with at least 256 nodes per layer providing rates within $\sim 30\%$ of the target rate as measured by the repeated 1024 node-per-layer fits.

\section{Discussion}\label{sec:discuss}

In this section, we discuss the implications of our results and possible avenues for future work on efficient SPS computation.

\subsection{Optimal Architecture}\label{sec:optarch}

The key question this work aims to answer is how to determine what ANN architecture one should use when fitting a given set of galaxy SEDs. We aim to find the fastest (i.e. simplest) emulator that achieves posterior accuracy goals; however, the only metrics we have available for evaluating an emulator's ability to recover accurate posteriors (without running the entire analysis performed in this work), such as emulation precision, are all in \emph{flux} space. Due to the complex and degenerate nature of the parameter-SED relationship, it is difficult to correlate a given parameter accuracy target with a flux precision target.

A natural target would be to compare the emulation precision to the precision of the observations being fit. If provided with only the results shown in Figure \ref{fig:prec}, one would conclude that all emulators except for the 32 node-per-layer emulator would potentially be sufficiently accurate and precise since the precision in some or all of their filters exceeds the precision of the observations (i.e. typical emulation errors of $\sim\!5\%$ or better). If we make a further constraint that \emph{all} emulated filters must provide precision better than the observational uncertainty floor of $5\%$, then the 64 node-per-layer emulator would also be excluded. In addition, one would note from Figure \ref{fig:cov} that correlated emulation errors are smaller in magnitude and predominantly concentrated to filters with similar effective wavelengths for wider architectures. Then, since we aim to choose the fastest architecture that meets our accuracy and precision targets (i.e. Figure \ref{fig:emulspeed}), we would conclude from these \emph{a priori} results that the $128$ node-per-layer emulator would be the correct choice for this physical model and set of observations. However, if one is not as concerned about the run time of the emulator, then increasing the width is desirable in order to minimize the emulator's contribution to the overall error budget.

We can then verify whether or not this decision to use the 128 node-per-layer emulator for SED-fitting would actually lead to accurate posterior distributions. We find in Figure \ref{fig:ecdf} that the three emulators consisting of at least 256 nodes-per-layer all converge to common distributions of $Q_{i;j,k}$, with catastrophic outlier rates approaching the ``ideal'' value found by running duplicate 1024 node-per-layer fits as demonstrated in Figure \ref{fig:threesigma}. At this level of flux precision, the discrepancies of the posteriors can be solely attributed to sampling variance. In other words, the improved flux precision that comes from using emulators with more than 256 nodes per layer is \emph{not} being met with a significant improvement in the recovery of accurate posterior distributions. Therefore, this \emph{a posteriori} analysis would conclude that the 256 node-per-layer emulator is the optimal choice for this task.

Thus, there is a small discrepancy between the conclusions found by these two avenues of inquiry: the \emph{a priori} results would indicate that the 128 node-per-layer emulator should be chosen due to it being simplest emulator whose precision exceeds the observational precision (see Figure \ref{fig:prec}), while the \emph{a posteriori} results would indicate that the 256 node-per-layer emulator should be chosen due to it being the simplest emulator that achieves similar distributions of discrepancies (see Figure \ref{fig:ecdf}) and similar rates of catastrophic outliers (see Figure \ref{fig:threesigma}) up to the limit imposed by sampling uncertainty. This demonstrates that the emulation precision acts as a good rule-of-thumb for choosing an appropriate architecture but fails to tell the full story. This is likely partly due to the long tails in emulators' errors (i.e. Figure \ref{fig:tails}), which result in the emulator producing significant errors (e.g. larger than $5\%$ in flux) more often than one would predict when assuming Gaussian errors. Furthermore, not all parameters are affected equally by this, and thus the choice of parameters being fit must also enter into consideration when selecting an appropriate architecture. Figure \ref{fig:ecdf} shows that $M_{*,\textrm{formed}}$ and $\tau_{\textrm{AGN}}$ are largely unaffected by the errors introduced by the 128 node-per-layer emulator, while $\tau_2$ and especially $Z_{\textrm{gas}}$ are more affected. For this reason, we deem the three emulators with at least 256 nodes per layer as being sufficient for use in fitting galaxy SEDs, the 128 node-per-layer emulator as being \emph{marginally} sufficient (e.g. useful for cases where $\tau_2$ and $Z_{\textrm{gas}}$ are not being fit), and the remaining two emulators (32 and 64 nodes per layer) as being insufficient.

Based on these results, we propose a method for choosing a \emph{minimal} emulator architecture for a given physical model and set of observations. First and foremost, more complex physical models require more complex emulator architectures for a given precision target due to the more complicated input/output relationship that these models pose. Thus, to prevent the emulator from adding significant noise to constraints on galaxy parameters, an architecture must be chosen that provides sufficient accuracy and precision in order for the emulator to exceed the observational uncertainties in all of the filters in question. Then, the covariance matrix must also be examined to ensure that an assumption of uncorrelated errors is valid so that one can convolve an emulator's $\sigma_i$ precision with the observational uncertainty for that filter. This work has shown these steps, when combined, provide emulators capable of producing posteriors that are as accurate as sampling variance will permit for most parameters. However, a slight improvement in the emulator's precision (e.g. further doubling the network width, which produces emulation precision that is approximately $\sim5$ times better than the observational uncertainties) can deliver a notable improvement in parameter recover for some parameters, and may provide more confidence in posterior accuracy in situations where absolute accuracy is desired and the fastest possible speed is not strictly necessary.

\subsection{Distributions for Training, Testing, and Priors}\label{sec:dists}

\begin{figure*}
    \plottwo{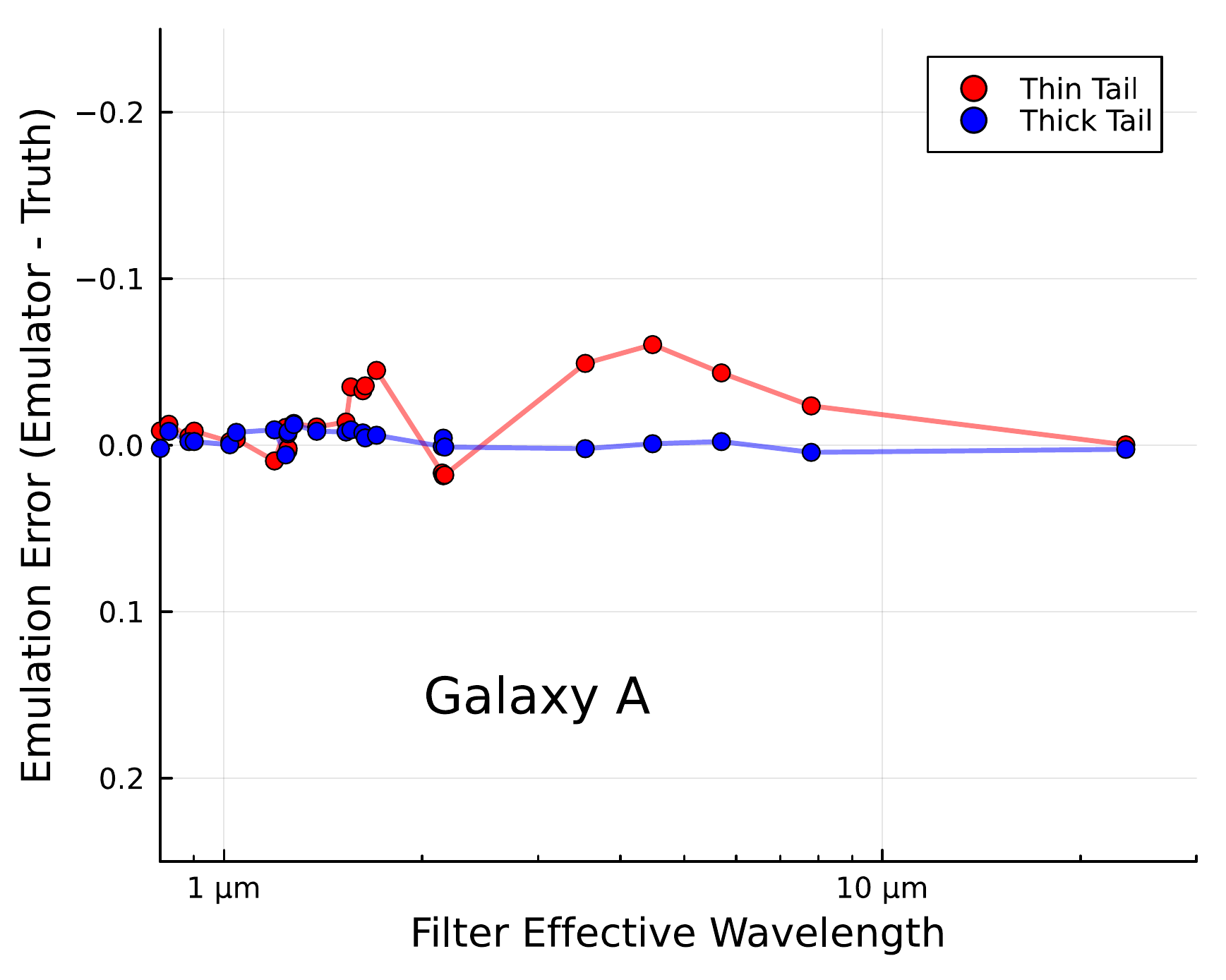}{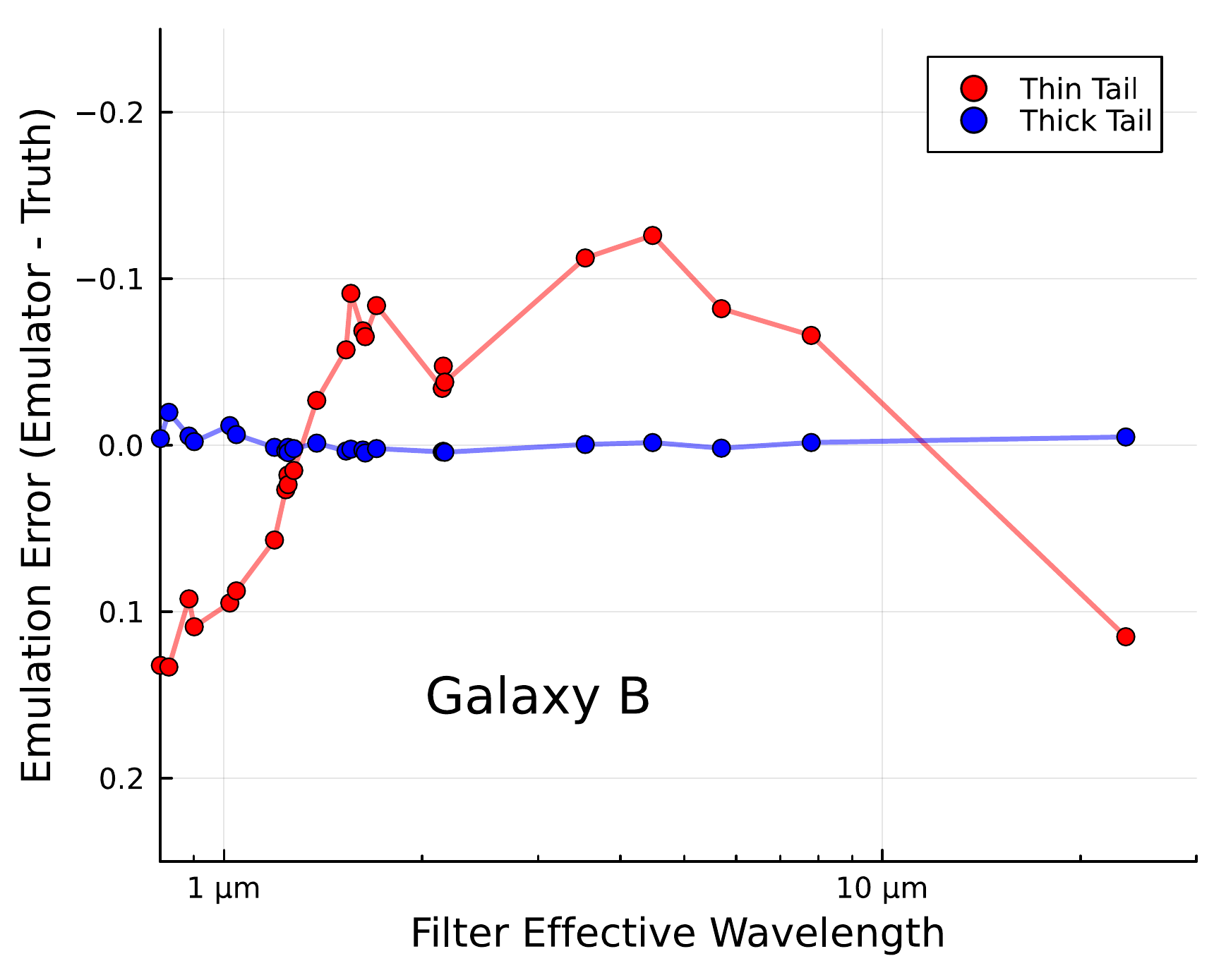}
    \caption{The emulation precision of the two emulators used to fit the two mock galaxies described in Section \ref{sec:dists}. Plotted in red is the emulation error for the emulator whose training set used the thinner prior distributions, and plotted in blue is the emulation error for the 1024 node-per-layer emulator described in this work. In both cases, the emulation error is evaluated using the true parameters values for the mock observations. We see that the precision of the thin tail emulator is dramatically worse for Galaxy B, whose SFH parameters are extreme, while the thick tail emulator does well for both galaxies.}
    \label{fig:extremephot}
\end{figure*}

In this work, we deviate from previous works \citep[e.g.][]{alsing20,kwon22} in not using our prior distributions as the probability distributions used for generating our training and test sets. This is a subtle change, but an important one -- while informative priors downweight the probability calculation in portions of parameter space that are either known or highly suspected to be rarely occupied by real galaxies, using these same non-uniform distributions for generating training sets will have the unintended consequence of also causing the emulator to maximize its errors for rare objects.

For example, the prior distribution used for the adjacent SFR ratios (which are SFH parameters controlling the ratio of SFR in adjacent time bins) is an extremely narrow distribution, providing almost no prior density near the upper and lower limits of the prior (which are values corresponding to objects which have rapidly increasing or decreasing SFRs in some portion of their SFHs). Specifically, only $\sim 0.2\% $ of all samples in the training set have $4 \leq \lvert \log_{10}\left(\textrm{SFR}_i/\textrm{SFR}_{i+1}\right) \rvert \leq 5$.

To demonstrate the effect this subtlety can have, we generate two additional mock galaxy observations (denoted Galaxy A and Galaxy B) using the same specifications as the mock catalog created in Section \ref{sec:sps} (including random $0.1\ \textrm{mag}$ observational errors) but with manually-specified SPS parameters. These two galaxies only differ by their SFH: Galaxy A represents a galaxy whose SFH parameters are located at the center of the training set and prior distributions (i.e. $\log_{10}(\textrm{SFR}_i/\textrm{SFR}_{i+1})=0$), thus producing a flat SFH with mass-weighted age $\langle t \rangle_\textrm{m} \approx 1.89\ \textrm{Gyr}$, while Galaxy B represents a galaxy whose SFH parameters are located far from the center of the distributions (i.e. $\log_{10}(\textrm{SFR}_i/\textrm{SFR}_{i+1})=-2.5$), in this case producing a galaxy with a rapidly declining SFH with $\langle t \rangle_\textrm{m} \approx 3.57\ \textrm{Gyr}$. Both galaxies are generated at $z=1.75$, where $t_\textrm{univ} \approx 3.77\ \textrm{Gyr}$.

\begin{figure*}
    \plottwo{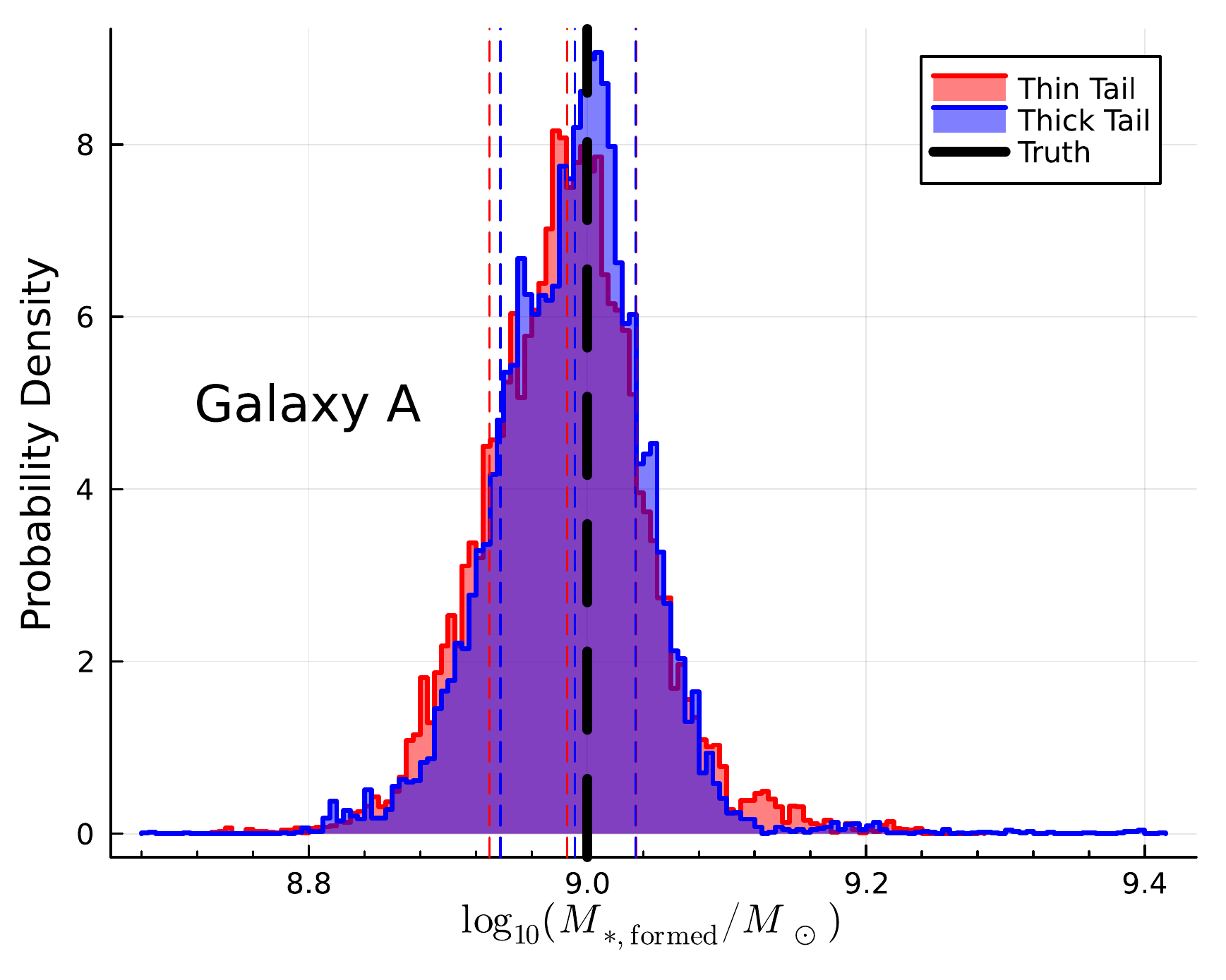}{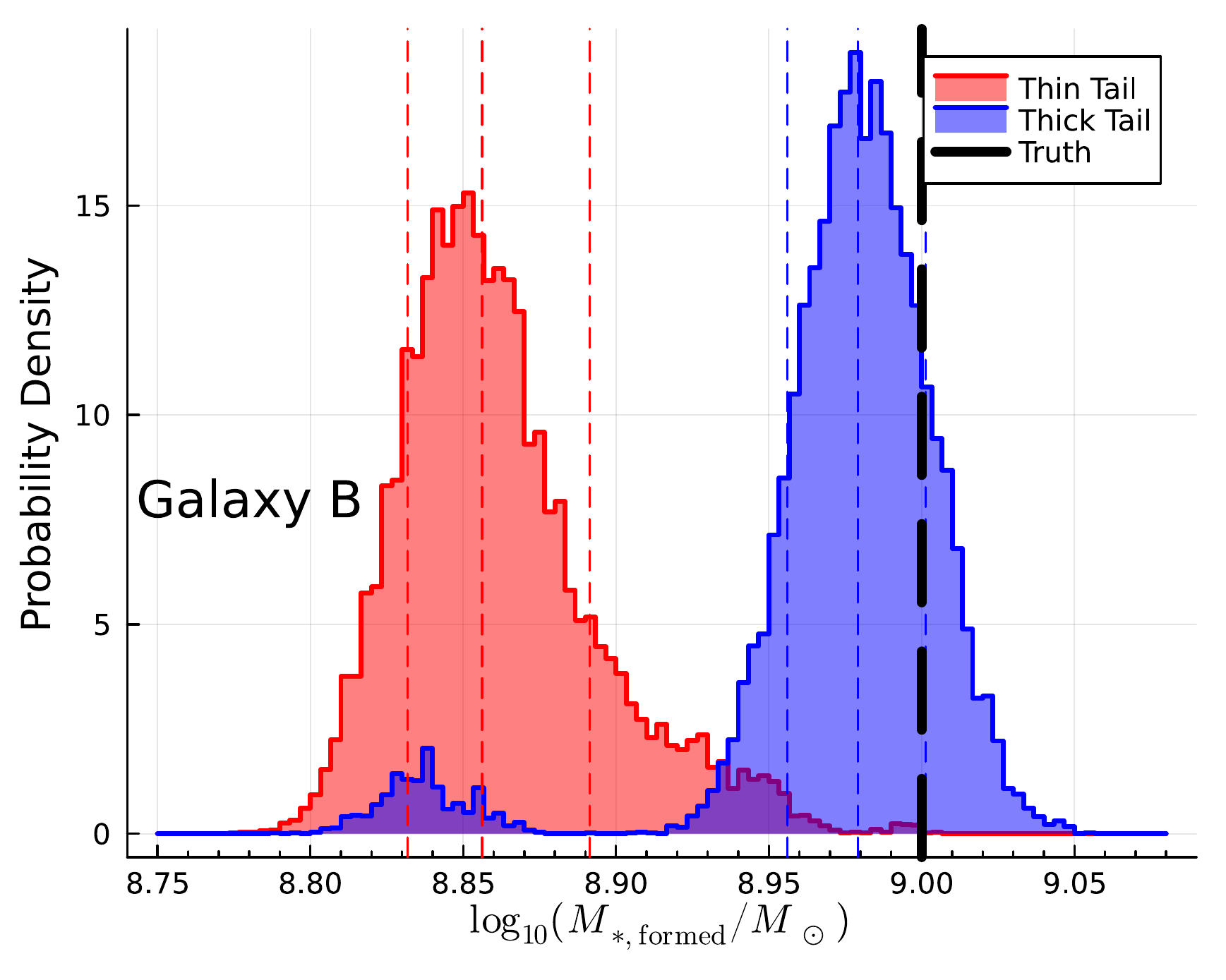}
    \plottwo{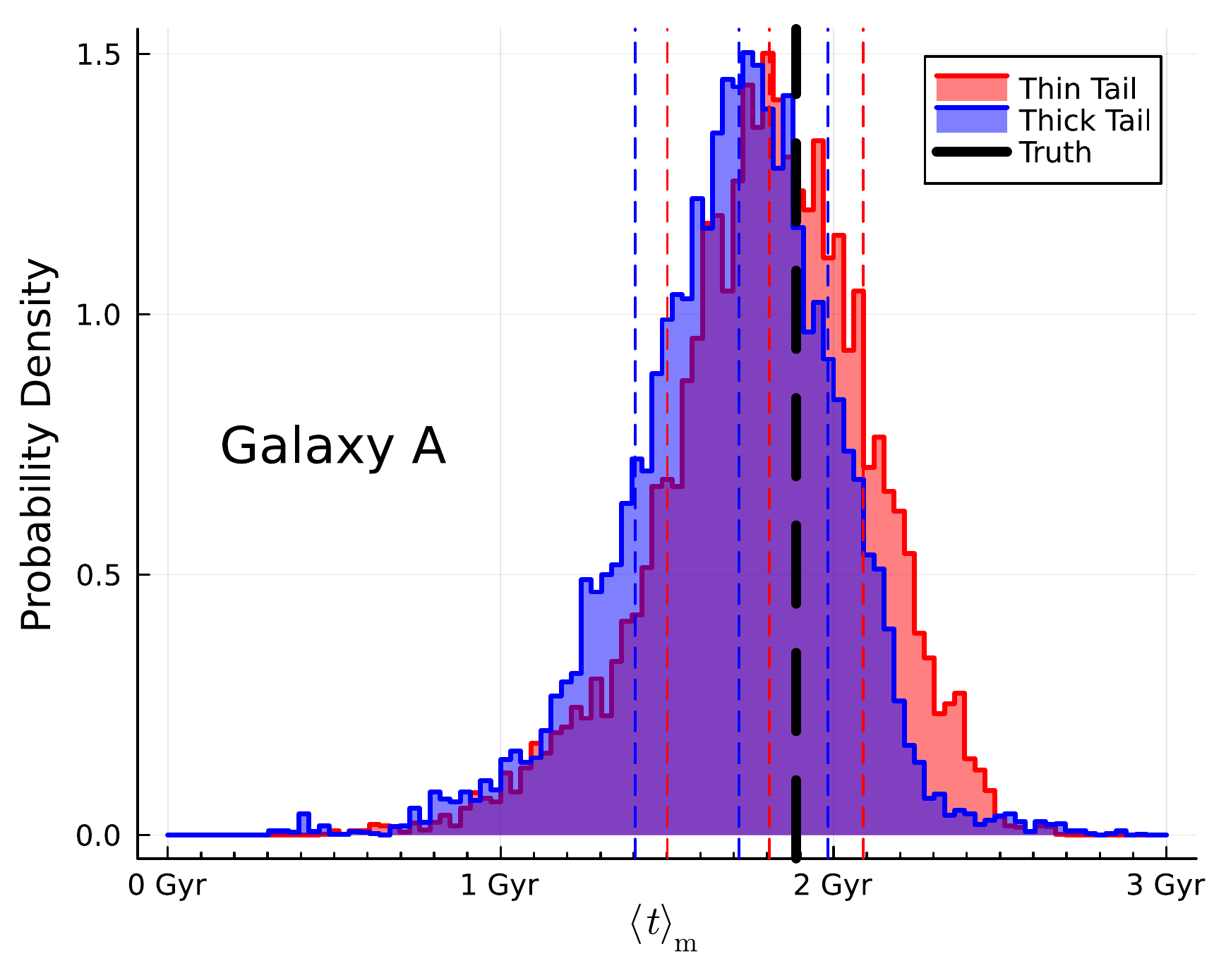}{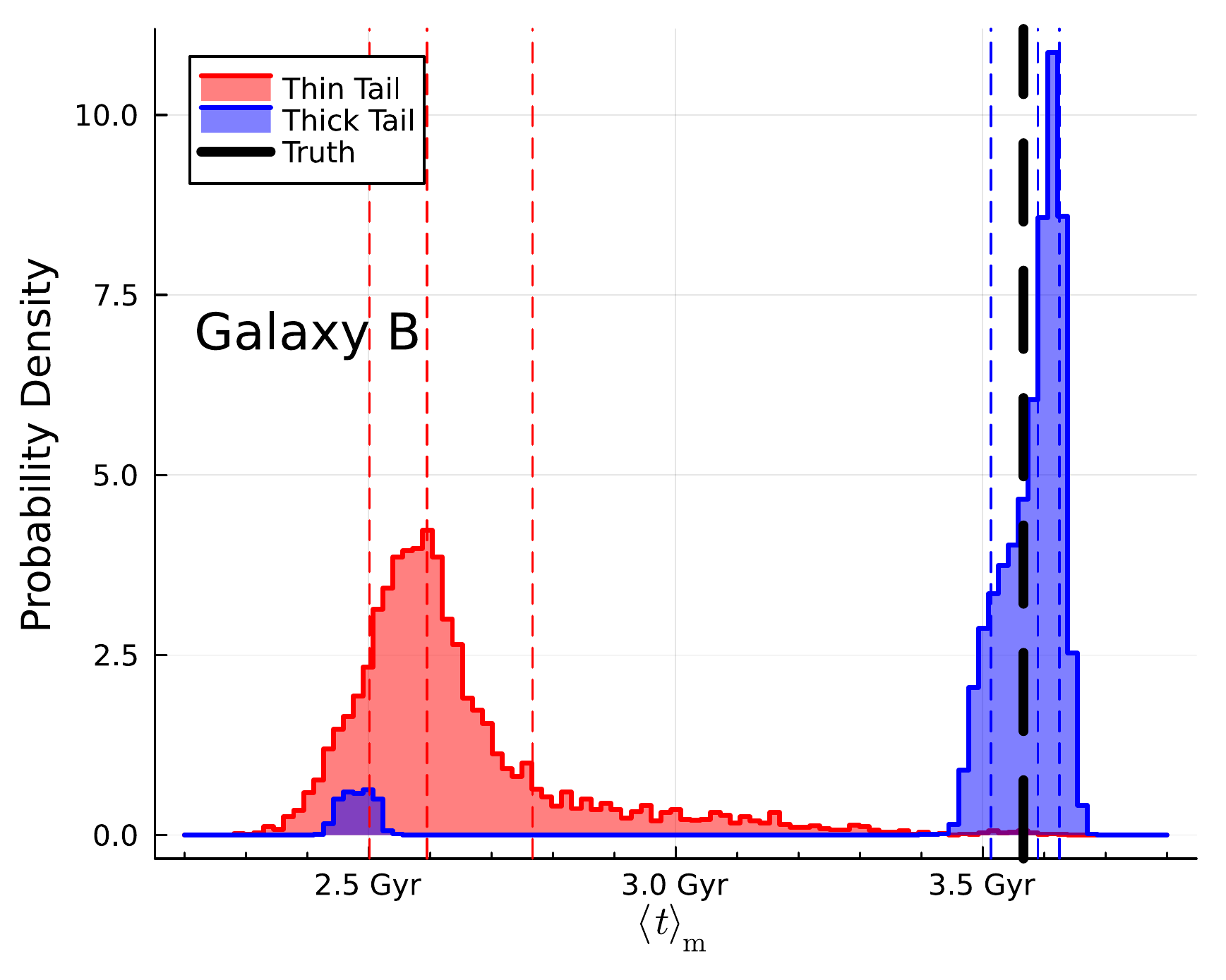}
    \caption{The posteriors for the fits to Galaxies A and B as described in Section \ref{sec:dists}. Left/right panels show the posteriors for the fits to Galaxy A/B for the total stellar mass formed $\log_{10}(M_{*,\textrm{formed}}/M_\odot)$ and the mass-weighted age $\langle t \rangle_\textrm{m}$. Thick dashed black line denotes the true parameter value, while the thin colored dashed lines show the median of the posterior along with the $\pm 1 \sigma$ quantiles.}
    \label{fig:extremepost}
\end{figure*}

We then fit these observations using the same procedure as before, but in this case we also fit for redshift using a prior distribution $\textrm{Uniform}(0.5, 3.0)$ to increase the sensitivity to small emulation errors due to the additional degeneracies redshift introduces. Furthermore, in addition to fitting the mock observations with the 1024 node-per-layer emulator described in this work (denoted as ``Thick Tail'' in Figures \ref{fig:extremephot} and \ref{fig:extremepost}), we also fit the observations with a similar 512 node-per-layer emulator\footnote{The reason the same architecture was not used for the ``Thin Tail'' emulator is due to the fact that adjusting the training set distributions will affect the emulation precision after training. In this case, this 512 node-per-layer emulator provided test set precision estimates that were similar to the $1024$ node-per-layer emulator.} whose training set used the narrower \emph{prior} distributions listed in Table \ref{tab:physmod} rather than the wider training set distributions (denoted as ``Thin Tail''). Figure \ref{fig:extremephot} shows the true emulation errors for each of these emulators when evaluated at the true mock parameters. For both galaxies, we see that the thin tail emulator is less precise than the thick tail emulator; however, this effect is far more pronounced for Galaxy B, which lies in a region of parameter space where the ``Thin Tail'' emulator's training set is poorly sampled, with emulation errors in excess of $0.1\ \textrm{mag}$.

We then see in Figure \ref{fig:extremepost} the effect this degradation in emulation precision can have on an SED fit. For Galaxy A, both emulators managed to accurately recover the galaxy's true stellar mass formed and mass-weighted age. However, the ``Thin Tail'' emulator did not fare as well with Galaxy B as the ``Thick Tail'' emulator did, with posteriors that exclude the mock truth with high probability. As such, we conclude that it is important to strongly consider the distributions used when generating a emulator's training set, and to be aware of the unintended consequences that using physical prior distributions can have on SED fitting results.

\subsection{Emulating Difficult Parameters}\label{sec:zred}

\begin{figure*}
    \plotone{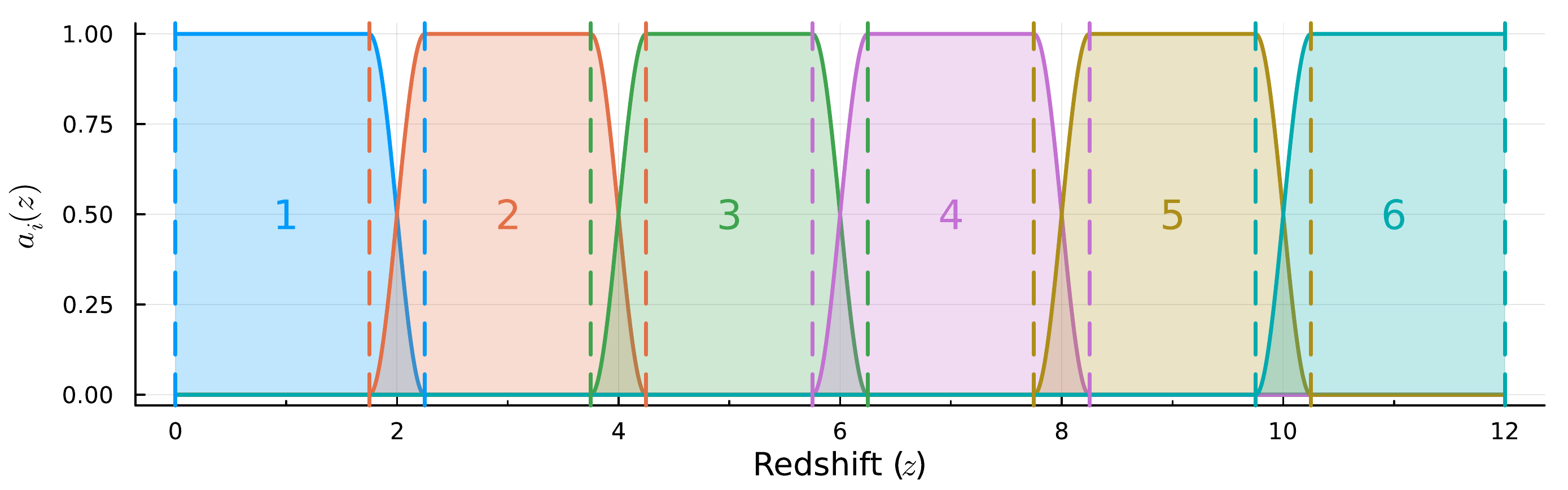}
    \caption{An example of the emulator ``stitching'' method proposed in Section \ref{sec:zred} to create a composite emulator spanning the redshift range $0 \leq z \leq 12$. To construct the composite emulator, 6 separate emulators are trained on slightly overlapping redshift domains (whose limits are represented by dashed lines), and they are then given a coefficient $a_i(z)$ to represent the contribution they should make to the overall flux estimate. Each emulator in this example covers a nominal redshift domain of $\Delta z \approx 2.5$, similar to the redshift domain covered by the emulator in this work and thus would likely have similar emulation precision and execution time.}
    \label{fig:stitch}
\end{figure*}

In our tests, one problem we encountered was the difficulty in expanding the redshift training/test set distributions. This doesn't necessarily come as a surprise -- an emulator trained on a training set where redshift is held to a fixed value only needs to learn how the rest-frame spectral flux density in each selected filter (with filter transmission curves blueshifted to the rest-frame) changes as a function of each of the input parameters, and regions of the rest-frame SED not well covered by the selected filters' transmission curves can even be ignored entirely.\footnote{This is by no means the exact logic that is used to train the emulator in either case; the logic used during training is whatever logic minimizes the loss function. However, is indicative of the simplifications that are present when training on a fixed redshift training set (or a training set sampled from a relatively narrow distribution of redshifts) and the complexities that are present on a training set where redshift is allowed to vary.} However, an emulator trained on a training set where redshift is allowed to vary must instead learn how the rest-frame spectrum (either in full or in large portions) changes as a function of parameters, along with the characteristics of each filter's transmission curve. This task necessarily becomes more difficult as the redshift domain an emulator is trained on is widened. Furthermore, this is a problem that is specific to photometric emulators, as most of the effects of redshift can be applied independently in the case of spectroscopic emulators (i.e. the emulator would predict the rest-frame spectrum, although the SFH definition used in our physical model is dependent on the age of the universe at a given redshift, and thus redshift would still need to be included as an input). Therefore, expanding the redshift domain will typically result in loss in emulation precision, or alternatively a longer execution time if a wider emulator architecture is selected to compensate for the loss in precision.

To combat these problems, we propose a method of ``stitching'' separate emulators trained on a variety of redshift domains into a single composite emulator. Suppose we have trained $K$ emulators across various overlapping redshift domains, such that emulator $i$ is trained on the redshift domain $z_{\textrm{L},i} \leq z \leq z_{\textrm{H},i}$, the emulators collectively cover the redshift domain $z_{\textrm{L}} \leq z \leq z_{\textrm{H}}$, and a maximum of two emulators overlap at any given redshift. Then, we can define a transition angle $\theta_{i}(z)$,
\begin{equation}
    \theta_i(z) = \frac{\pi}{2}\frac{z - z_{\textrm{L},i+1}}{z_{\textrm{H},i} - z_{\textrm{L},i+1}},
\end{equation}
where we also define $z_{\textrm{H},0}=-\infty$ and $z_{\textrm{L},K}=+\infty$ in order to handle the cases at the edges of the redshift domain. We then wish to generate a coefficient $a_i(z)$ representing the contribution emulator $i$ should give to the total spectral flux density estimate,
\begin{equation}
    a_{i}(z) = \begin{cases}
    0 & (z > z_{\textrm{H},i}) \parallel (z < z_{\textrm{L},i}),  \\
    1 & z_{\textrm{H},i-1} \leq z \leq z_{\textrm{L},i+1}, \\
    \sin^2{\theta_{i-1}(z)} & z_{\textrm{L},i} \leq z < z_{\textrm{H},i-1}, \\
    \cos^2{\theta_{i}(z)} & z_{\textrm{L},i+1} < z \geq z_{\textrm{H},i}. \\
    \end{cases}
\end{equation}
This coefficient scheme has a few desirable characteristics for use with emulators. Firstly, it provides the ability to gracefully handle overlapping emulation domains, where the square sine and cosine functions are used to let two emulators ``vote'' against one another such that $\sum_{i=1}^{K} a_i(z) = 1$. Furthermore, it provides a continuous derivative as a function of redshift, which can be useful with gradient-based samplers (see Section \ref{sec:gradsamp}). However, it must be stressed that the overlaps must be selected such that no more than 2 coeffients are nonzero at any given redshift (i.e. only two emulators may overlap at any given redshift). Futhermore, due to the requirement to run two emulators for redshifts within overlaps, this method produces a slight increase in execution times, which increases as the overlap widths increases and the number of overlaps increases.

\begin{deluxetable}{cccc}[hb!]\label{tab:examplestitch}
\tablecaption{Example $0 \leq z \leq 12$ Composite Emulator}
\tablehead{\colhead{Emulator ($i$)} & \colhead{Lower Limit ($z_{\textrm{L},i}$)} & \colhead{Upper Limit ($z_{\textrm{H},i}$)} & \colhead{Width ($\Delta z_i$)}}
\startdata
1 & 0.00 & 2.25 & 2.25 \\
2 & 1.75 & 4.25 & 2.50 \\
3 & 3.75 & 6.25 & 2.50 \\
4 & 5.75 & 8.25 & 2.50 \\
5 & 7.75 & 10.25 & 2.50 \\
6 & 9.75 & 12.00 & 2.25 \\
\enddata
\end{deluxetable}

The total spectral flux density $E_\nu(z,\ldots)$ can then be computed as a linear combination of the individual emulator outputs $E_{\nu,i}(z,\ldots)$,
\begin{equation}
    E_{\nu}(z,\ldots) = \sum_{i=1}^{K} a_i E_{\nu,i}(z,\ldots),
\end{equation}
and in Figure \ref{fig:stitch}, we demonstrate one potentially useful application of this technique. Following the launch of JWST, there has been a deluge of candidate high-redshift, high-mass galaxies discovered \citep[e.g.][]{naidu22,atek23,labbe23}. This has led for a demand to rapidly parse incoming photometric observations for new candidates, and to distinguish high-redshift, high-mass candidates from low-redshift, low-mass interlopers (e.g. by incorrectly interpreting the Balmer break as the Lyman break). This task would require an emulator that covers a far wider redshift range than the one used in this work (i.e. $\Delta z = 3.0 - 0.5 = 2.5$), and would thus be a good candidate for constructing a ``stitched'' emulator. Shown in Figure \ref{fig:stitch} is an example of a composite emulator which collectively covers a redshift domain of $0 \leq z \leq 12$, which would allow it to fit a large number of the candidate high-redshift, high-mass candidates found to date. Six individual emulators are used in this example (with limits and widths shown in Table \ref{tab:examplestitch}), each of which is trained on a $\Delta z \approx 2.5$ redshift domain, with overlap widths of $\Delta z = 0.5$. Assuming a sampler uniformly explores all redshifts within the composite domain, this choice of overlaps would result in a typical increase in execution time of $\sim\!21\%$ compared to running any of the individual emulators. Additionally, since the width of each emulator's redshift domain is similar to the emulator used throughout this work, it is likely each emulator's emulation precision and execution time would be similar to those found in Section \ref{sec:apriori}. For demonstration purposes, the maximal redshift covered by this scheme is $z=12$, but this could be extended to any higher redshift as desired by appending additional emulators covering higher redshift domains.

\subsection{Gradient-Based Sampling}\label{sec:gradsamp}

As described in Section \ref{sec:photdata}, the sampler used in this work to perform the SED fits with \texttt{Prospector} was \texttt{dynesty}, a nested sampler that uses MCMC sampling. While it has been demonstrated that this is a suitable sampling method for fitting galaxy SEDs \citep[e.g.][]{leja19}, it is expected that the computational burden of SED fitting when using an emulator may be reduced by using a gradient-based sampler instead \citep{neal11,betancourt17}, such as Hamiltonian Monte Carlo \citep[HMC,][]{duane87} or the No-U-Turn Sampler \citep[NUTS,][]{hoffman11}. ANNs are typically implemented in libraries compatible with automatic differentiation (AD), which can provide exact evaluations of an emulator's derivative with a computational cost comparable to a hard-coded derivative, making them a prime candidate for being paired with a gradient-based sampler. However, one should also approach this method with caution -- gradients are not accounted for in the loss function used to train the emulators in this work.

Alternatively, when using a gradient-based sampler, it may not be necessary to resort to an ANN emulator to handle SPS. Instead, there has been recent work to create a \emph{differentiable} SPS code, one which produces identical predictions to a code like \texttt{FSPS} (unlike an emulator, which is approximate) but does so in a way that can exploit the AD capabilities of packages like \texttt{JAX} \citep{bradbury18} in Python or \texttt{ForwardDiff.jl} \citep{revels16} in Julia. One such implementation is \texttt{DSPS} \citep{hearin23}, which can generate predicted SEDs with roughly the same execution time as ANN emulators. However, \texttt{DSPS} is currently only capable of generating SEDs using a precomputed SSP spectral grid, and bringing it to feature-parity with \texttt{FSPS} will likely result in either a slower execution time (in the case of using stellar spectral libraries and isochrones as the base) or an extremely high memory requirement (in the case of using high-fidelity precomputed SSP grids). Regardless, the ability to explore a posterior distribution more efficiently by pairing a differentiable SPS code with a gradient-based sampler would likely prove to provide a net improvement in SED-fitting execution times.

\section{Conclusions}\label{sec:conclusion}

In this work, we have evaluated the performance of SPS emulators with six different ANN architectures, with evaluations in both flux space and parameter space, to assess the relationship between emulator architecture, precision, and execution time, including testing each emulator's ability to recover physical parameters from observed galaxy SEDs. From our analyses, we have drawn a number of conclusions:

\begin{enumerate}
    \item ANN emulators dramatically reduce the computational burden of SPS, with an execution time improvement upper limit of $\sim 10^5$ compared to their native SPS counterparts on a CPU. Actual improvement will be less than this, dependent on the ANN architecture required to reproduce fluxes to within the target precision (with more complex architectures requiring longer execution times), but most architectures that provide reasonably accurate and precise emulators are at least $\sim\! 10^3$ times faster. In addition, we find that emulation execution time scales with respect to network width as $\propto N^2$. Simpler architectures are preferred in this respect.
    \item ANN emulators can accurately, precisely, and quickly match the predictions given by existing SPS codes to well within the typical observational uncertainties encountered in real-world data, with emulation precision improving as the number of nodes per layer in the ANN emulator increases. Emulators in this work have precisions ranging from $\sim\!0.2\ \textrm{mag}$ at worst to $\sim\!0.005\ \textrm{mag}$ at best. Furthermore, we find that emulation precision tends to scale with respect to network width as $\propto N^{-1}$.  More complex architectures are preferred in this respect.
    \item ANN emulators typically have less correlated emulation errors when using more complex architectures, allowing users to safely convolve observational uncertainties with estimated emulation precision without requiring a full covariance matrix.
    \item Emulators are capable of recovering posterior medians that are consistent with fits run with traditional SPS codes, with no apparent bias and typical differences only of order $25-40\%$ for stellar mass, stellar metallicity, star formation rate, and stellar age.
    \item Based on a set of $\sim\!10^4$ SED fits to observed galaxies in the deep extragalactic COSMOS field, we find that posteriors tend to coalesce to a common result (within sampling variance) as the emulators' precision increases. Convergence begins to occur roughly at the same architecture where we see the scale of observational errors exceeds the scale of typical emulation errors and where we see emulation errors' correlations between photometric filters become insignificant, setting a good benchmark for a minimal architecture for future studies that require the use of emulators.
    \item By fitting a sample of $10^3$ mock observations using both an emulator and \texttt{FSPS}, we show that both emulators and \texttt{FSPS} are able to recover the true physical parameters for a galaxy to within the same level of precision.
    \item The distributions chosen for generating emulator training sets can have large effects on an emulator's ability to fit rare objects. Choosing a non-uniform distribution leads to inaccurate flux predictions in regions of parameter space that are poorly sampled in the training set, which then leads to inaccurate posteriors when fitting SEDs. Using the same distributions for the training set as are used as a prior leads to this exact scenario, and is especially problematic since the most interesting objects tend to reside in rare regions of parameter space.
\end{enumerate}

In sum, these conclusions pave the way for ANN-based emulators to be used in real-world SED-fitting applications. We demonstrate that not only can photometric emulators dramatically speed up the rate at which we can perform SED-fitting, but we can also trust the results that these emulator-based fits provide. Useful avenues of further inquiry include optimizing the architectures used for SPS emulation (e.g. convolutional layers) in addition to extending this analysis to high-resolution spectroscopic emulators. These results open the door to fitting large photometric surveys with SPS emulators to reduce computational costs, all the while being able to trust the results we receive from these emulators.

\begin{acknowledgments}

Based on observations with the NASA/ESA Hubble Space Telescope obtained from the Space Telescope Science Institute, which is operated by the Association of Universities for Research in Astronomy, Incorporated, under NASA contract NAS5-26555. Support for Program number AR-16146.008-A was provided through a grant from the STScI under NASA contract NAS5-26555. Computations for this research were performed on the Pennsylvania State University's Institute for Computational and Data Sciences' Roar supercomputer. KW acknowledges support for this work from the Astrophysics Data Analysis Program under NASA grant 80NSSC20K0416.

\end{acknowledgments}

\software{
Astropy \citep{astropy13},
dynesty \citep{speagle20,skilling04,skilling06,speagle22},
Flux.jl \citep{innes18a,innes18b},
FSPS \citep{conroy09,conroy10},
JAX \citep{bradbury18},
NumPy \citep{harris20},
Plots.jl \citep{breloff22},
Prospector \citep{leja17,johnson21b},
python-fsps \citep{johnson21c},
sedpy \citep{johnson21a},
}

\bibliography{bib}
\bibliographystyle{aasjournal}

\end{CJK*}
\end{document}

%% file: authors.tex
\author[0000-0003-0384-0681]{Elijah P. Mathews}
\affiliation{Department of Astronomy \& Astrophysics, The Pennsylvania State University, University Park, PA 16802, USA}
\affiliation{Institute for Computation \& Data Sciences, The Pennsylvania State University, University Park, PA 16802, USA}
\affiliation{Institute for Gravitation and the Cosmos, The Pennsylvania State University, University Park, PA 16802, USA}

\author[0000-0001-6755-1315]{Joel Leja}
\affiliation{Department of Astronomy \& Astrophysics, The Pennsylvania State University, University Park, PA 16802, USA}
\affiliation{Institute for Computation \& Data Sciences, The Pennsylvania State University, University Park, PA 16802, USA}
\affiliation{Institute for Gravitation and the Cosmos, The Pennsylvania State University, University Park, PA 16802, USA}

\author[0000-0003-2573-9832]{Joshua S. Speagle (\begin{CJK*}{UTF8}{gbsn}沈佳士\ignorespacesafterend\end{CJK*})}
\affiliation{Department of Statistical Sciences, University of Toronto, 9th Floor, Ontario Power Building, 700 University Ave, Toronto, ON M5G 1Z5, Canada}
\affiliation{David A. Dunlap Department of Astronomy \& Astrophysics, University of Toronto, 50 St George Street, Toronto ON M5S 3H4, Canada}
\affiliation{Dunlap Institute for Astronomy \& Astrophysics, University of Toronto, 50 St George Street, Toronto, ON M5S 3H4, Canada}
\affiliation{Data Sciences Institute, University of Toronto, 17th Floor, Ontario Power Building, 700 University Ave, Toronto, ON M5G 1Z5, Canada}

\author[0000-0002-9280-7594]{Benjamin D. Johnson}
\affiliation{Center for Astrophysics $\vert$ Harvard \& Smithsonian, 60 Garden Street, Cambridge, MA 02138, USA}

\author[0000-0003-1903-9813]{Justus Gibson}
\affiliation{Department for Astrophysical and Planetary Science, University of Colorado, Boulder, CO 80309, USA}

\author[0000-0002-7524-374X]{Erica J. Nelson}
\affiliation{Department for Astrophysical and Planetary Science, University of Colorado, Boulder, CO 80309, USA}

\author[0000-0002-1714-1905]{Katherine A. Suess}
\affiliation{Department of Astronomy and Astrophysics, University of California, Santa Cruz, 1156 High Street, Santa Cruz, CA 95064, USA}
\affiliation{Kavli Institute for Particle Astrophysics and Cosmology and Department of Physics, Stanford University, Stanford, CA 94305, USA}

\author[0000-0002-8224-4505]{Sandro Tacchella}
\affiliation{Kavli Institute for Cosmology, University of Cambridge, Madingley Road, Cambridge, CB3 0HE, UK}
\affiliation{Cavendish Laboratory, University of Cambridge, 19 JJ Thomson Avenue, Cambridge, CB3 0HE, UK}

\author[0000-0001-7160-3632]{Katherine E. Whitaker}
\affiliation{Department of Astronomy, University of Massachusetts, Amherst, MA 01003, USA}
\affiliation{Cosmic Dawn Center (DAWN), Copenhagen, Denmark}

\author[0000-0001-9269-5046]{Bingjie Wang (王冰洁)}
\affiliation{Department of Astronomy \& Astrophysics, The Pennsylvania State University, University Park, PA 16802, USA}
\affiliation{Institute for Computation \& Data Sciences, The Pennsylvania State University, University Park, PA 16802, USA}
\affiliation{Institute for Gravitation and the Cosmos, The Pennsylvania State University, University Park, PA 16802, USA}